\documentclass{article}

\usepackage{arxiv}

\usepackage[utf8]{inputenc} % allow utf-8 input
\usepackage[T1]{fontenc}    % use 8-bit T1 fonts
\usepackage{hyperref}       % hyperlinks
\usepackage{url}            % simple URL typesetting
\usepackage{booktabs}       % professional-quality tables
\usepackage{amsfonts}       % blackboard math symbols
\usepackage{nicefrac}       % compact symbols for 1/2, etc.
\usepackage{microtype}      % microtypography
\usepackage{lipsum}
\usepackage{graphicx}
\usepackage{amssymb, amsmath}
\usepackage{color}
\usepackage{url}
\usepackage{comment}
\usepackage{afterpage}
\usepackage{here}
\usepackage{mathrsfs}
\newcommand{\reffig}[1]{Figure~\ref{#1}}

\newcommand{\reftab}[1]{Table~\ref{#1}}
\newcommand{\refeq}[1]{\eqref{#1}}
\newcounter{alnum}

\newcounter{rmnum}

\newcounter{muni}

\renewcommand{\hat}[1]{\widehat{#1}}

\renewcommand{\Re}{\ensuremath{\mathbb{R}}}

\newcommand{\bi}[1]{\ensuremath{\boldsymbol{#1}}}

\newcommand{\NC}{\ensuremath{\mathcal{N}}}

\title{Generating various airfoil shapes with required lift coefficient using conditional variational autoencoders}

\author{
	Kazuo Yonekura \thanks{\texttt{yonekura@struct.t.u-tokyo.ac.jp}}\\
	Department of Systems Innovasion\\
	The University of Tokyo\\
	Tokyo, JAPAN 113-8656 \\
	\And
	Kazunari Wada \\
	Department of Systems Innovasion\\
	The University of Tokyo\\
	Tokyo, JAPAN 113-8656 \\
	\AND
	Katsuyuki Suzuki \\
	Department of Systems Innovasion\\
	The University of Tokyo\\
	Tokyo, JAPAN 113-8656 \\
}

\begin{document}
	\maketitle
	
	\begin{abstract}
		Multiple shapes must be obtained in the mechanical design process to satisfy the required design specifications. 
		The inverse design problem has been analyzed in previous studies to obtain such shapes. However, finding multiple shapes in a short computation period is difficult while using the conventional methods.
		This paper proposes the use of the conditional variational autoencoders (CVAE) with normal distribution, denoted by $\mathcal{N}$-CVAE, along with the von Mises-Fischer distribution, denoted by $\mathcal{S}$-CVAE,  to find multiple solutions for the inverse design problems. 
		Both the CVAE models embed shapes into a latent space. The $\mathcal{S}$-CVAE  enables the separation of data in the latent space, whereas the $\mathcal{N}$-CVAE embeds the data in a narrow space. 
		These different features are used for various tasks in this study. 
		In one of the tasks, the dataset consists of only one type of data and generates similar airfoils. Here, $\mathcal{S}$-CVAE outperforms $\mathcal{N}$-CVAE because it can separate the data. 
		Another task involves combining different types of airfoils and generating new types of data. $\mathcal{N}$-CVAE is useful in this instance since it embeds different shapes in the same latent area, due to which, the model outputs intermediate shapes of different types.
		The shape-generation capability of $\mathcal{S}$-CVAE and $\mathcal{N}$-CVAE are experimentally compared in this study.
	\end{abstract}

	% keywords can be removed

\section{Introduction}
A mechanical design must primarily satisfy the given design specifications. The mechanical design process can be divided into two phases: rough design and tuning.
In the rough design process, rough drawings are produced by human designers, often considering trade-off relationships. One candidate is then chosen and the process of tuning the shape is initiated. 
The rough design process incurs a large amount of labor cost, and the result depends on the experience of the designer. 
A candidate shape must be obtained in a short period of time. 
This task is known as an inverse design problem. 

Inverse analysis and design have gained considerable attention in the recent years for application in fluid machinery such as turbines, pumps, and wings \cite{Bonaiuti10,Goto02,Kennon85,Cao05,Zhengming85}. 
The numerical sensitivity analysis is one of the methods of inverse design \cite{Narducci95,Tortorelli94}. However, the sensitivity-based methods require a long computation time. 
The principal component analysis (PCA) method requires less computation time and can be categorized as a data-driven approach. 
The PCA-based dimension reduction approach is used for design exploration in fields such as aerospace \cite{YW14} and marine engineering \cite{Gaggero19}.
Bui-Thanh et al. \cite{Thanh04} also utilized the PCA method for the inverse design of flow fields.
Lee et al. \cite{Lee10} utilized a dimension-reduction method for reliability analysis and optimization.
This method analyzes and extracts the features from various shapes and designs. However, it is difficult to consider the correlations such as the airfoil shape and the lift coefficient, between the shape features and the aerodynamic performance. 
A performance-driven design was therefore proposed in the architecture \cite{Brown19} to consider the correlations between the performance and the design parameters. 
However, the methods mentioned above do not provide an exact design which satisfies the specifications, due to which, multiple designs must be explored based on the additional analyses. 
Therefore, the shapes which satisfy the specified requirements must be directly obtained. 

A new type of shape can be obtained by using the PCA-based methods even though machine learning generally performs the interpolation of data. Nita et al. \cite{Nita14} constructed a dataset with different shapes and generated a low-dimensional linear space by applying the PCA method to the dataset. A new type of shape \cite{Patent} was then obtained from the linear space. 
In this example, the dataset consists of different types of data located far away from each other in the design space. The new type of data is located between these data. It is essentially a combination of different types of data. 
It is desirable to obtain novel shapes from an application point of view, due to which, the design methods must be capable of handling multiple types of data. 

The use of deep neural networks to generate shapes from data has gained interest in the recent years.
\cite{Achour20} proposed the generative adversarial network (GAN) model to generate shapes with low or high lift coefficient. 
\cite{Chen21} proposed the B{\'e}zierGAN to obtain smooth airfoils. 
\cite{Yonekura21a} proposed the conditional variational autoencoder (CVAE) to output smooth shapes with specific lift coefficient. 
However, the variations in the generated shapes and the combination of different types of airfoils have not been analyzed in the previous studies. 
In this study, two types of CVAE models have been used to generate various types of airfoil shapes. 
The first is a CVAE model with a normal distribution of $\mathcal{N}$-CVAE, which is an ordinal CVAE model. 
The other is a CVAE model with a von Mises-Fischer (vMF) distribution, which is called $\mathcal{S}$-CVAE. 

In the recent years, machine learning, particularly deep learning, has considerably improved in various fields, such as image recognition and signal processing. 
The capability of the feature extraction methods have also been improved based on deep learning.
Deep neural network has applied to mechanical design in several studies \cite{Li20,Yonekura18,Yu19,Kai20,Zhang19a}. 
One of the deep neural network models is the autoencoder (AE), which is a dimension reduction method, and has been successfully used for image processing \cite{Hinton06}.
An AE and a variational AE (VAE) \cite{Kingma13} 
are composed of neural networks; they extract the features from the input data and represent the data as small-dimensional latent vectors. 
The VAEs are used in various fields, such as image processing \cite{Pu16} and anomaly detection \cite{HXu18}. 
A conditional VAE \cite{Sohn14} adds nodes to the VAE architecture used to feed the labels with input; the CVAE extracts the features while considering the labels. 
The VAEs and the CVAEs are referred to as generative models because they generate new data and images that do not exist in the real world. 
These generative models are mainly studied in computer science and are used for image processing \cite{Tolstikhin17}. 

The ordinal VAE model has been observed to produce the same result even if the latent variables are different. This issue is called the posterior collapse or the Kullback-Leibler (KL) collapse. 
\cite{Hasnat17,HSVAE} proposed a hyperspherical VAE ($\mathcal{S}$-VAE) to resolve the KL collapse issue. $\mathcal{S}$-VAE uses a von Mises-Fisher distribution rather than a normal distribution in the latent space. 

The proposed method can be used by a designer to compare the airfoils with different specifications but with the same loss coefficient to obtain a rough idea of the design or to consider the trade-off relationships. Generally, a considerable amount of effort is required to  design each shape. This issue can be resolved by using the CVAE; the designers can also produce different shapes in a short period of time with this method.
The designers require a wide variety of shapes to consider the trade-off relationships. Additionally, novel shapes which are different from the training dataset are required to obtain a novel concept of a product. 

Two design tasks are discussed in this paper. 
Firstly, a single type of airfoil dataset, called the NACA airfoil is used, and a wide variety of shapes that indicate the required $C_{\rm L}$ are obtained. 
Secondly, two types of airfoil data, that is, the NACA and the Joukowski airfoils, are used to obtain shapes that are different from both the NACA and Joukowski airfoils. 
$\mathcal{S}$-CVAE and the $\mathcal{N}$-CVAE model are used to perform these tasks.

The remainder of this paper is organized as follows. Section 2 introduces the CVAE model Section 3 describes the shape generation method. The proposed method is applied to the NACA airfoil data in Section 4. In Section 5, the proposed method is applied to a combination of the NACA and the Joukowski airfoils. Lastly, Section 6 concludes the paper.

\section{Machine learning model}\label{sec:VAEmodel}
\subsection{Conditional variational autoencoder}
\begin{figure}[h]
	\begin{center}
		\begin{minipage}[h]{0.5\textwidth}
			\begin{center}
				\includegraphics[width=\linewidth]{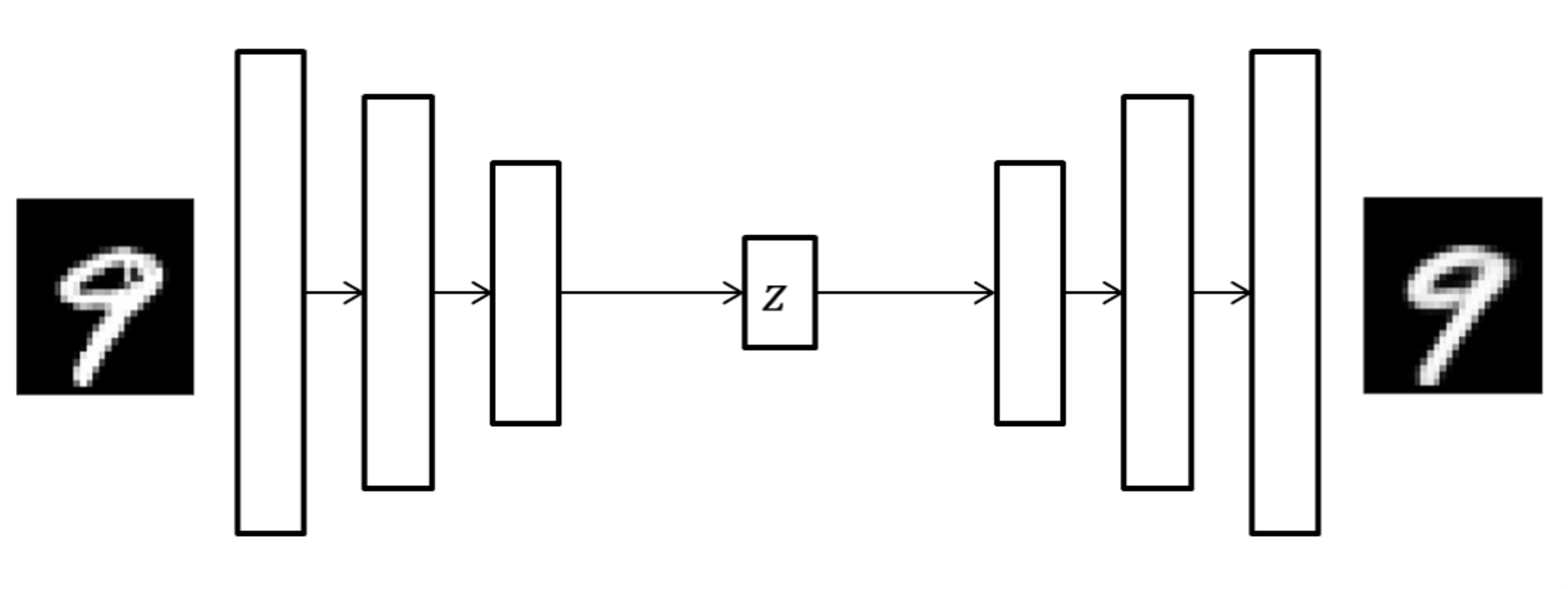}\par {(a) AE}
			\end{center}
		\end{minipage}%
		\par
		\begin{minipage}[h]{0.5\textwidth}
			\begin{center}
				\includegraphics[width=\linewidth]{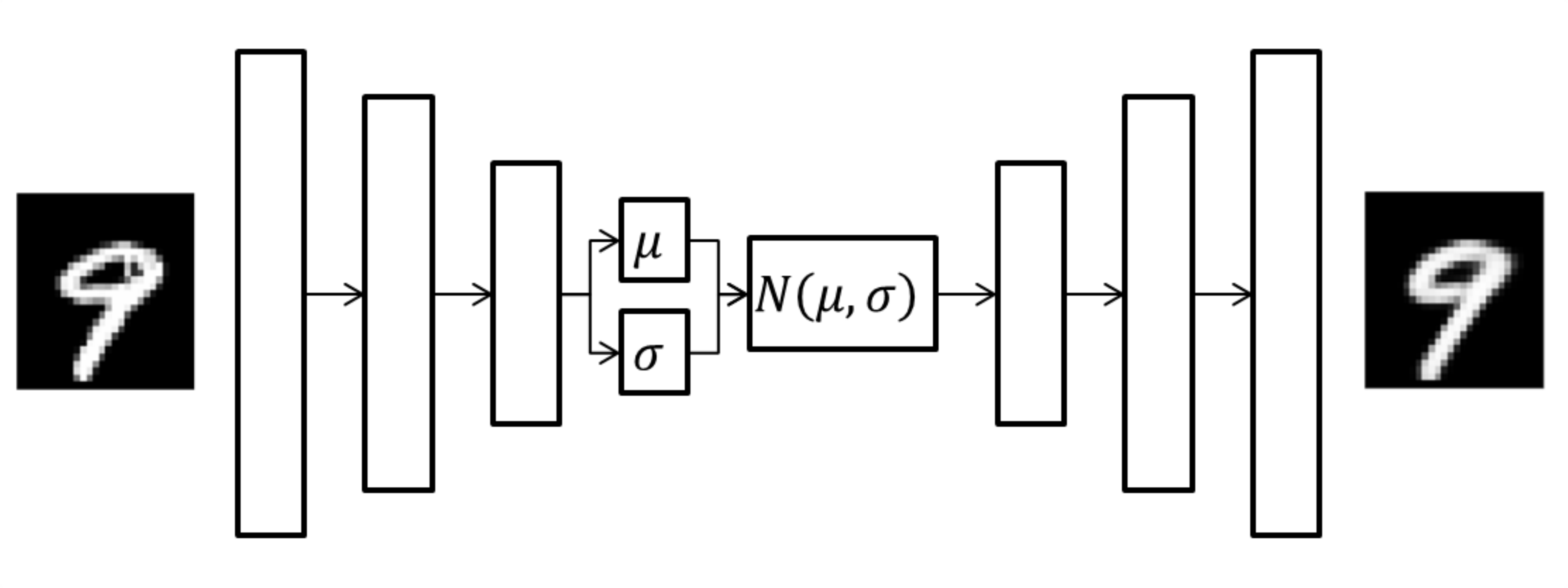}\par  {(b) VAE ($\mathcal{N}$-VAE)}
			\end{center}
		\end{minipage}%
		\par
		\begin{minipage}[h]{0.5\textwidth}
			\begin{center}
				\includegraphics[width=\linewidth]{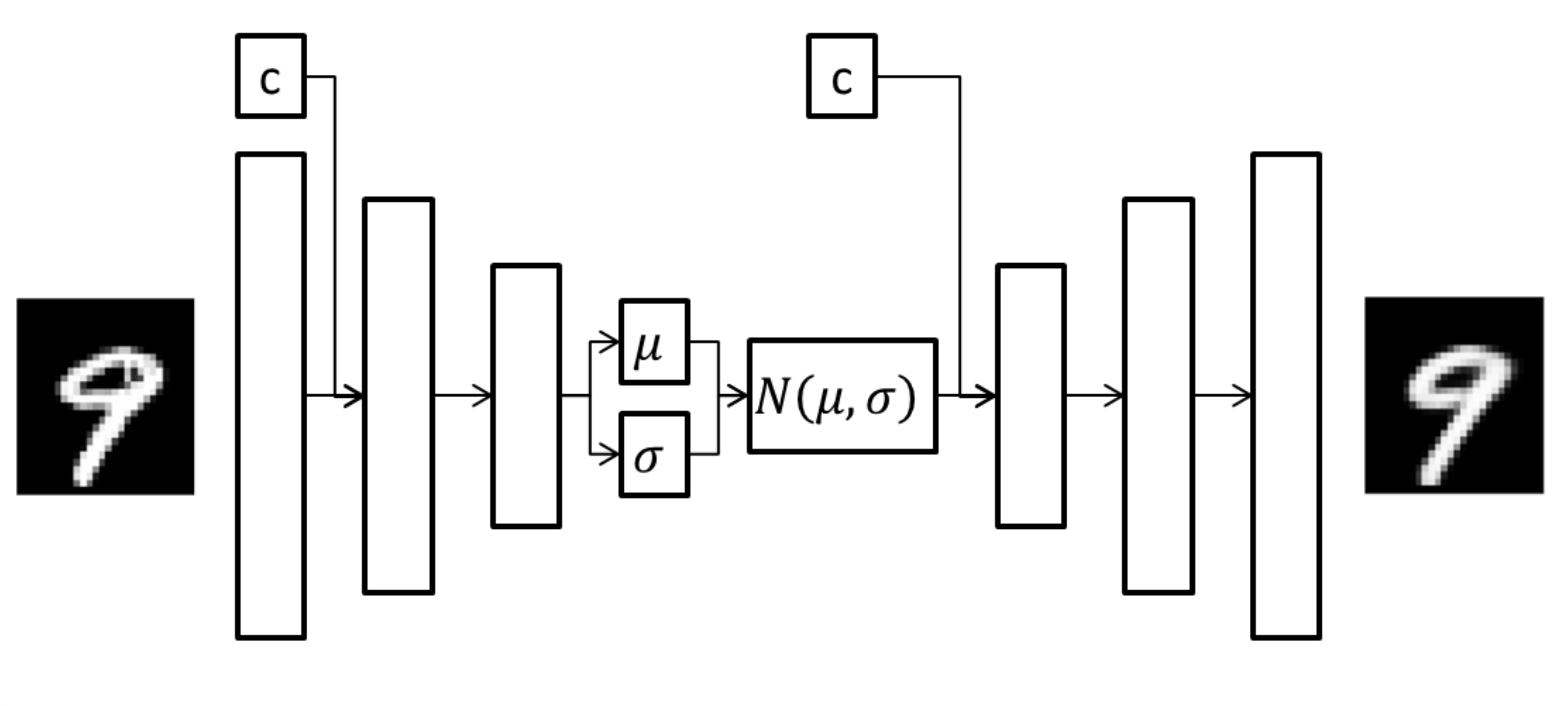}\par	{(c) CVAE ($\mathcal{N}$-CVAE)}
			\end{center}
		\end{minipage}%
		\caption{Schematics of AE, VAE, and CVAE.}
		\label{fig:AE}
	\end{center}
\end{figure}

The CVAE is a neural network model, which is an extension of an AE and a VAE, as illustrated in \reffig{fig:AE}. 
The AE consists of an encoder and a decoder; the encoder extracts the features from the input data and the decoder reconstructs the data from these features. 
Essentially, the encoder reduces the dimension of the data.
This feature is also known as a latent vector, and the space of the latent vector is called the latent space. 
The latent space in an AE is usually a Euclidean space, i.e., $\Re^d$. 
An AE model is trained to output the same data as the input. 
The objective function of the training is to minimize the reconstruction loss, $\mathcal{L}_{\rm rec} = \| \bi{x} - \bi{y} \|^2 $, where $\bi{x}$ and $\bi{y}$ represent the input and the output vectors, respectively.

The VAE uses probability distribution in the latent space and the ordinal VAE uses a normal distribution in the Euclidean space; therefore, the ordinal VAE is represented as $\mathcal{N}$-VAE. 
The normal distribution is implemented using the equation, $ \bi{x} = \bi{\mu} + \NC_d( \bi{0}, I ) \bi{\sigma} $, where $\NC_d( \bi{0}, I )$ is a $d$-dimensional normal distribution with the variance-covariance matrix, $I$.
$\bi{\mu} \in \Re^d$ is a mean vector, $\bi{\sigma} \in \Re^d$ is a variance vector, and $I$ is an identity matrix.
The VAE model is also trained to output the same data as the input, but its objective function differs from that of the AE. 
The objective function, $\mathcal{L}$, is the sum of the reconstruction loss and the KL divergence, $\mathcal{L}_{\rm KL} (p||q)$.
\begin{align}
	\mathcal{L} = \mathcal{L}_{\rm rec} + \mathcal{L}_{\rm KL}(p||q). \label{eq.loss}
\end{align} 
The KL divergence measures the difference between the two probability distributions, $p$ and $q$. In the VAE model, $p$ is the distribution of the latent vector and $q$ is the standard normal distribution.

In the CVAE model, a label node is added to the input of the encoder and the decoder of the VAE model. 
The same value is input at both the nodes when the model is trained.
The encoder reduces the dimension of input data with label, $c$, 
and the decoder reconstructs the data with the information of the label, $c$. 
The objective function of the CVAE is identical to that of the VAE, Eq.\eqref{eq.loss}.

\subsection{Hyperspherical conditional variational autoencoder ($\mathcal{S}$-CVAE)}
The Hyperspherical VAE has been proposed in \cite{HSVAE}. 
In this study, the hyperspherical VAE is used for the CVAE, which is termed as the hyperspherical CVAE ($\mathcal{S}$-CVAE). 
The network architecture of the $\mathcal{S}$-VAE is nearly identical to that of the $\mathcal{N}$-VAE; the difference is that the $\mathcal{S}$-VAE uses a vMF distribution rather than a normal distribution.
The probability density function of the vMF distribution is $f(x) = C \exp( \kappa \bi{\mu}^{\top} \bi{x})$, where the variable, $\bi{x}$, is defined as $\bi{x} \in \mathcal{S}$, $\mathcal{S} = \{ \bi{x} \in \Re^{d+1} \mid \| \bi{x} \| = 1 \}$.
$\bi{\mu} \in \mathcal{S} $ is the mean of the distribution, $\kappa \geq 0$ is a concentration parameter, and $C$ is a constant term used to normalize the distribution.
For a fixed $\bi{\mu}$, $\bi{x}$ is concentrated at $\bi{\mu}$, and $\kappa$ corresponds to the variance of the distribution.
$\mathcal{S}$ represents a $d$-dimensional hypersphere. 
Hence, the vMF distribution can be considered as a probability distribution in the hypersphere. 
The objective function of the $\mathcal{S}$-VAE is represented by Equation \eqref{eq.loss}. 
The uniform distribution of the von Mises Fischer distribution, that is,  ${\rm vMF} (\cdot, 0)$, is used as the prior; ${\rm vMF} (\mu, \kappa)$ is used as the posterior. 
The KL divergence, $\mathcal{L}_{KL}$, is then calculated as follows:
\begin{align*}
	\mathcal{L}_{KL}( {\rm vMF} (\mu, \kappa) || {\rm vMF} ( \cdot, 0)) 
	&= \kappa \frac{I_{{d+1}/2} (\kappa)}{I_{{d+1}/2-1} (\kappa)}
	+ \left( \frac{{d+1}}{2} -1 \right) \log{\kappa} 
	- \frac{{d+1}}{2} \log{\left(2 \pi \right) }\\
	&- \log{I_{{d+1}/2-1} (\kappa)}
	+ \frac{{d+1}}{2} \log{ \pi }
	+ \log{2}
	- \log{\Gamma \left( \frac{{d+1}}{2} \right)}
\end{align*}
The KL divergence is a function of $\kappa$, but is independent of $\mu$. This is because the prior is uniformly distributed on $\mathcal{S}$. 
Conversely, in a normal distribution, the prior is $ \mathcal{N}(0,1)$, because the mean of a normal distribution is located at the origin in a normal VAE, and the KL divergence is a function of $\mu$.

One of the advantages of using the $\mathcal{S}$-VAE is avoiding the KL collapse. 
The KL collapse was reported in \cite{Song19,Zhang19b} in the $\mathcal{N}$-VAE. The KL collapse is a situation in which the decoder outputs identical or similar data if a different latent vector is input. 
Davidson et al. \cite{HSVAE} reported that the $\mathcal{S}$-VAE prevents the KL collapse, and outperforms the $\mathcal{N}$-VAE, especially when the input data has a circular structure.

\begin{figure}[h]
	\begin{center}
		\begin{minipage}[h]{0.99\textwidth}
			\begin{center}
				\includegraphics[width=60mm]{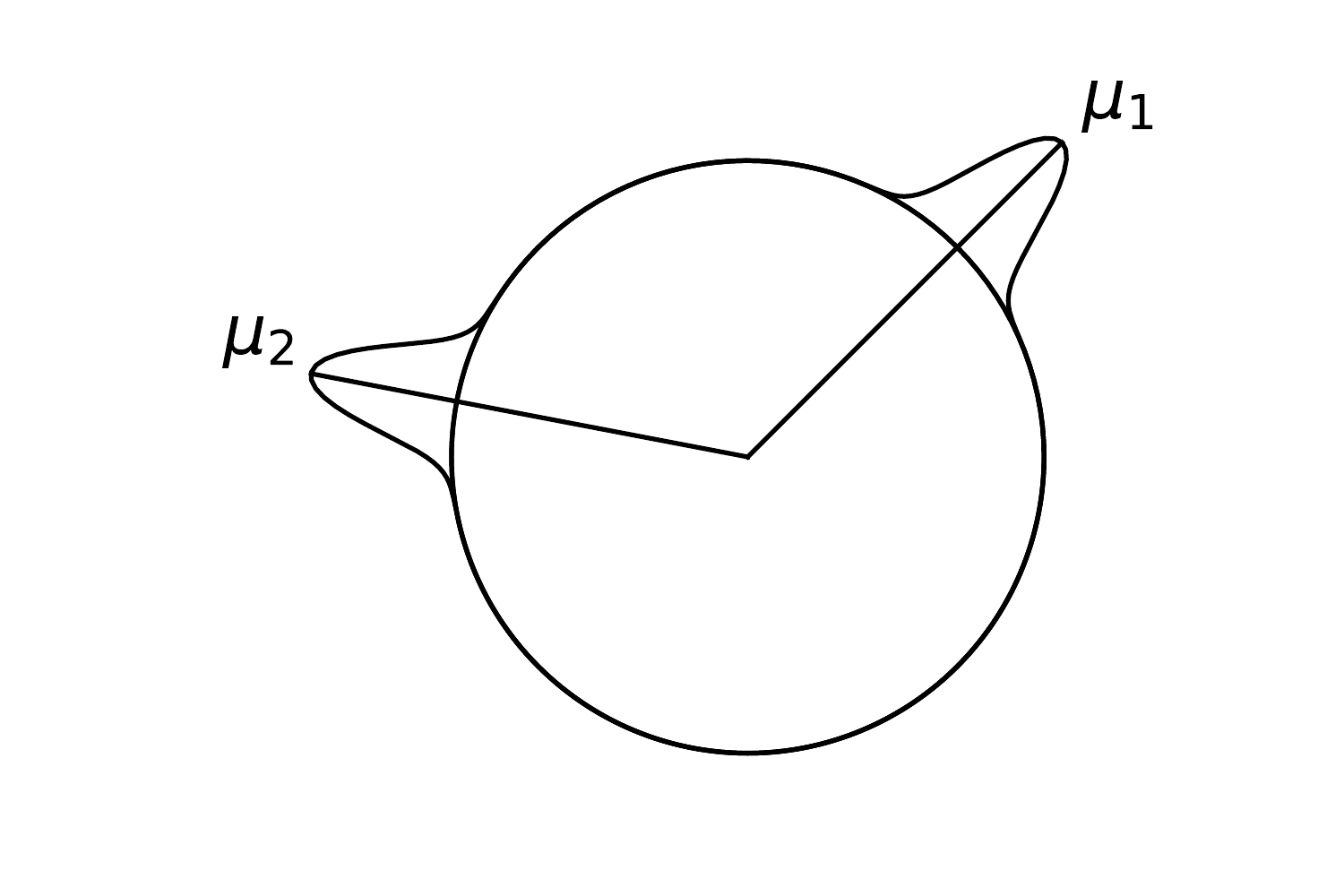}
				%{(a) AE}
			\end{center}
		\end{minipage}%
		\caption{von Mises Fischer distribution ${\rm vMF} \left( \mu, \kappa=100 \right)$.}
		\label{fig:vMF}
	\end{center}
\end{figure}

\section{CVAE based shape generation method}
\subsection{Methodology}
\begin{figure}[h]
	\begin{center}
		\begin{minipage}[h]{0.99\textwidth}
			\begin{center}
				\scalebox{0.45}{ % size should be (a) / 0.657807
					\includegraphics{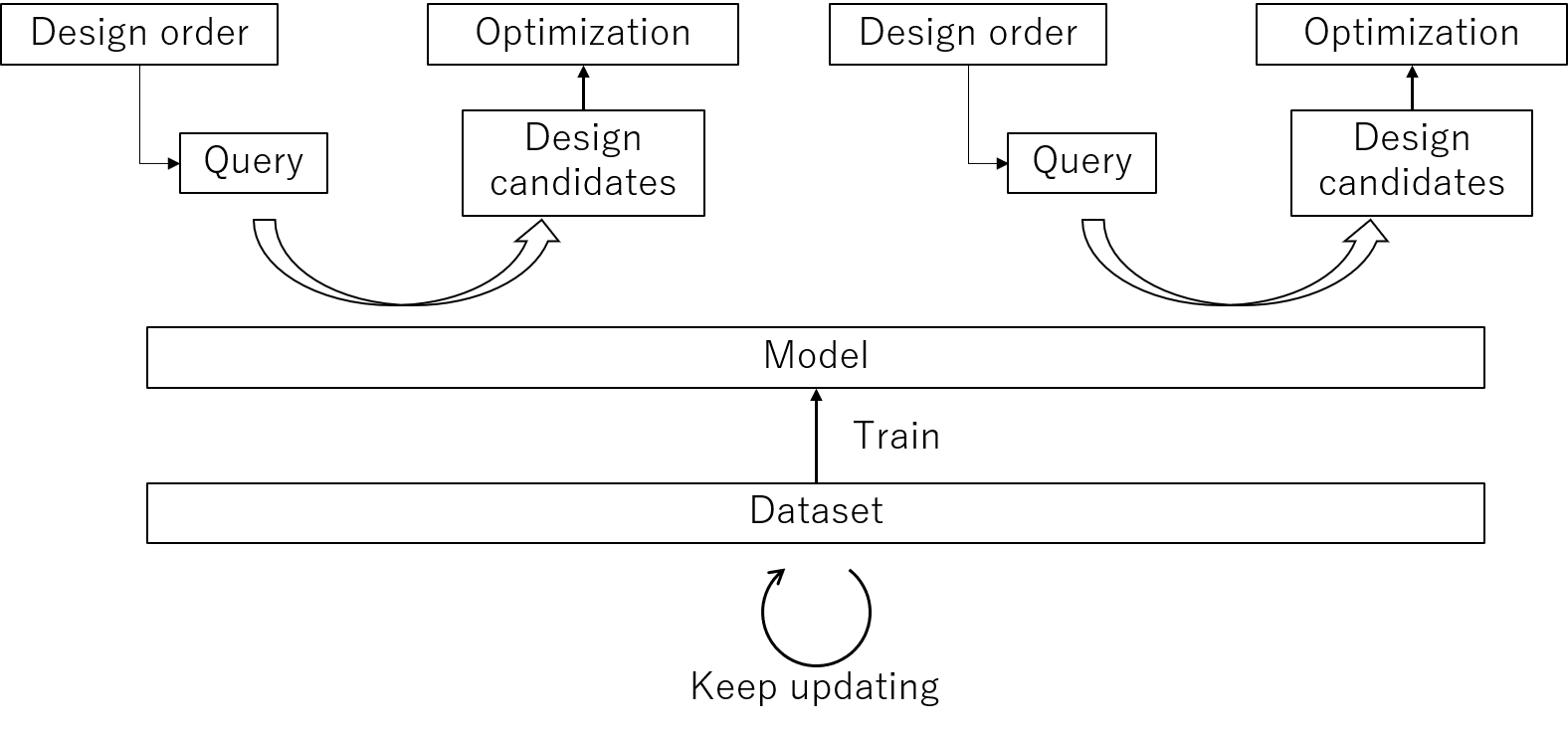}
				}
			\end{center}
		\end{minipage}%
		\caption{Schematics of the design system.}
		\label{fig:System}
	\end{center}
\end{figure}

\begin{figure}[h]
	\begin{center}
		\begin{minipage}[h]{0.99\textwidth}
			\begin{center}
				\scalebox{0.45}{ % size should be (a) / 0.657807
					\includegraphics{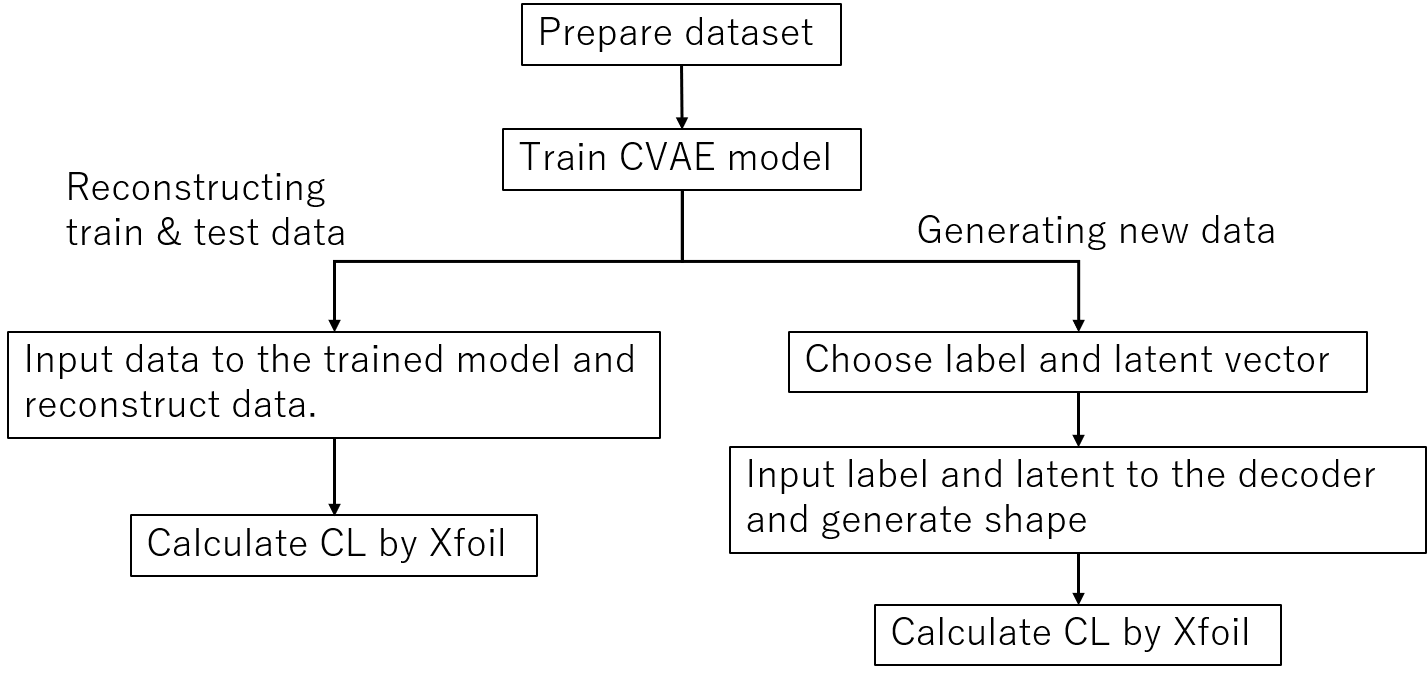}
				}
			\end{center}
		\end{minipage}%
		\caption{Flowchart of the proposed method.}
		\label{fig:flowchart}
	\end{center}
\end{figure}

This study uses the CVAE model to analyze and generate shapes. 
The airfoil design task is shown in \reffig{fig:System}. 
The machine learning model is trained once, and is used for multiple design queries. Once the model is trained, the model is re-used for many times. 
Both the $\mathcal{N}$-VAE and the $\mathcal{S}$-VAE are used in the numerical examples, with the same procedures for both models. 
The flowchart of the proposed method is presented in \reffig{fig:flowchart}. 
Initially, the dataset consists of the shape and its aerodynamic performance. 
In the numerical example, the airfoil shape data and its lift coefficient ($C_{\rm L}$) are used.

The aim of this method is handle the different types of airfoils, that is, the NACA and the Joukowski airfoils. The shape is represented by using a set of points to handle the different types of shapes. If the free curves, e.g., B{\'e}zier curves, are used, the method of choosing the control points suitable for all the different types of shapes is not important. 
The outline of the airfoil is represented by a set of points, $(x_i, y_i)^\top$, and the shape is represented by $s = (x_1, x_2, ..., x_n, y_1, y_2, ..., y_n)^\top$. 
The lift coefficient is calculated by using Xfoil in Section 3.2, and the computation of Xfoil requires more than 120 points \cite{XFOIL}. 
The number of points, $n$, was set as $248$ in the following numerical example. 

The dataset is split into the training and the test data with a ratio of $9:1$. The training data are used to train the model, whereas the test data are used to evaluate the extent to which the trained model can reconstruct the unseen data.
The training data is fed into the CVAE models to train the model. 
The reconstructed shapes are output after the training the model with the training and the test data. 
The reconstruction of the test data shows the capability to detect unseen shapes. 

As explained in Section \ref{sec:VAEmodel}, the objective function of the training is the sum of the reconstruction loss and the KL loss. 
Hence, the reconstructed shapes may not show the same $C_{\rm L}$ that is required by the label. 
To evaluate the error of $C_{\rm L}$, the numerical calculations of $C_{\rm L}$ and $C_{\rm D}$ are conducted for the reconstructed shapes. 
The error between the label and the recalculated $C_{\rm L}$ is then calculated based on the mean squared error (MSE): 
\begin{align*}
	\mathcal{L}_{C_{\rm L}} = \frac{1}{N} \sum_{i=1}^N \left( c^l - C_{\rm L}^r \right) ^2, 
\end{align*}
where $C_{\rm L}^r$ represents the recalculated $C_{\rm L}$. The error is defined for the training, testing, and the generated data, which are denoted by $\mathcal{L}_{C_{\rm L}}^{\rm train}$, $\mathcal{L}_{C_{\rm L}}^{test}$, and $\mathcal{L}_{C_{\rm L}}^{gen}$, respectively.

The new shapes are generated by using only the decoder section.
Firstly, a latent vector, $\bi{z}$, and a desired lift coefficient, $c^{l}$, are chosen. 
In the $\mathcal{S}$-VAE, the latent space is embedded in the hyperphere, and hence, the norm of the latent vector must be equal to $1$, that is, $\|\bi{z}\|=1$. 
$\bi{z}$ and $c^{l}$ are then fed into the decoder, and the decoder outputs the shapes.
$C_{\rm L}$ is calculated again for the generated shapes to check and evaluate the error of the label, $c^{l}$, and the actual $C_{\rm L}$ of the reconstructed shape.
The difference between the label and the recalculated data is evaluated by using the MSE. 

The variation of the generated shapes is evaluated by the following process.
Firstly, it is assumed that $\{ x_i \mid (i=1,2,\dots, N) \}$ is the set of generated shapes. The set includes the shapes which indicate a large error of $C_{\rm L}$. These shapes are eliminated as they should not be considered as design candidates. More precisely, $ \mathcal{G} = \left \{ x_i \mid \left(c^l|_{x_i} - C_{\rm L}|_{x_i}\right)^2 \leq \varepsilon \right \}$ is set, where $\varepsilon$ represents the tolerance of the error that is used for $\varepsilon=0.02$ in the following examples. 
The mean of the shapes, $\bi{\mu} = \frac{1}{|\mathcal{G}|} \sum_{\bi{x}_i \in \mathcal{G}} \bi{x}_i$ is then calculated, and the variation in the shape, $ v $ is calculated as the mean of the norm of the shape deviation.   
$ v = \frac{1}{|\mathcal{G}|} \sum_{\bi{x}_i \in \mathcal{G} } \| \bi{x}_i - \bi{\mu} \| $.

\subsection{DataSet} \label{sec.data}
\subsubsection{NACA airfoils}
\begin{figure}[htb]
	\begin{center}
		\begin{minipage}[h]{0.5\textwidth}
			\begin{center}
				\includegraphics[width=\linewidth]{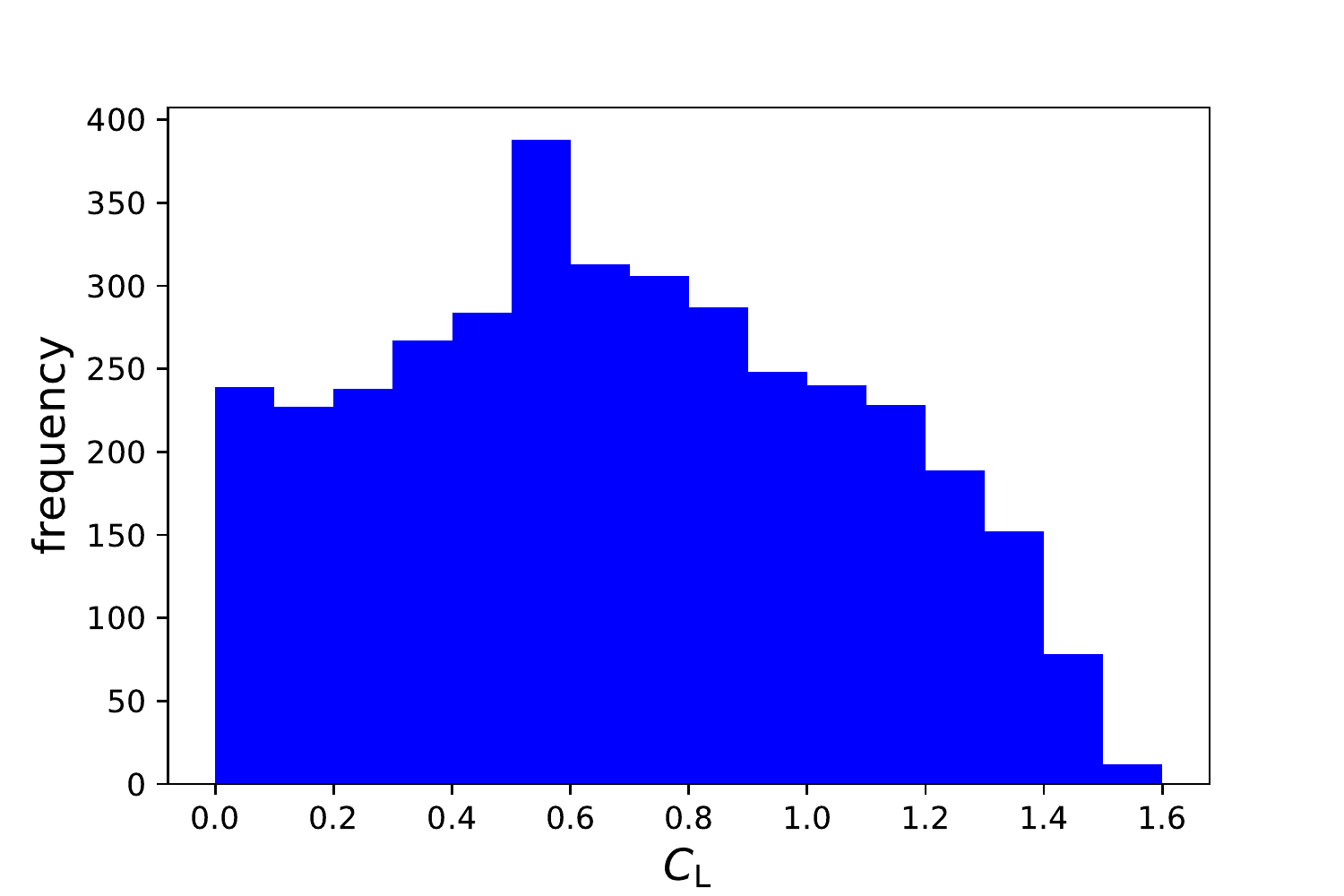}
				\par
				{(a) NACA airfoil. }
			\end{center}
		\end{minipage}%
		\begin{minipage}[h]{0.5\textwidth}
			\begin{center}
				\includegraphics[width=\linewidth]{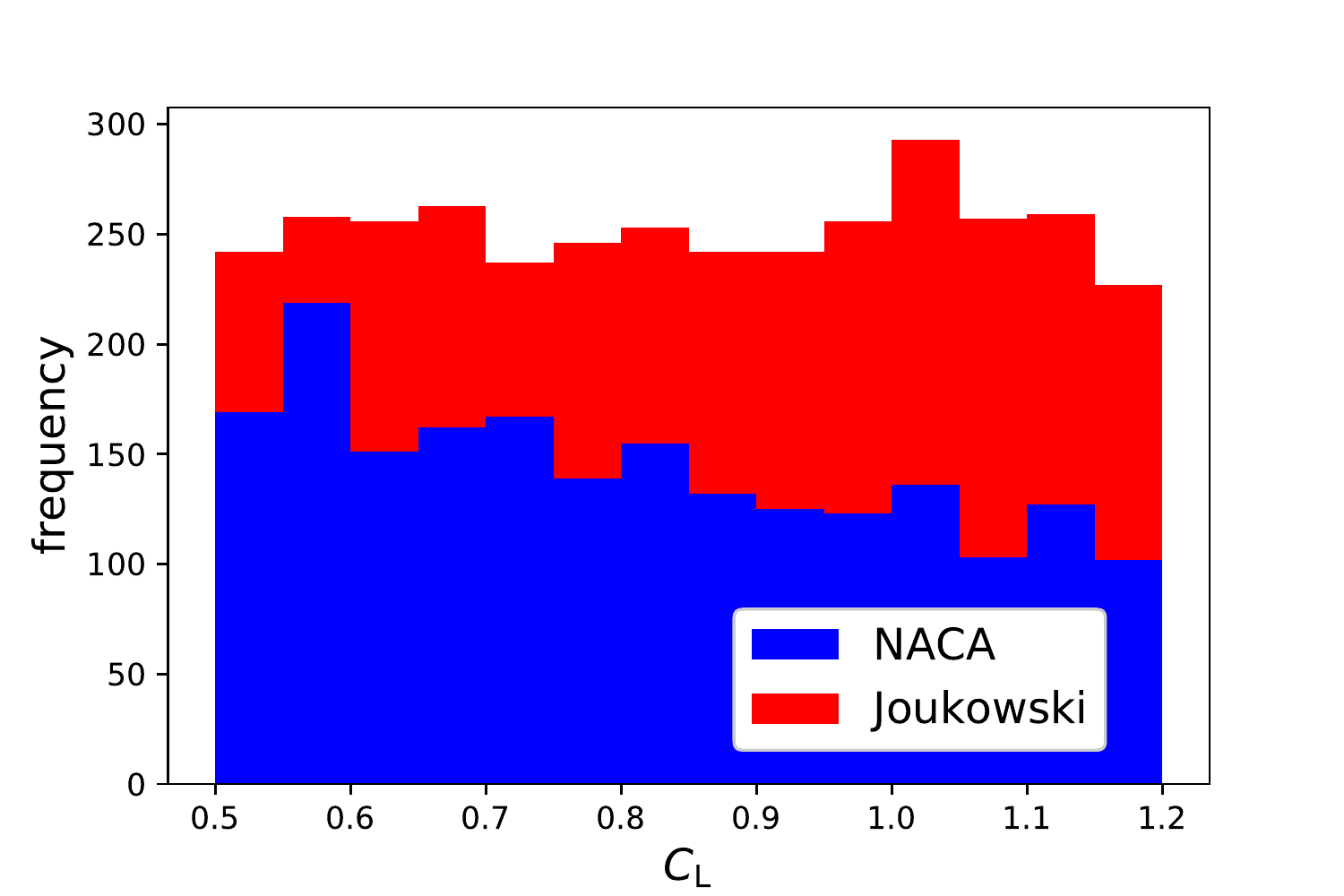}
				\par
				{(b) Mixture of the NACA and Joukowski airfoil. }
			\end{center}
		\end{minipage}%
		\caption{Histogram of $C_{\rm L}$.}
		\label{fig:CVAE_hist}
	\end{center}
\end{figure}

\begin{figure}[htb]
	\begin{center}
		\begin{minipage}[h]{0.5\textwidth}
			\begin{center}
				\includegraphics[width=\linewidth]{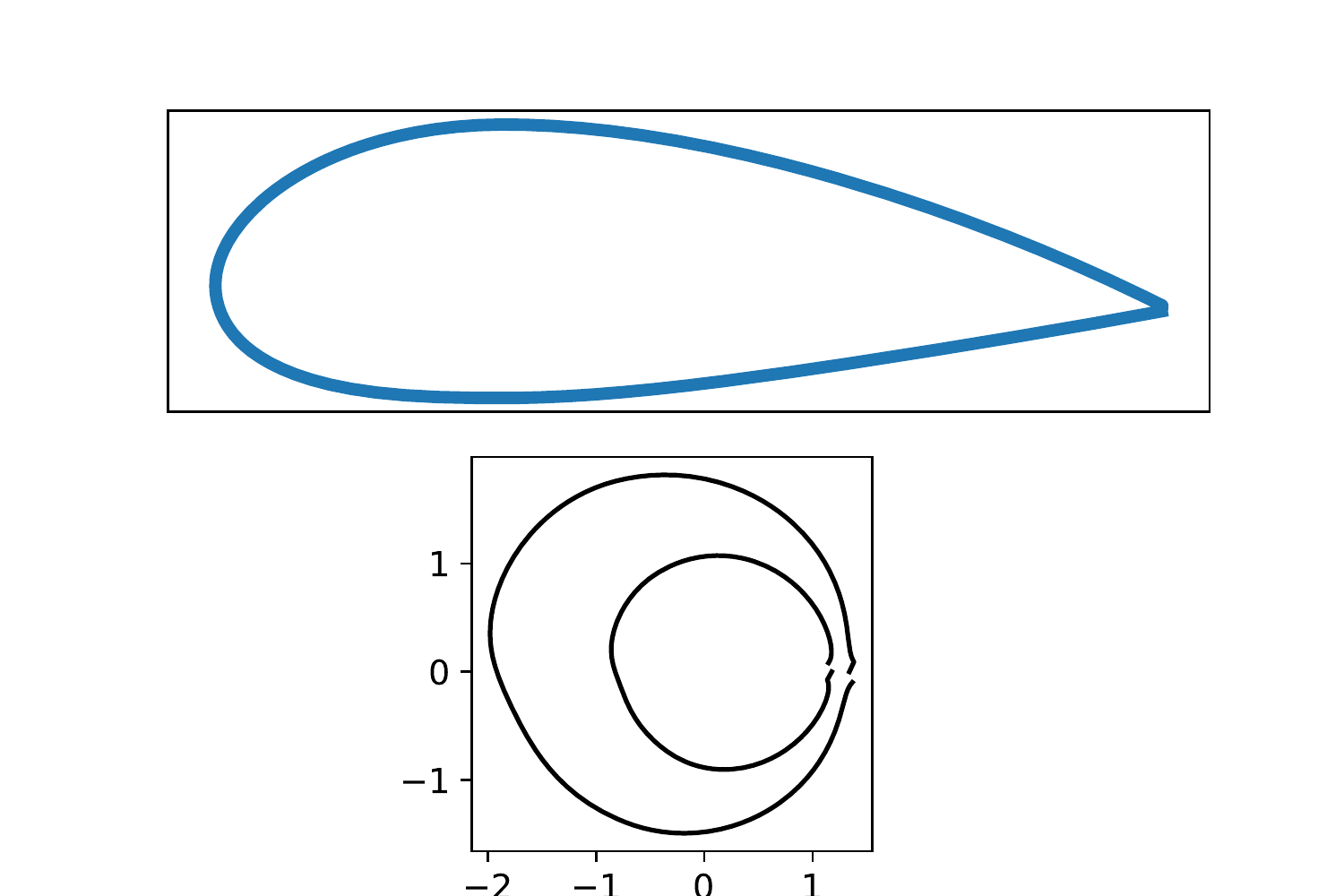}
				\par
				{(a) NACA airfoil ($w=0.478$). }
			\end{center}
		\end{minipage}%
		\begin{minipage}[h]{0.5\textwidth}
			\begin{center}
				\includegraphics[width=\linewidth]{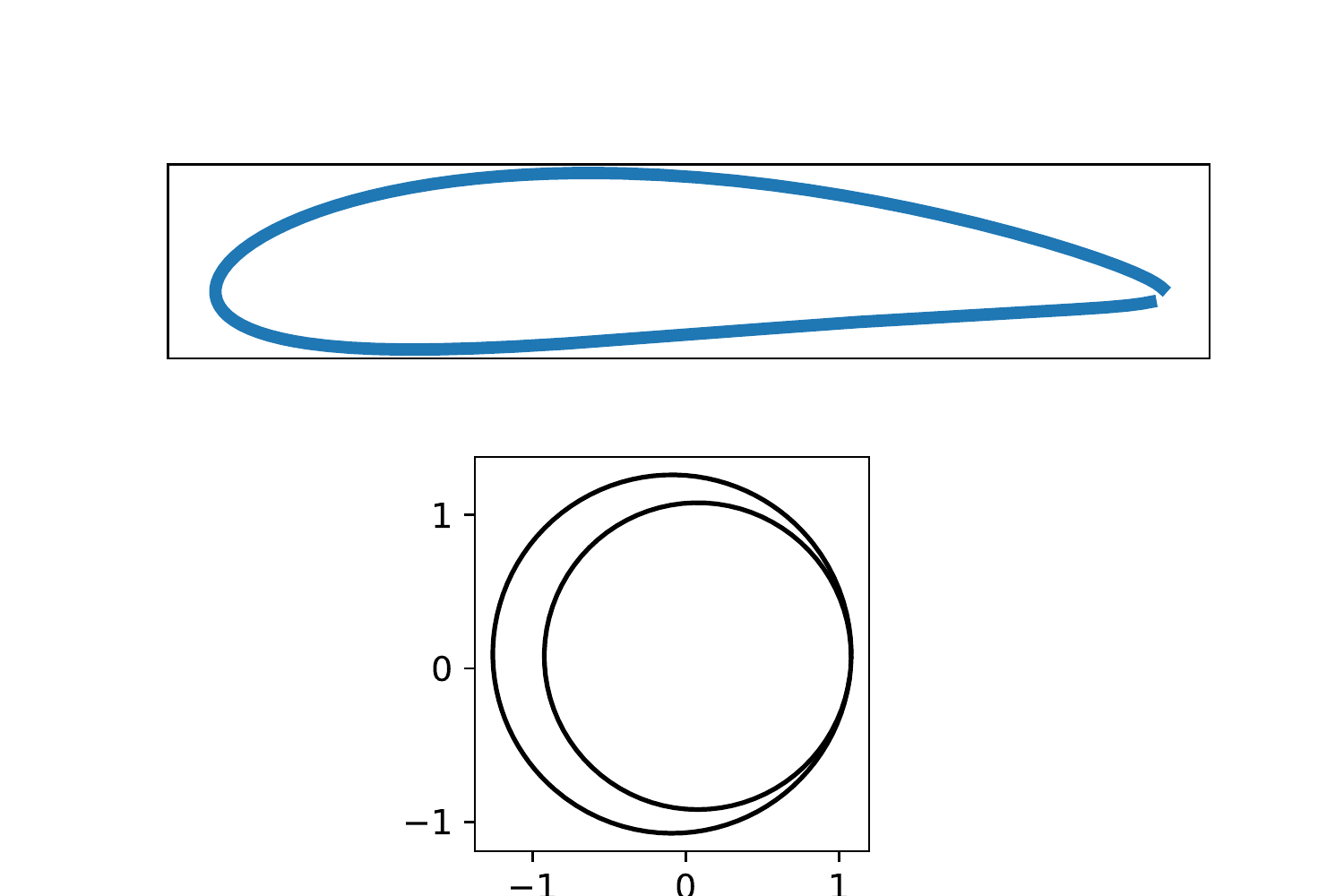}
				\par
				{(b) Joukowski airfoils ($w=6.22\times 10^{-7}$). }
			\end{center}
		\end{minipage}%
		\caption{Examples of airfoils and Joukowski inverse transformation.}
		\label{fig:CVAE_ex}
	\end{center}
\end{figure}

The NACA 4-digit airfoils  are used in this study \cite{Abbot}. The total number of data points is $3696$. 
The lift coefficient $C_{\rm L}$ is calculated by using the XFoil \cite{XFOIL}, which is a flow calculation toolbox based on the panel method.
The distribution of $C_{\rm L}$, shown in \reffig{fig:CVAE_hist} (a), is similar to a uniform distribution. 
The mean of the distribution is $\mu_{C_{\rm L}} = 0.6866$, and the standard deviation is $\sigma_{C_{\rm L}} = 0.3848$. 
An example of an NACA airfoil is shown in \reffig{fig:CVAE_ex} (a).

\subsubsection{Joukowski airfoils}\label{sec.J}
The Joukowski airfoil is a series of airfoils proposed by Joukowski. 
A set of points representing an airfoil shape, $ \zeta \in \mathbb{C}$ is generated by the Joukowski transformation of a circle $z \in \mathbb{C}$. 
The Joukowski transformation is defined as follows:
\begin{align}
	\zeta = z + \frac{c_{\rm J}^2}{ z }, \label{eq.j}
\end{align}
where $z$ is a circle and $c_{\rm J}$ is a variable defined by 
\begin{align*}
	& \left| z-(a+bi) \right| = r, \\
	& (c_{\rm J}-a)^2 = r^2 - b^2.
\end{align*}
where $ a, b, c_{\rm J} \in \mathbb{R} $. 
Various types of airfoils can be generated by changing $ a, b, r$,. 
The parameters from $a \in \{ 0, 0.002, \dots, 0.2 \}$, and $b \in \{ 0, 0.001, \dots, 0.2 \}$ are chosen, and $r$ is fixed at $1.1$. 

The generated shape, $\zeta$ is scaled such that its chord length becomes $1.0$. 
\begin{align}
	{\rm Re}(\hat{\zeta}) = \frac{ {\rm Re}(\zeta) - \ell}{m}, \quad	{\rm Im}(\hat{\zeta}) = \frac{{\rm Im}(\zeta)}{m}. \label{eq.scale}
\end{align}
$C_{\rm L}$ is calculated in the same manner as the NACA airfoils. 
In this study, a dataset of the NACA airfoil and the Joukowski airfoil is constructed. The histogram of $C_{\rm L}$ of the NACA and the Joukowski airfoils is shown in \reffig{fig:CVAE_hist} (b). 
An example of the Joukowski airfoil is shown in \reffig{fig:CVAE_ex} (b). 

The Joukowski airfoil is transformed into two circles by using the Joukowski inverse transformation. The mathematical formulation of the Joukowski inverse transformation is presented in Section A.1. 
If the Joukowski inverse transformation is applied to the NACA airfoil, the result is not a circle. 
Hence, by calculating the roundness of the inverse transformation, it can be determined whether the shape is similar to a Joukowski airfoil or not. 
Examples of the Joukowski inverse transformation to the NACA and the Joukowski airfoils are shown in \reffig{fig:CVAE_ex} (a) and (b), with the roundness, $w$. The value of $w$ of the NACA airfoil is not zero, whereas the $w$ of the Joukowski airfoil is sufficiently small.

\section{NACA airfoil generation using ${\mathcal S}$-CVAE and ${\mathcal N}$-CVAE}
In this section, the ${\mathcal S}$-CVAE and the ${\mathcal N}$-CVAE models are  trained by using the NACA airfoil data. The aim of this task is to generate a wide variety of shapes which meet the required $C_{\rm L}$. The program is implemented in Python using Tensorflow \cite{tf}. The computation was conducted on Intel Core i5-10210U 1.6 GHz CPU with 8GB memory.  

\subsection{Training and shape generation}
The NACA airfoil data is fed to both ${\mathcal N}$-CVAE and ${\mathcal S}$-CVAE and new airfoil shapes are generated. 
The model is trained by using the same NACA data presented in section \ref{sec.data}. 
The data are split in the ratio of $9:1$ and fed as the training and the test datasets, respectively. 
The number of nodes in each layer is $n_1=500$ and $n_2=500$, where $n_1$ is used for the first layer of the encoder and the second layer of the decoder, and $n_2$ is used for the second layer of the encoder and the 1st layer of the decoder. 
The latent dimension, $d$, is also determined by comparing $\mathcal{L}_{C_{\rm L}}^{gen}$, which is listed in \reftab{tab:latentNodes}. $d=2$ is chosen as it indicates the smallest $\mathcal{L}_{C_{\rm L}}^{gen}$. 

\begin{table}[htpb]
	\begin{center}
		\caption{$\mathcal{L}_{\rm C_{\rm L}}^{gen}$ comparison with respect to different latent dimensions.}
		\label{tab:latentNodes}       % Give a unique label
		\begin{tabular}{crrrrr}
			\hline\noalign{\smallskip}
			Lat. dim. & 2 & 4 & 8 & 16 & 32 \\
			\noalign{\smallskip}\hline\noalign{\smallskip}
			%$\mathcal{L}_{\rm KL}$ &  &  &  &  &   \\
			$\mathcal{S}$-CVAE & $\bold{0.03019}$ & 0.03736 & 0.07091 & 0.04014 & 0.02213   \\
			$\mathcal{N}$-CVAE & $\bold{0.05172}$ & 0.05282 & 0.05949 & 0.07300 & 0.06565 
			\\
			\noalign{\smallskip}\hline
		\end{tabular}%
	\end{center}%
\end{table}%

\begin{table}[htpb]
	\begin{center}
		\caption{$\mathcal{L}_{\rm C_{\rm L}}$ of reconstructed and generated shapes.}
		\label{tab:CVAE_gen}       % Give a unique label
		\begin{tabular}{cccc}
			\hline\noalign{\smallskip}
			Case & Train & Test & Generated from Random $\bi{z}$  \\
			& ($\mathcal{L}_{\rm C_{\rm L}}^{\rm train}$) & ($\mathcal{L}_{\rm C_{\rm L}}^{\rm test}$) & ($\mathcal{L}_{\rm C_{\rm L}}^{\rm gen}$) \\
			\noalign{\smallskip}\hline\noalign{\smallskip}
			${\mathcal N}$-CVAE & 0.02364 & 0.05140	 & 0.05172 \\
			${\mathcal S}$-CVAE & 0.02829 & 0.04842  & 0.01965  \\
			\noalign{\smallskip}\hline
		\end{tabular}%
	\end{center}%
\end{table}%

\begin{figure}[htb]
	\begin{center}
		\begin{minipage}[h]{0.5\textwidth}
			\begin{center}
				\includegraphics[width=\linewidth]{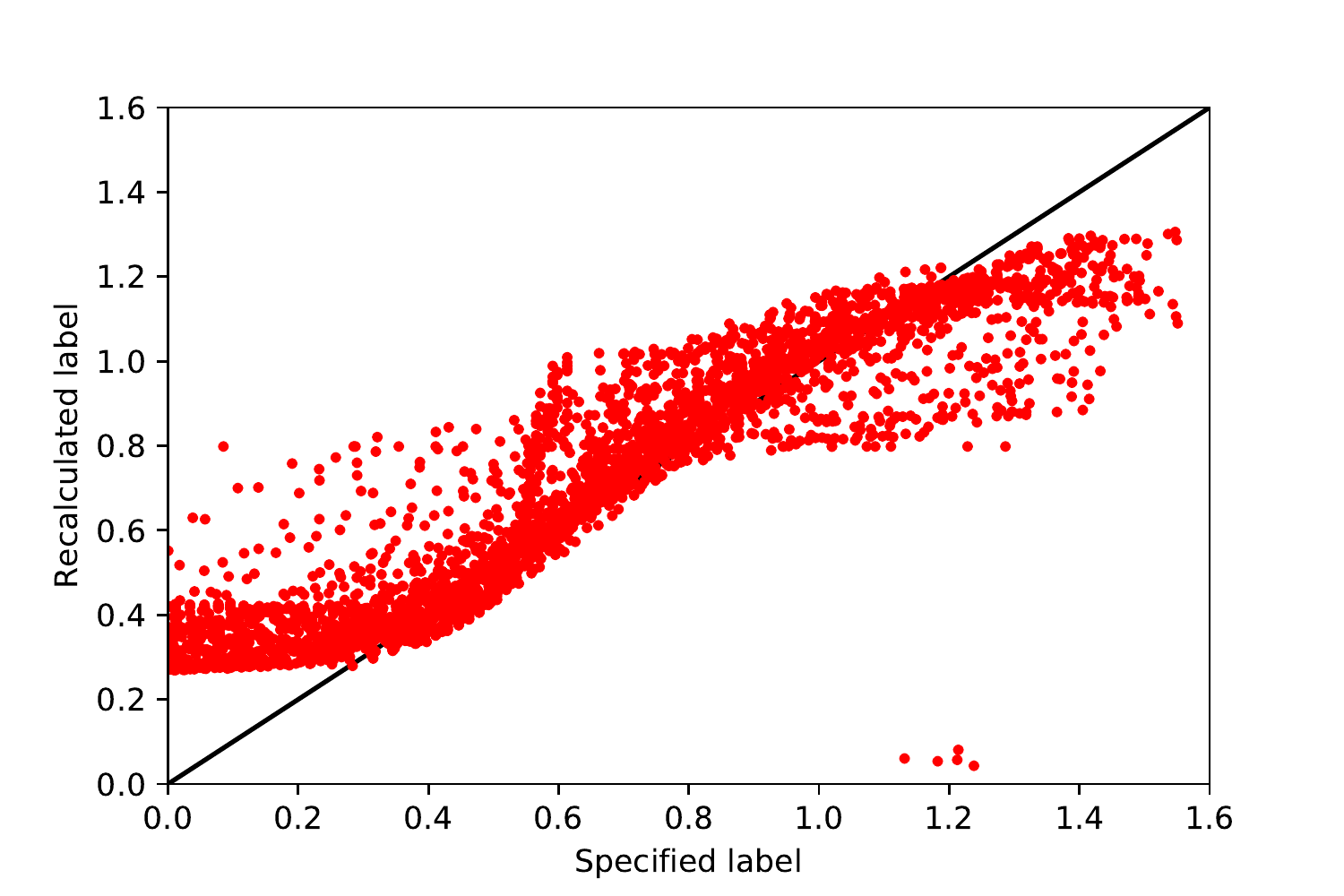}
				\par
				{(a) Training data reconstruction using ${\mathcal S}$-CVAE.}
			\end{center}
		\end{minipage}%
		\begin{minipage}[h]{0.5\textwidth}
			\begin{center}
				\includegraphics[width=\linewidth]{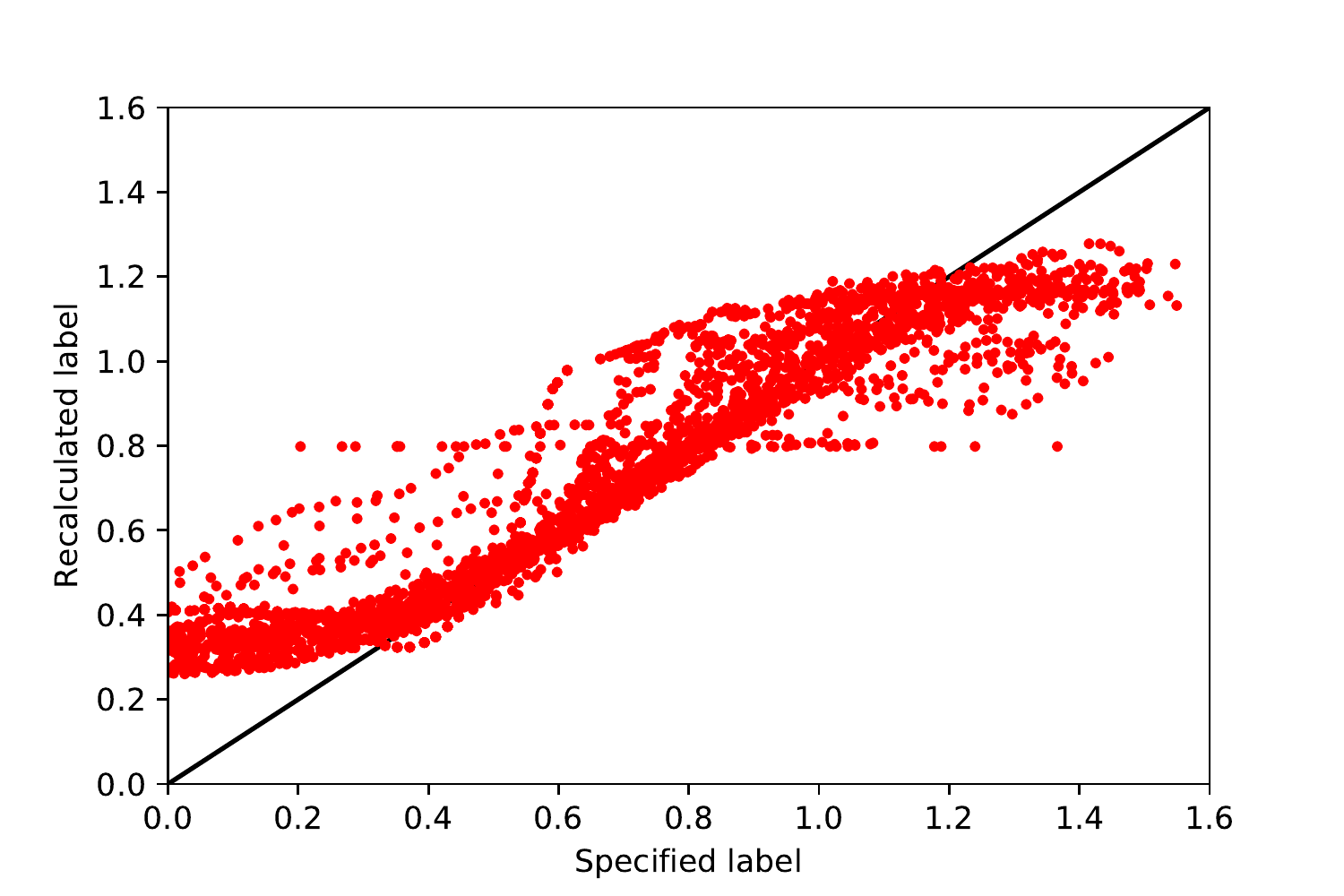}
				\par
				{(b) Training data reconstruction using ${\mathcal N}$-CVAE.}
			\end{center}
		\end{minipage}%
		\par
		\begin{minipage}[h]{0.5\textwidth}
			\begin{center}
				\includegraphics[width=\linewidth]{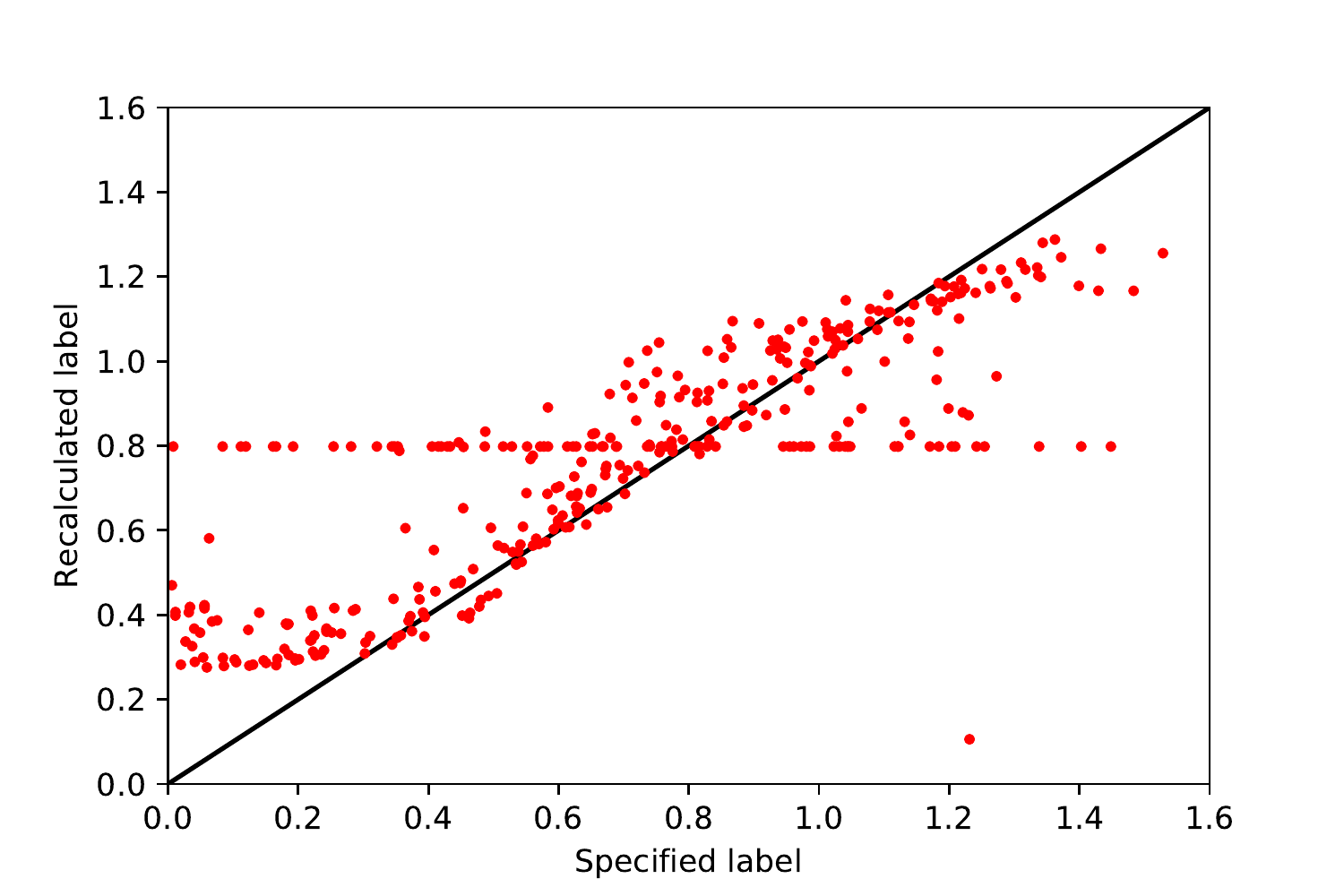}
				\par
				{(c) Test data reconstruction using ${\mathcal S}$-CVAE.}
			\end{center}
		\end{minipage}%
		\begin{minipage}[h]{0.5\textwidth}
			\begin{center}
				\includegraphics[width=\linewidth]{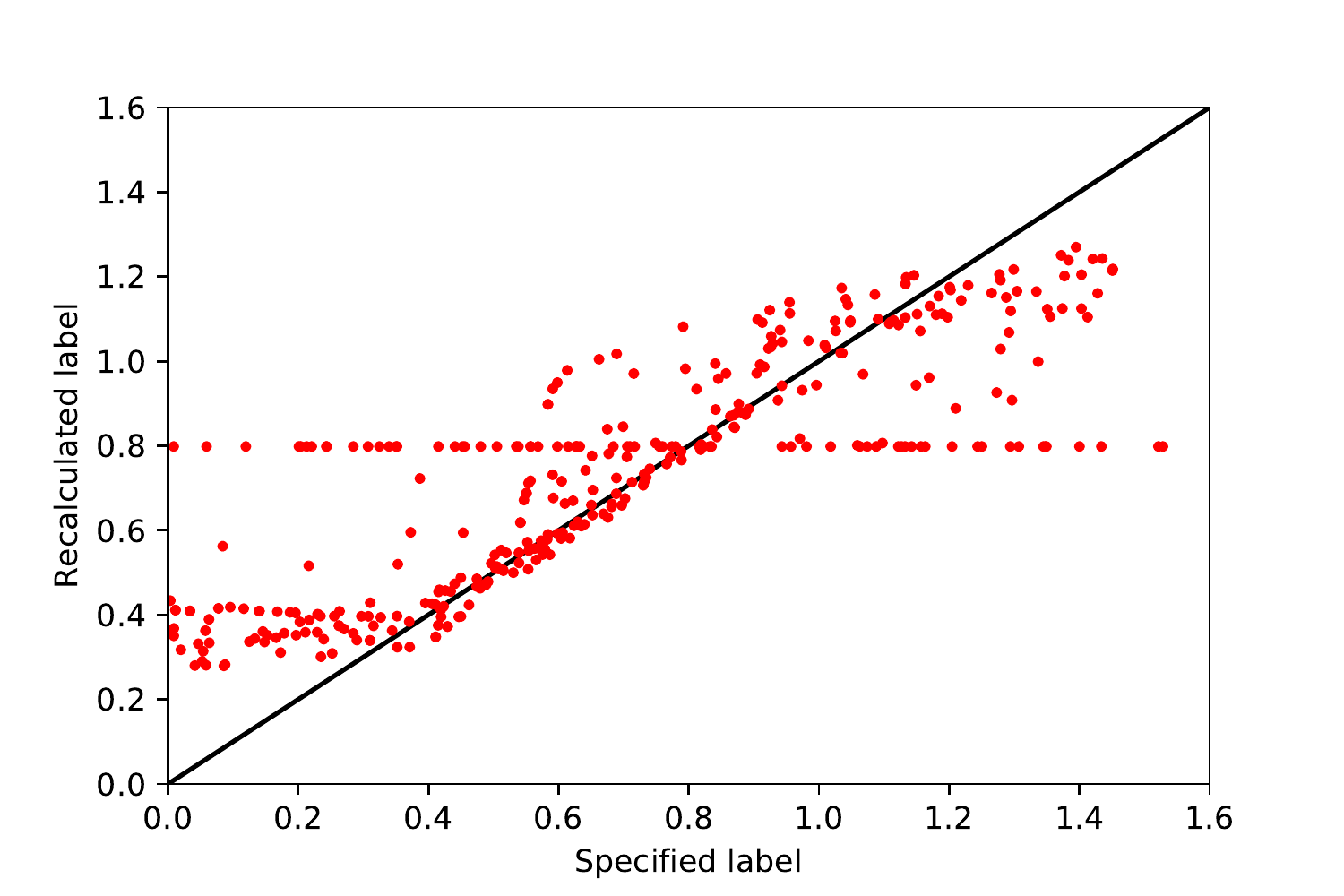}
				\par
				{(d) Test data reconstruction using ${\mathcal S}$-CVAE.}
			\end{center}
		\end{minipage}%
		\par
		\begin{minipage}[h]{0.5\textwidth}
			\begin{center}
				\includegraphics[width=\linewidth]{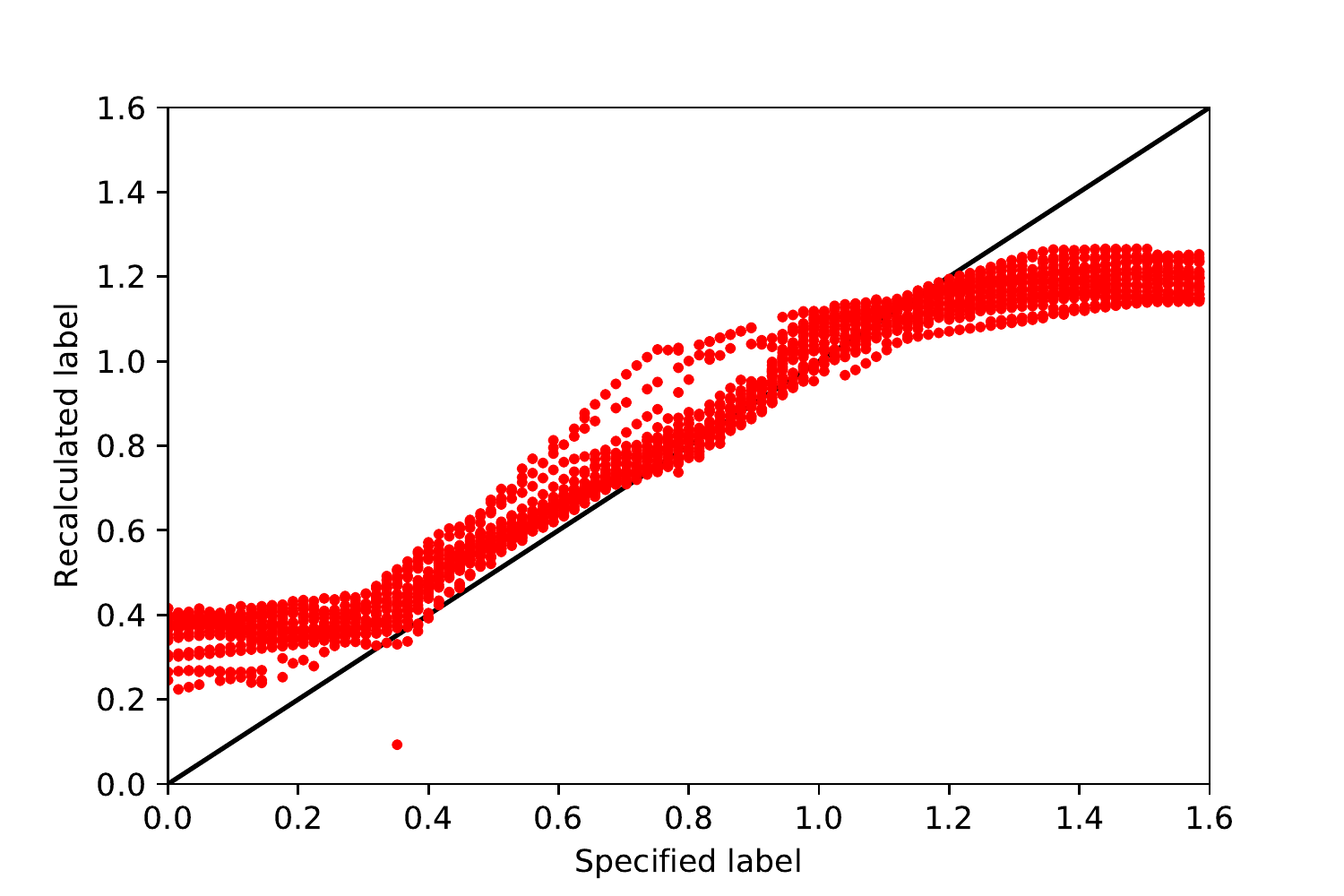}
				\par
				{(e) Random sampling with ${\mathcal S}$-CVAE.}
			\end{center}
		\end{minipage}%
		\begin{minipage}[h]{0.5\textwidth}
			\begin{center}
				\includegraphics[width=\linewidth]{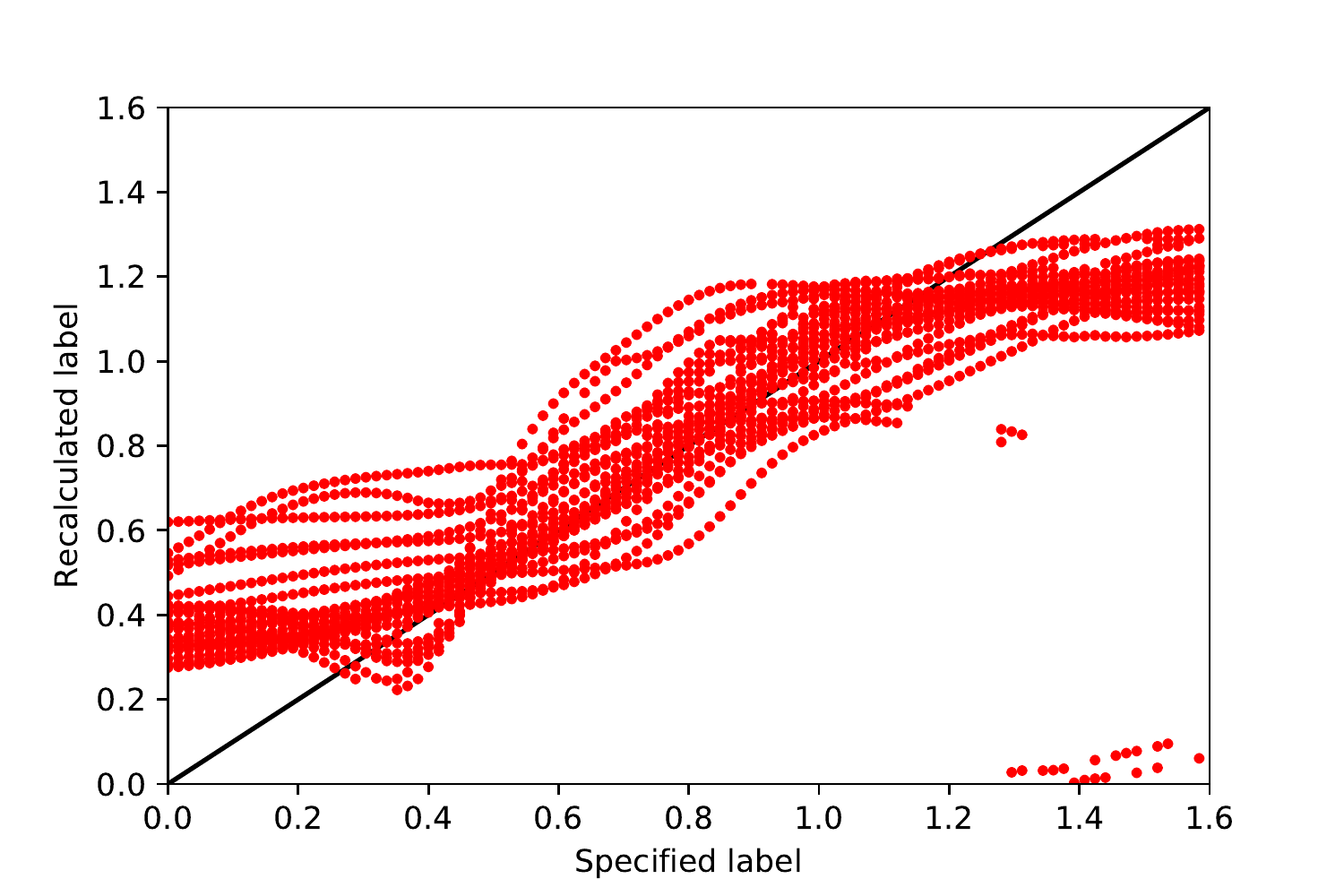}
				\par
				{(f) Random sampling with ${\mathcal N}$-CVAE.}
			\end{center}
		\end{minipage}%
		\par
		\begin{minipage}[h]{0.5\textwidth}
			\begin{center}
				\includegraphics[width=\linewidth]{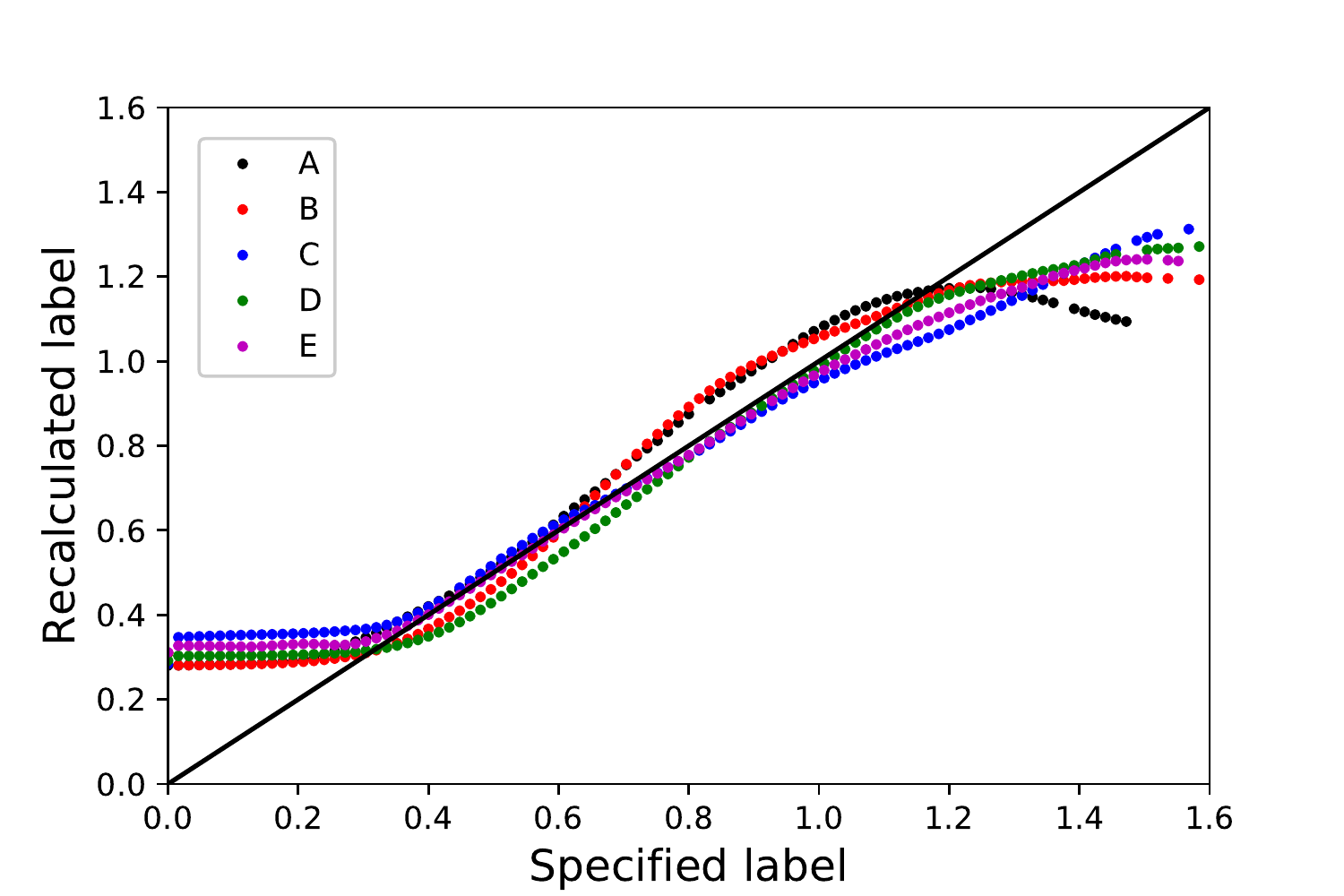}
				\par
				{(g) Fixed latent sampling with ${\mathcal S}$-CVAE.}
			\end{center}
		\end{minipage}%
		\begin{minipage}[h]{0.5\textwidth}
			\begin{center}
				\includegraphics[width=\linewidth]{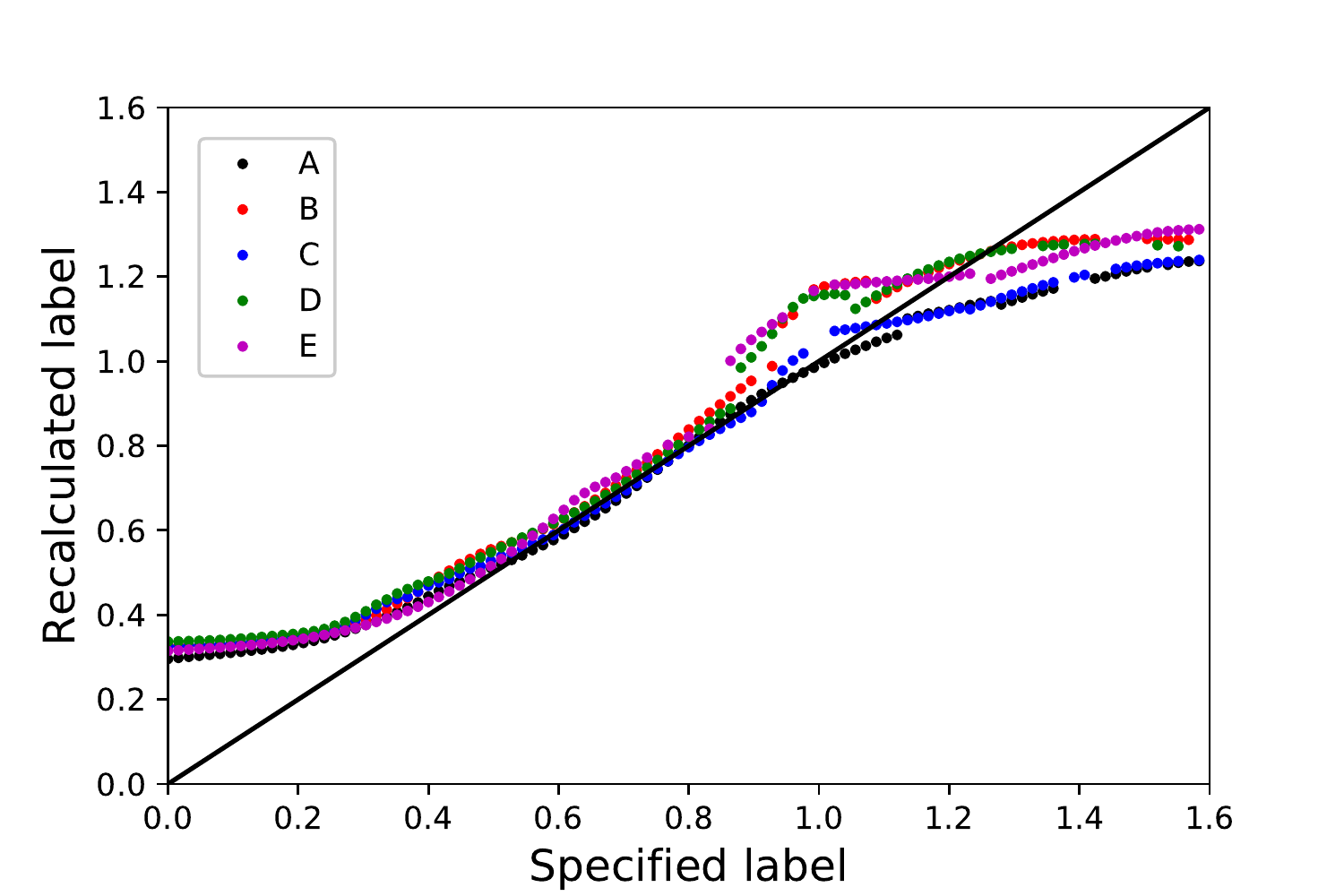}
				\par
				{(h) Fixed latent sampling with ${\mathcal N}$-CVAE.}
			\end{center}
		\end{minipage}%
		\caption{Error of $C_{\rm L}$ in generated airfoils.}
		\label{fig:CVAE_gen}
	\end{center}
\end{figure}

The accuracy of the reconstructed and the generated shapes of both $\mathcal{S}$-CVAE and $\mathcal{N}$-CVAE are evaluated.
$\mathcal{L}_{\rm C_{\rm L}}$ are shown in \reftab{tab:CVAE_gen}. 
${\mathcal N}$-CVAE and ${\mathcal S}$-CVAE exhibit similar values corresponding to 
$\mathcal{L}_{\rm C_{\rm L}}^{\rm train}$ and $\mathcal{L}_{\rm C_{\rm L}}^{\rm test}$, and the error distributions shown in \reffig{fig:CVAE_gen} (a)--(d) are also similar. 

The new shapes are then generated. 
In $\mathcal{S}$-CVAE, $30$ latent vectors are randomly sampled from the entire latent space, that is, $\{ \bi{z} \mid  \| \bi{z} \| =1, \bi{z} \in \Re^{d+1} \} $. 
In $\mathcal{N}$-CVAE, $30$ latent vectors are sampled from the envelope of the latent vectors corresponding to the training data, and $ \{ \bi{z} \mid L_i \leq z_i \leq  U_i \} $, where $L_i$ and $U_i$ are the minimum and the maximum values of the latent vectors corresponding to the training data. 
When $\mathcal{L}_{\rm C_{\rm L}}^{\rm gen}$ is calculated, the labels are set as $ \mathcal{C} = \{ 0.0, 0.016, ..., 1.584 \}$. 
The results are shown in random z in \reftab{tab:CVAE_gen}, and \reffig{fig:CVAE_gen} (e) and (f).  
$\mathcal{L}_{\rm C_{\rm L}}^{\rm gen}$ in ${\mathcal S}$-CVAE is $0.01965$, which is less than half of that of ${\mathcal N}$-CVAE ($0.05172$). 
The error distribution is wider in ${\mathcal N}$-CVAE than in ${\mathcal S}$-CVAE. 
Consequently, ${\mathcal S}$-CVAE generates shapes more accurately corresponding to $C_{\rm L}$. 
In both the $\mathcal{S}$-CVAE and $\mathcal{N}$-CVAE models, the error of $C_{\rm L}$ is small when the label, $c^l$, is in the middle of its range, that is, $0.4 < c^l < 1.2$, and is relatively large for other $c^l$. 
Recalling the histogram of $C_{\rm L}$ in \reffig{fig:CVAE_hist}, the number of data points is smaller in the range, $1.2 < C_{\rm L}$. This data imbalance may cause a larger error in this range.

\begin{figure}[htb]
	\begin{center}
		\begin{minipage}[h]{0.5\textwidth}
			\begin{center}
				\includegraphics[width=\linewidth]{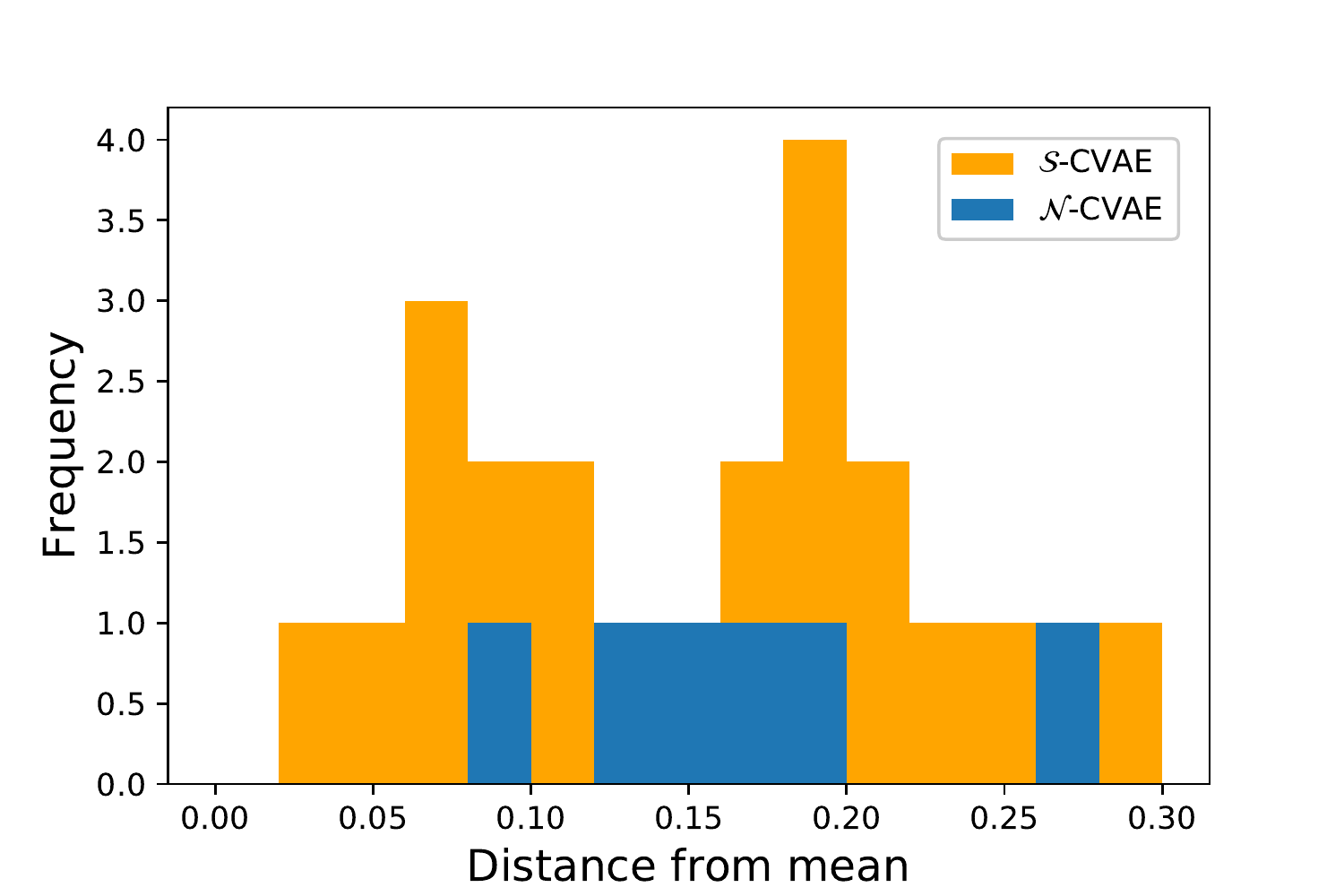}
			\end{center}
		\end{minipage}%
		\caption{Histogram of shape variation $v$.}
		\label{fig:CVAE_var}
	\end{center}
\end{figure}

\begin{figure}[htb]
	\begin{center}
		\begin{minipage}[h]{0.5\textwidth}
			\begin{center}
				\includegraphics[width=\linewidth]{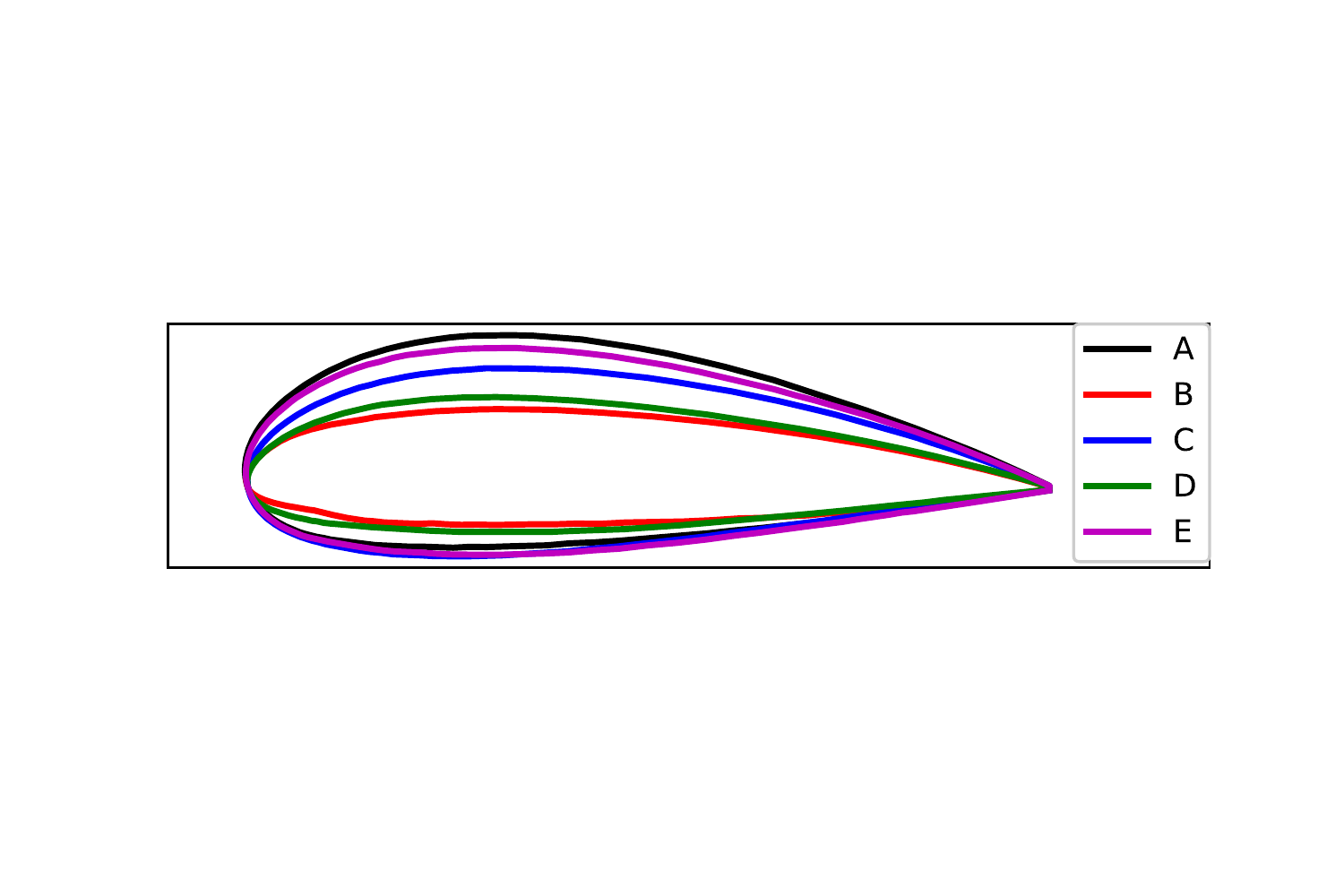}
				\par
				{(a) ${\mathcal S}$-CVAE.}
			\end{center}
		\end{minipage}%
		\begin{minipage}[h]{0.5\textwidth}
			\begin{center}
				\includegraphics[width=\linewidth]{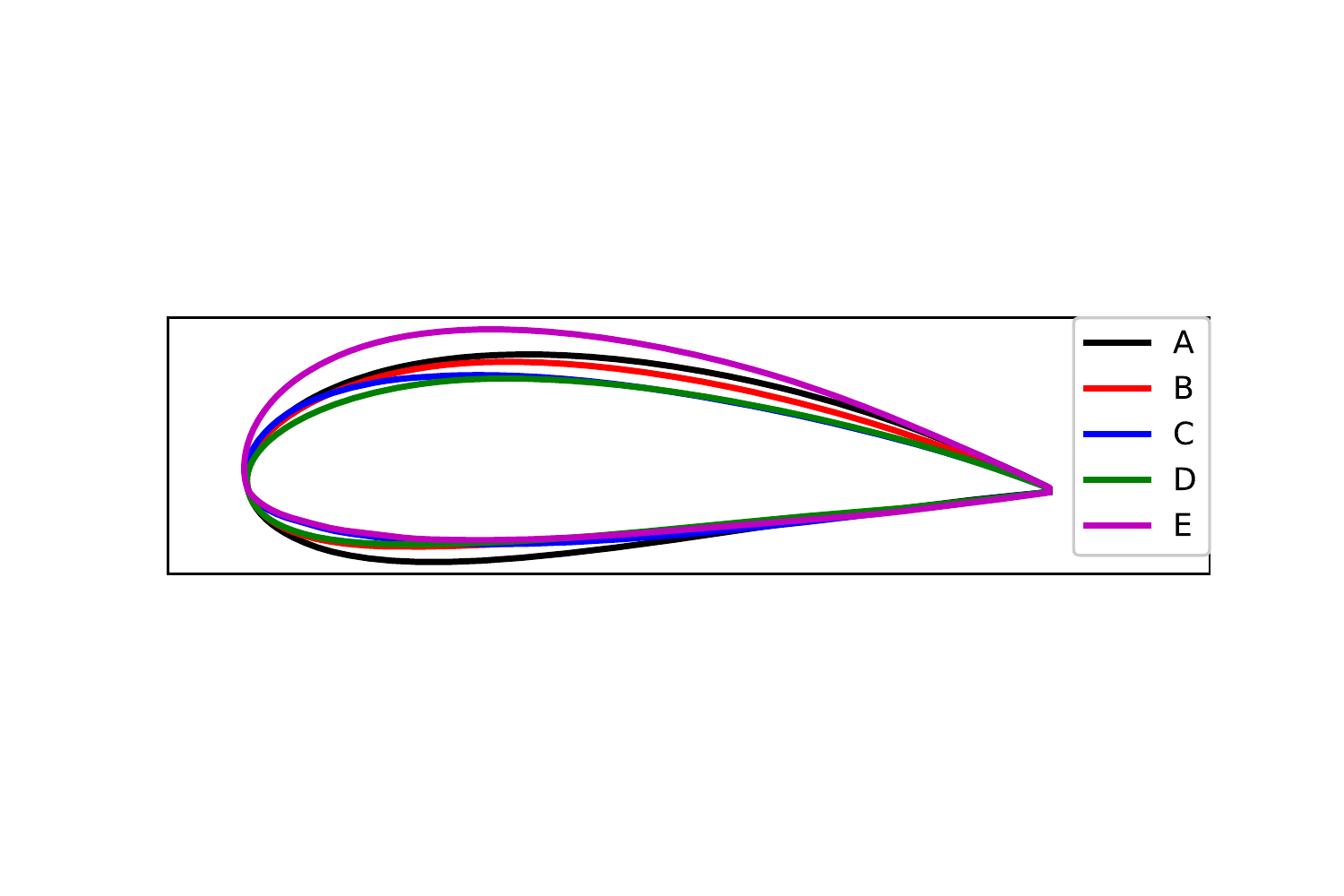}
				\par
				{(b) ${\mathcal N}$-CVAE.}
			\end{center}
		\end{minipage}%
		\caption{Generated shapes with $C_{\rm L} = 0.687$.}
		\label{fig:CVAE_mult}
	\end{center}
\end{figure}

In order to evaluate the shape variation, the shapes with sufficiently small error are first selected. The latent vectors which satisfy $\mathcal{L}_{C_{\rm L}} < 0.02$ are chosen from $30$ latent vectors. The number of selected latent vectors is 23 in $\mathcal{S}$-CVAE and 6 in $\mathcal {N}$-CVAE. 
From these shapes, five samples, A, B, C, D, and E, are selected, as shown in \reffig{fig:CVAE_mult}. Additionally, the error of $C_{\rm L}$ is shown in \reffig{fig:CVAE_gen} (g) and (h). 
\reffig{fig:CVAE_mult} shows that the $\mathcal{S}$-CVAE generates a wider variety of shapes.
This is also verified by calculating $v$, as explained in section 3.1. 
The histogram of $v$ is shown in \reffig{fig:CVAE_var}. 
The histogram indicates that the generated shapes of $\mathcal{S}$-CVAE demonstrate a wider variation than $\mathcal{N}$-CVAE. 
Consequently, ${\mathcal S}$-CVAE generates various types of shapes. 

\afterpage{\clearpage}
\subsection{Data embedding in the latent space}
\begin{figure}[htb]
	\begin{center}
		\begin{minipage}[h]{0.5\textwidth}
			\begin{center}
				\includegraphics[width=\linewidth]{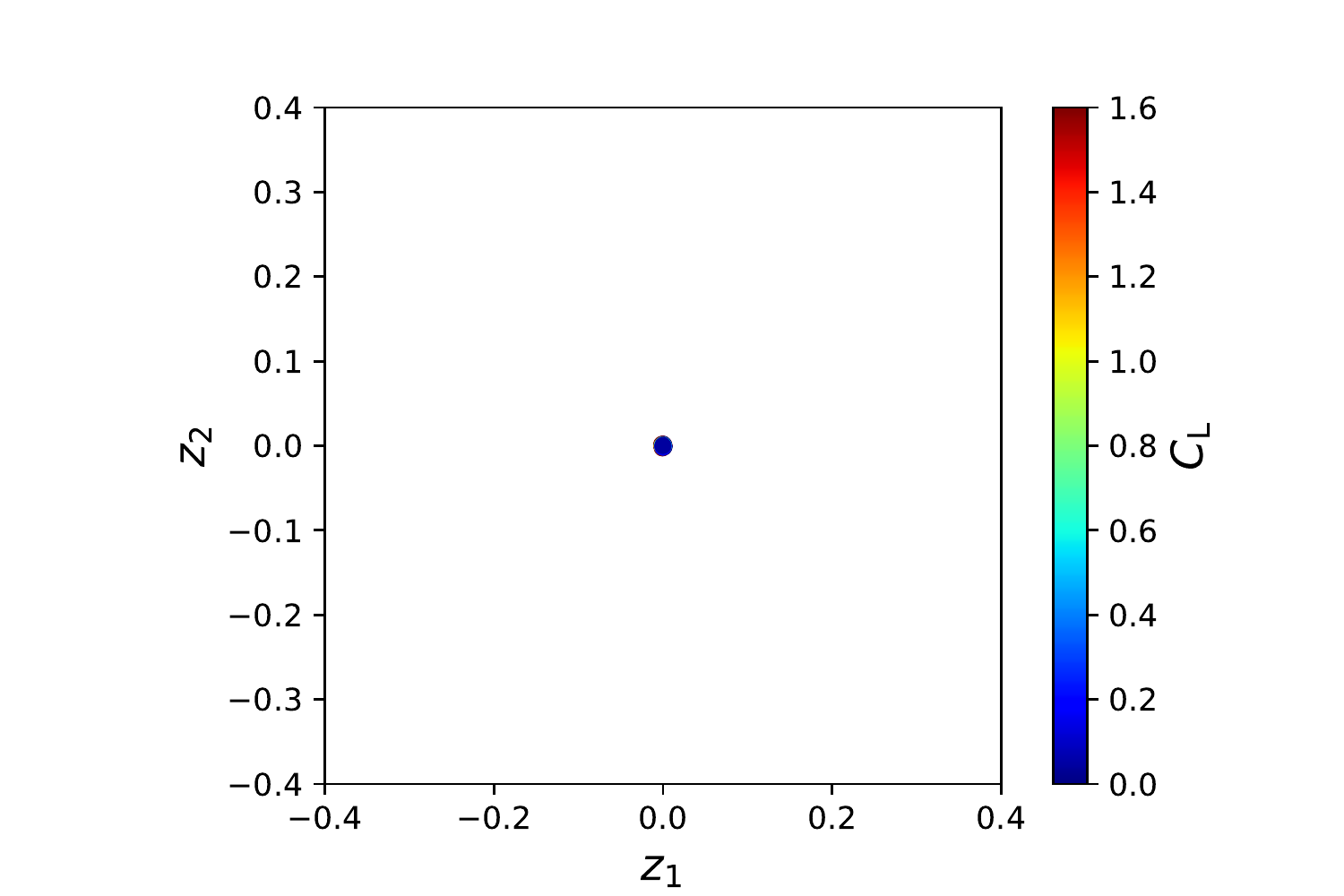}
				\par
				{(a) ${\mathcal N}$-VAE.}
			\end{center}
		\end{minipage}%
		\begin{minipage}[h]{0.5\textwidth}
			\begin{center}
				\includegraphics[width=\linewidth]{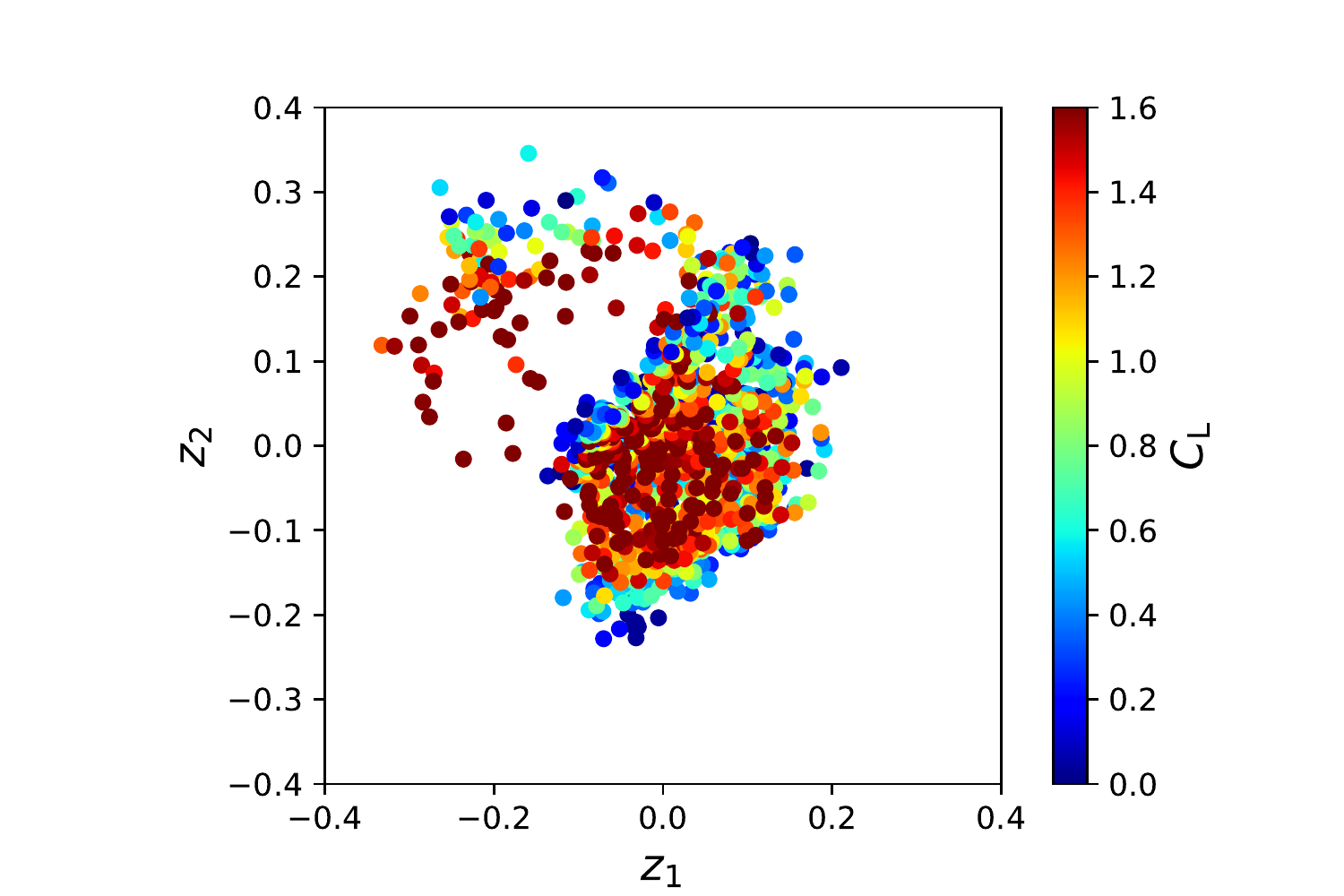}
				\par
				{(b) ${\mathcal N}$-CVAE.}
			\end{center}
		\end{minipage}%
		\par
		\begin{minipage}[h]{0.5\textwidth}
			\begin{center}
				\includegraphics[width=\linewidth]{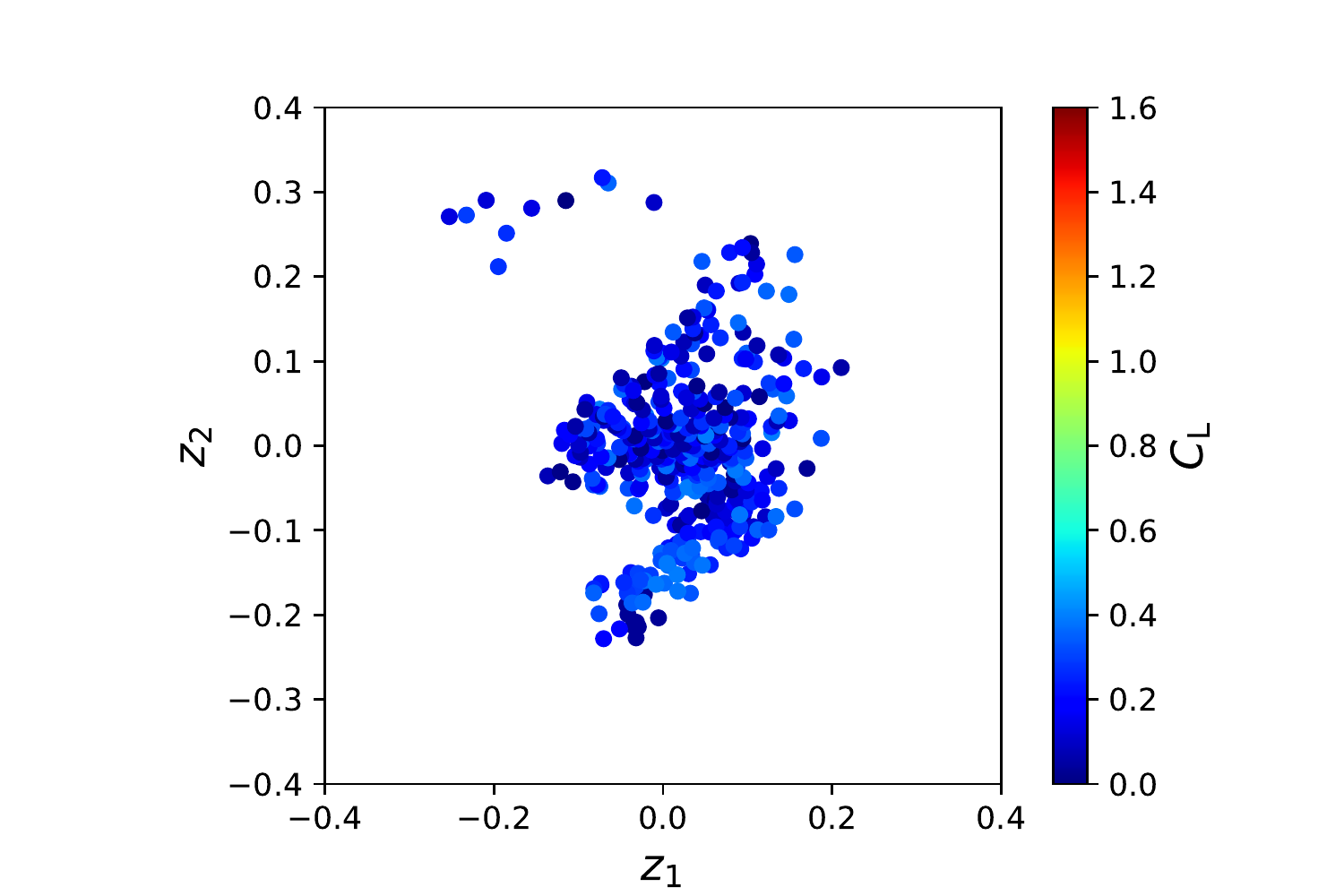}
				\par
				{(c) ${\mathcal N}$-CVAE ($C_{\rm L}<0.4$).}
			\end{center}
		\end{minipage}%
		\begin{minipage}[h]{0.5\textwidth}
			\begin{center}
				\includegraphics[width=\linewidth]{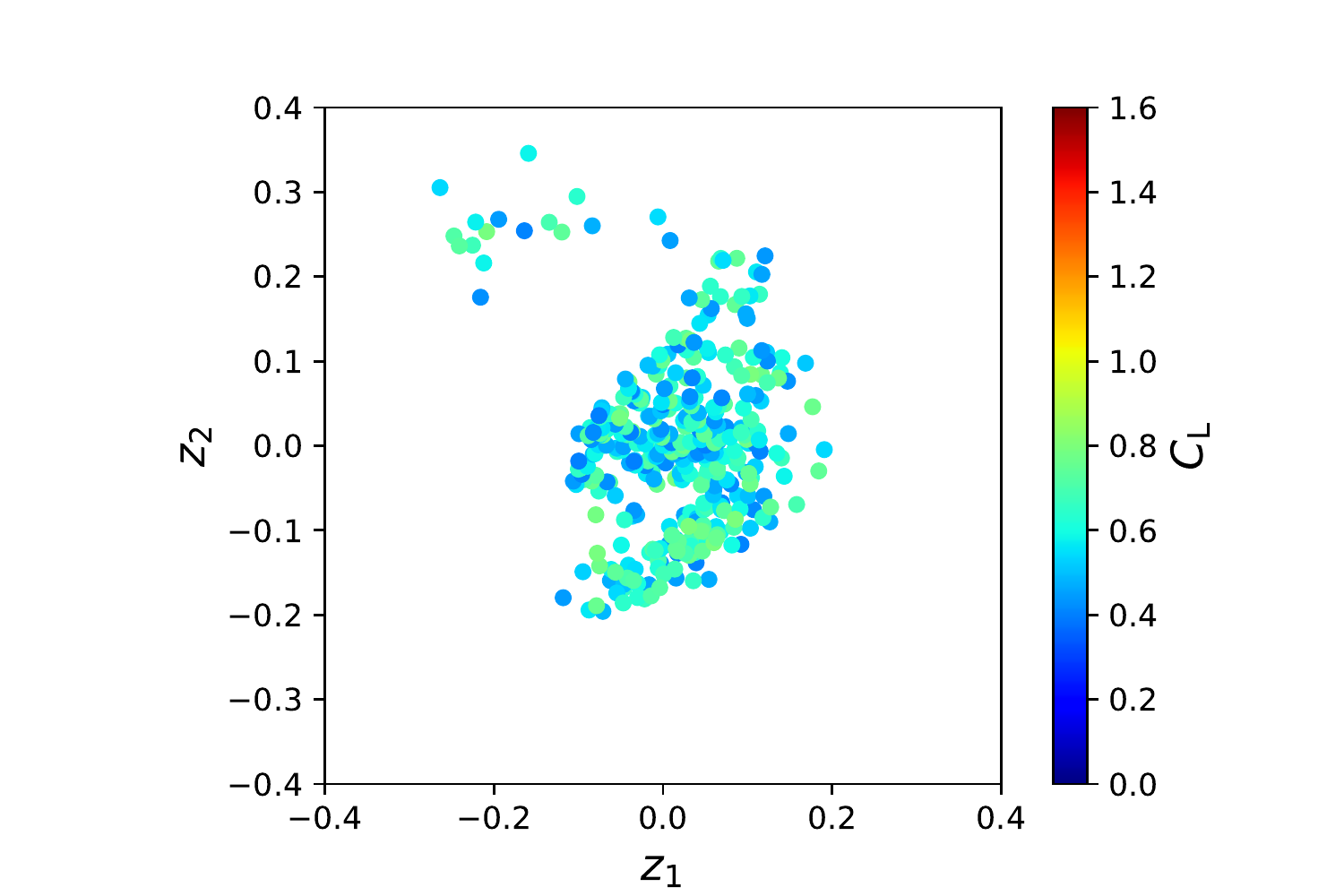}
				\par
				{(d) ${\mathcal N}$-CVAE ($0.4 \leq C_{\rm L}<0.8$).}
			\end{center}
		\end{minipage}%
		\par
		\begin{minipage}[h]{0.5\textwidth}
			\begin{center}
				\includegraphics[width=\linewidth]{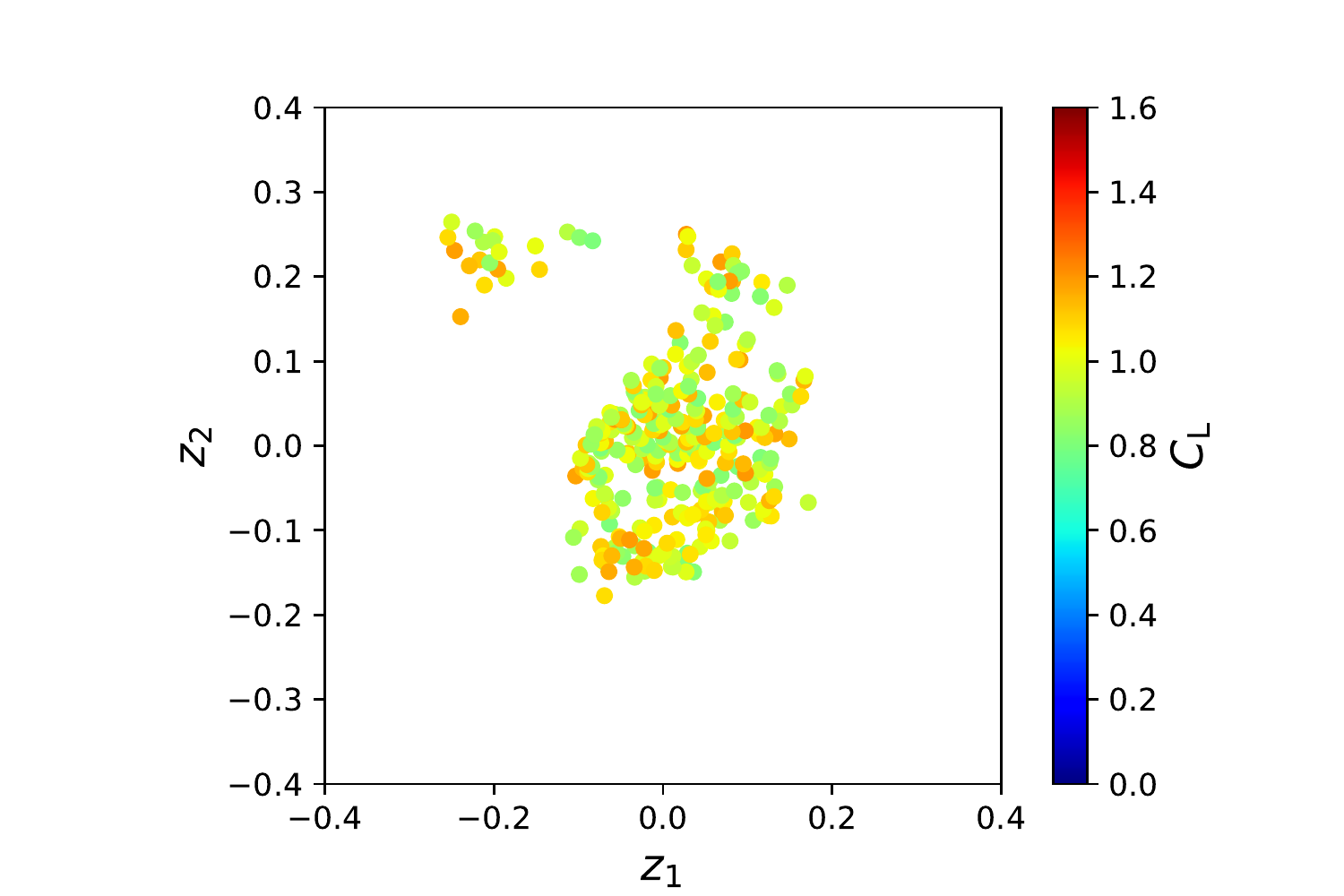}
				\par
				{(e) ${\mathcal N}$-CVAE ($0.8 \leq C_{\rm L}<1.2$).}
			\end{center}
		\end{minipage}%
		\begin{minipage}[h]{0.5\textwidth}
			\begin{center}
				\includegraphics[width=\linewidth]{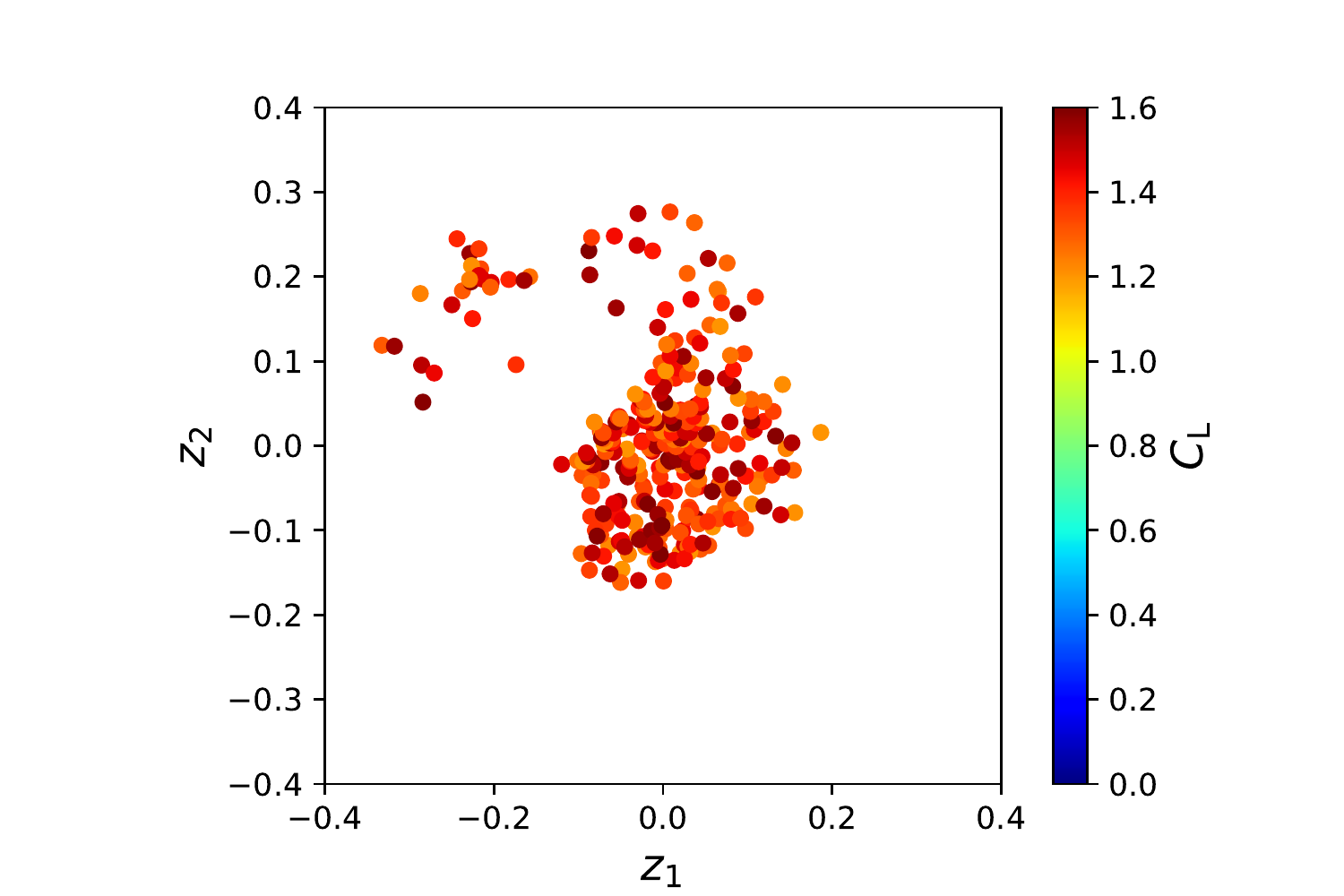}
				\par
				{(f) ${\mathcal N}$-CVAE ($1.2\leq C_{\rm L}$).}
			\end{center}
		\end{minipage}%
		\caption{Latent map of ${\mathcal N}$-VAE and ${\mathcal N}$-CVAE with respect to $C_{\rm L}$.}
		\label{fig:VAE_lat_vis}
	\end{center}
\end{figure}

\begin{figure}[htb]
	\begin{center}
		\begin{minipage}[h]{0.5\textwidth}
			\begin{center}
				\includegraphics[width=\linewidth]{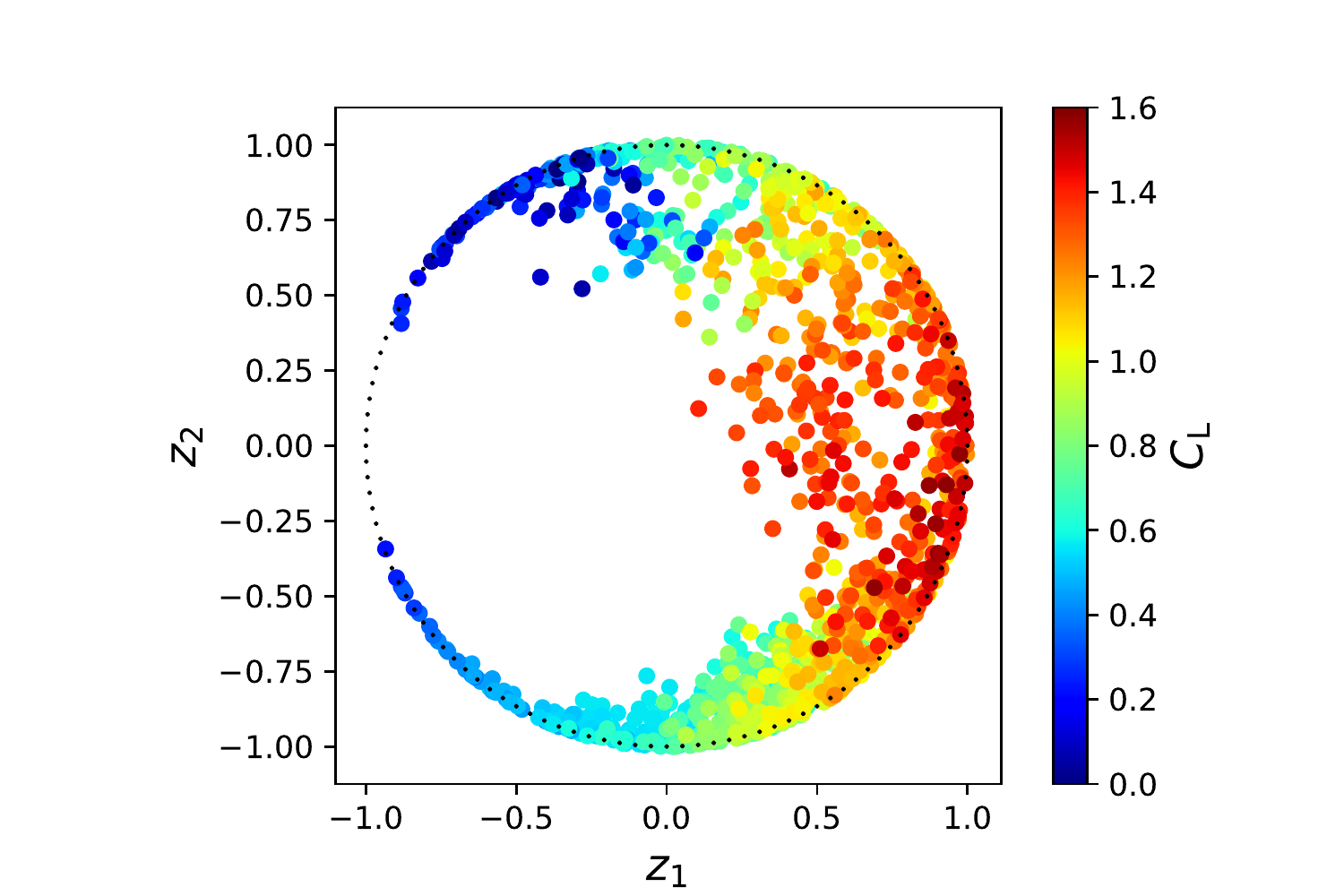}
				\par
				{(a) $z_3 \geq 0$.}
			\end{center}
		\end{minipage}%
		\begin{minipage}[h]{0.5\textwidth}
			\begin{center}
				\includegraphics[width=\linewidth]{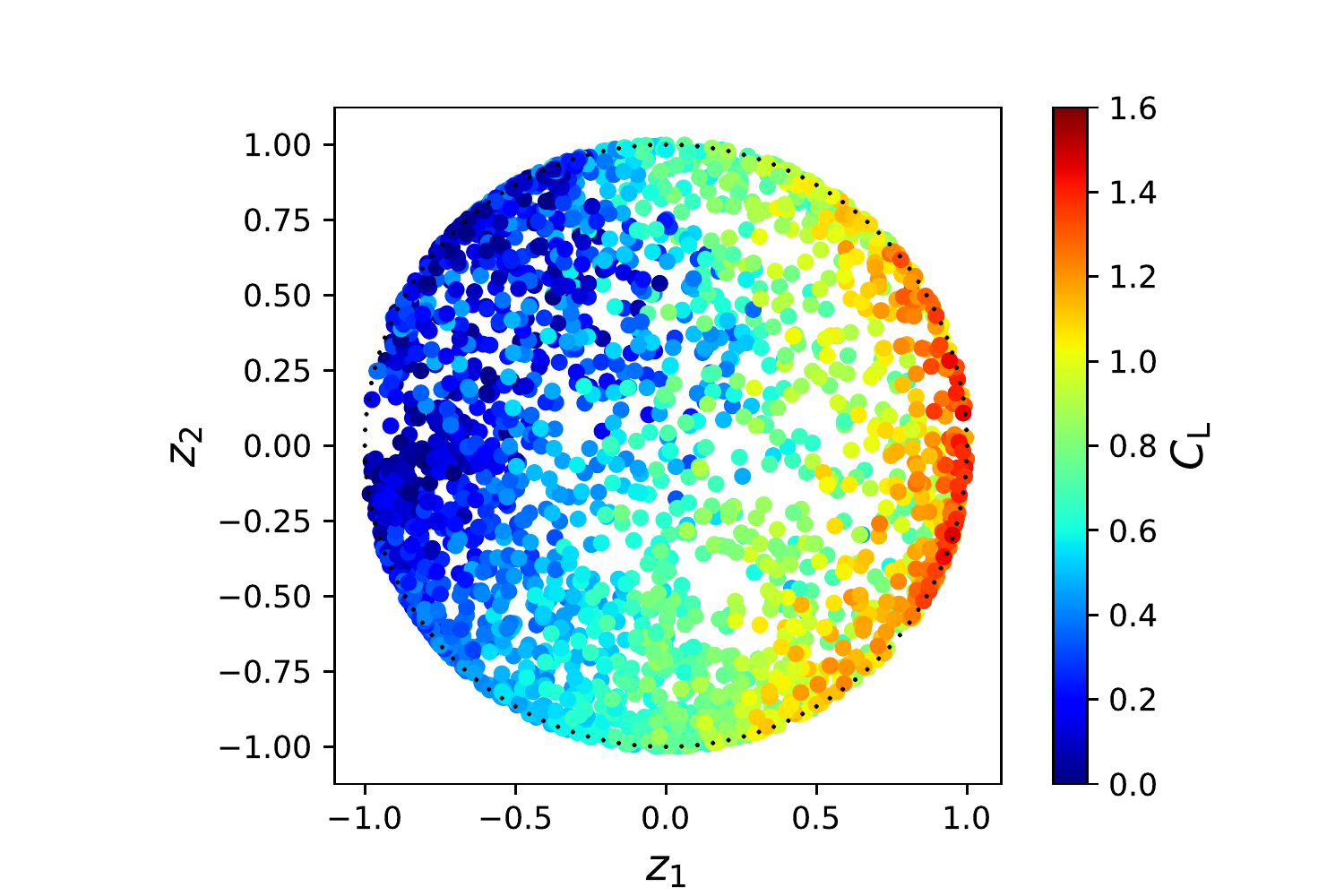}
				\par
				{(b) $z_3 < 0$.}
			\end{center}
		\end{minipage}%
		\caption{Latent map of ${\mathcal S}$-VAE with respect to  $C_{\rm L}$.}
		\label{fig:HSVAE_lat_vis_CL}
	\end{center}
\end{figure}

\begin{figure}[htb]
	\begin{center}
		\begin{minipage}[h]{0.5\textwidth}
			\begin{center}
				\includegraphics[width=\linewidth]{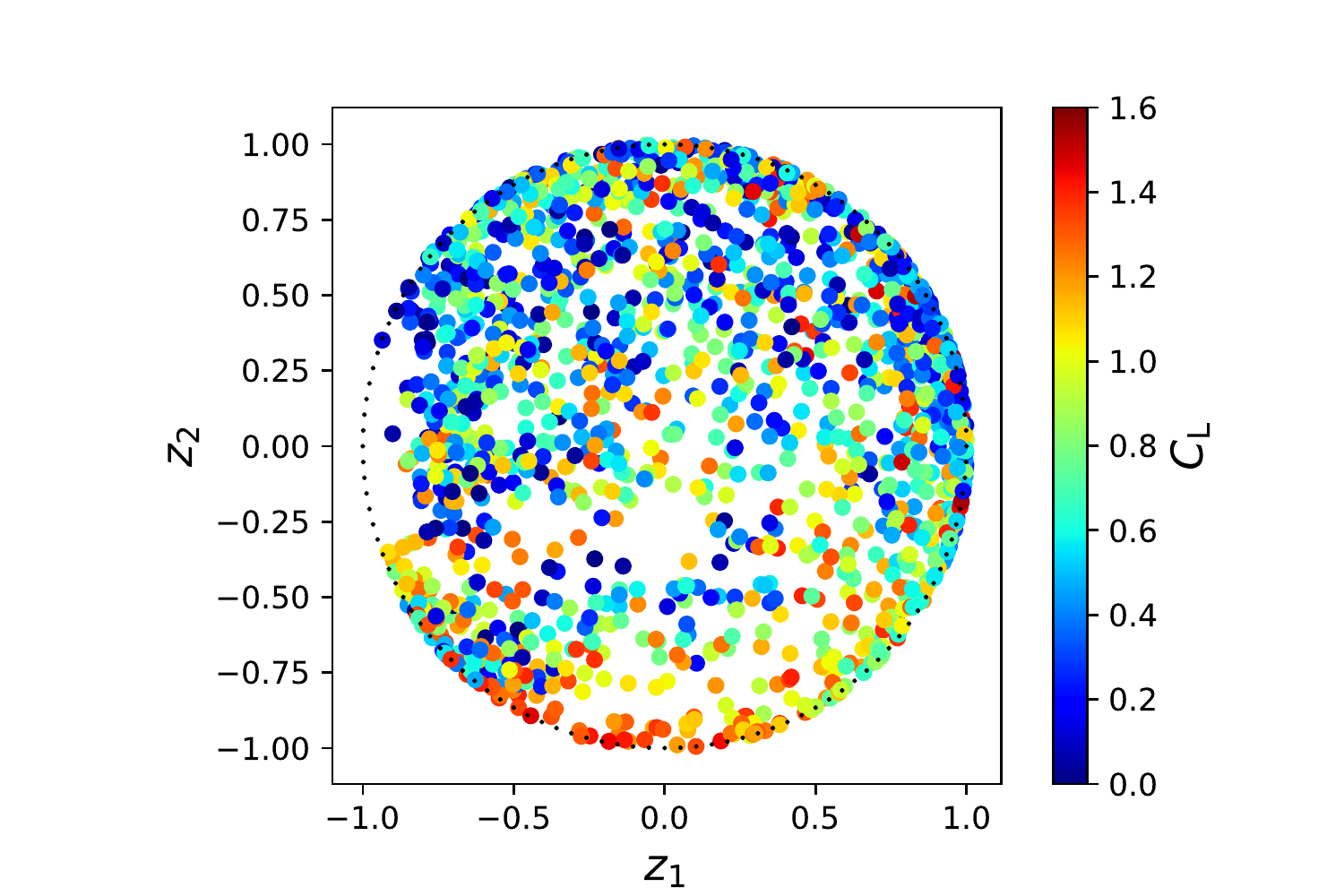}
				\par
				{(a) $z_3 \geq 0$.}
			\end{center}
		\end{minipage}%
		\begin{minipage}[h]{0.5\textwidth}
			\begin{center}
				\includegraphics[width=\linewidth]{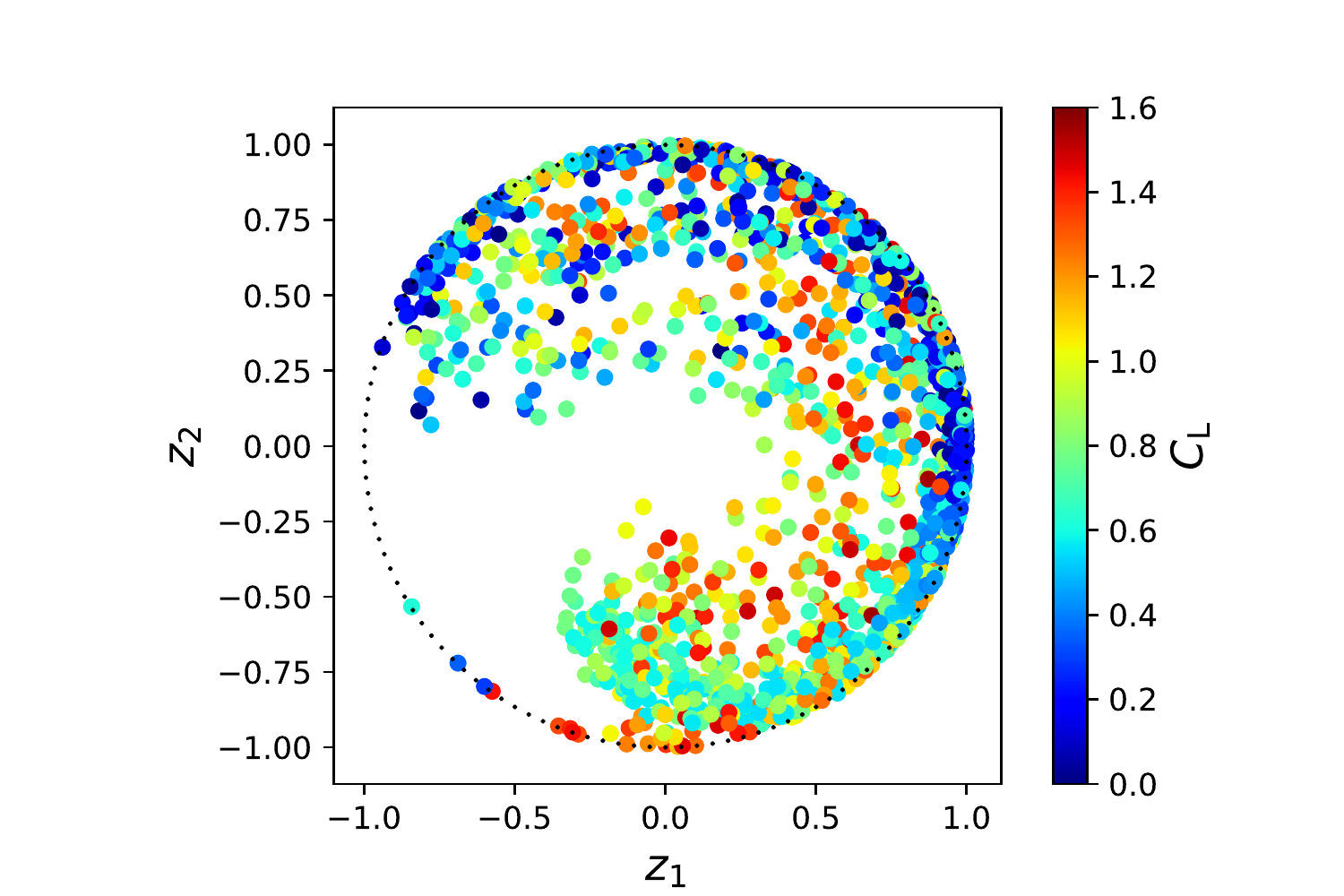}
				\par
				{(b) $z_3 < 0$.}
			\end{center}
		\end{minipage}%
		\par
		\begin{minipage}[h]{0.25\textwidth}
			\begin{center}
				\includegraphics[width=1.2\linewidth]{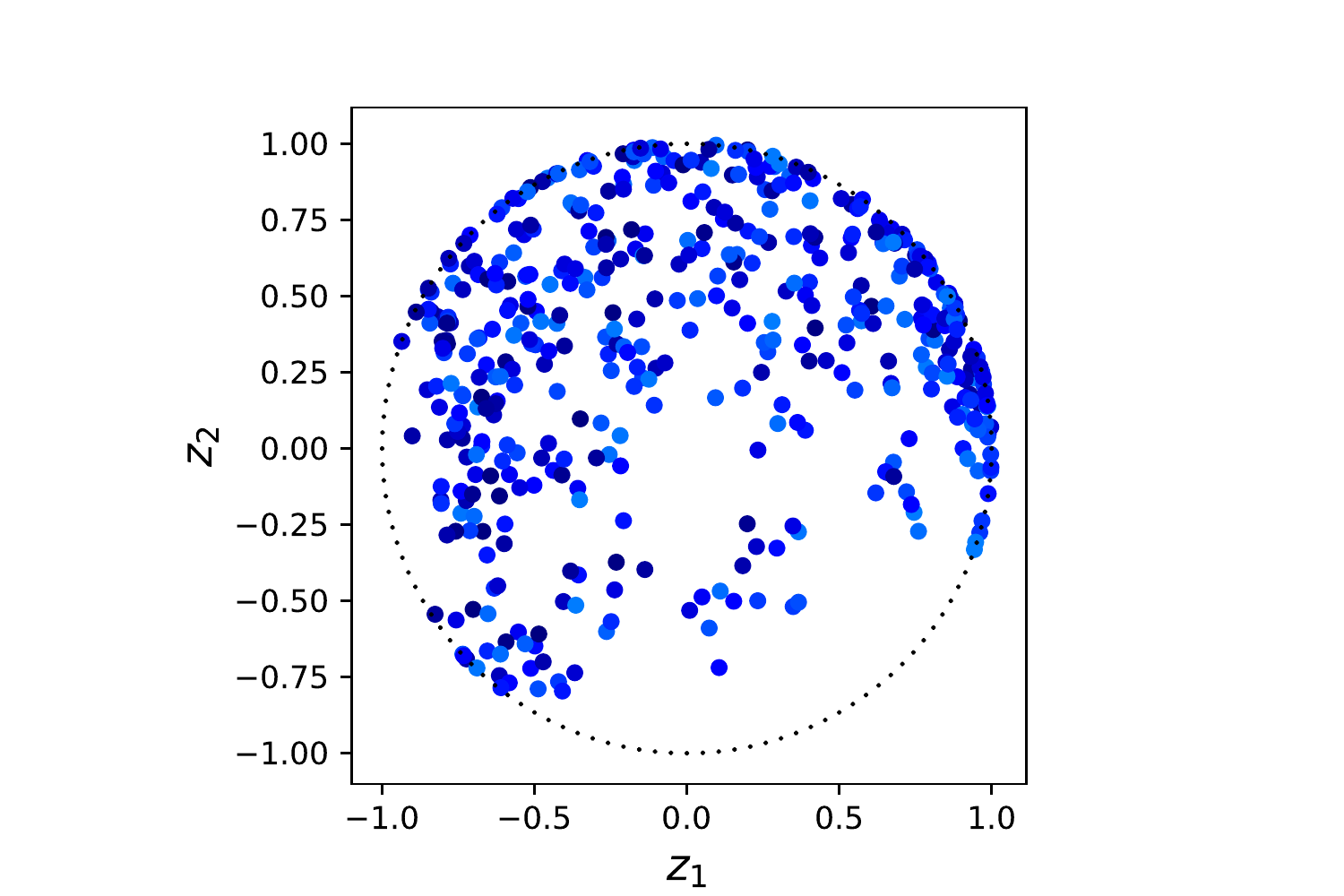}
				\par
				{(c) $z_3 \geq 0$ and $C_{\rm L}<0.4$.}
			\end{center}
		\end{minipage}%
		\begin{minipage}[h]{0.25\textwidth}
			\begin{center}
				\includegraphics[width=1.2\linewidth]{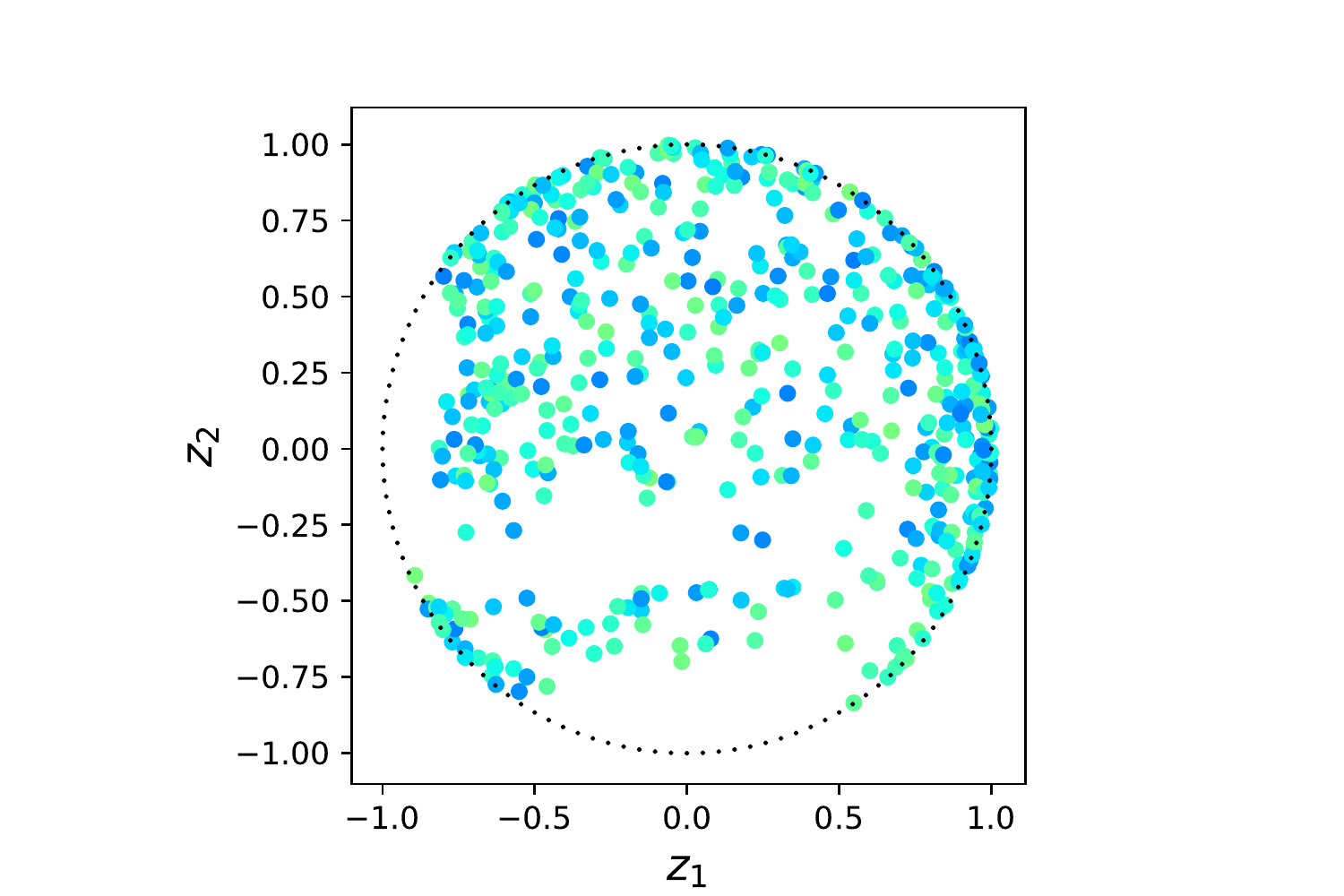}
				\par
				{(d) $z_3 \geq 0$ and $0.4\leq C_{\rm L}<0.8$.}
			\end{center}
		\end{minipage}%
		\begin{minipage}[h]{0.25\textwidth}
			\begin{center}
				\includegraphics[width=1.2\linewidth]{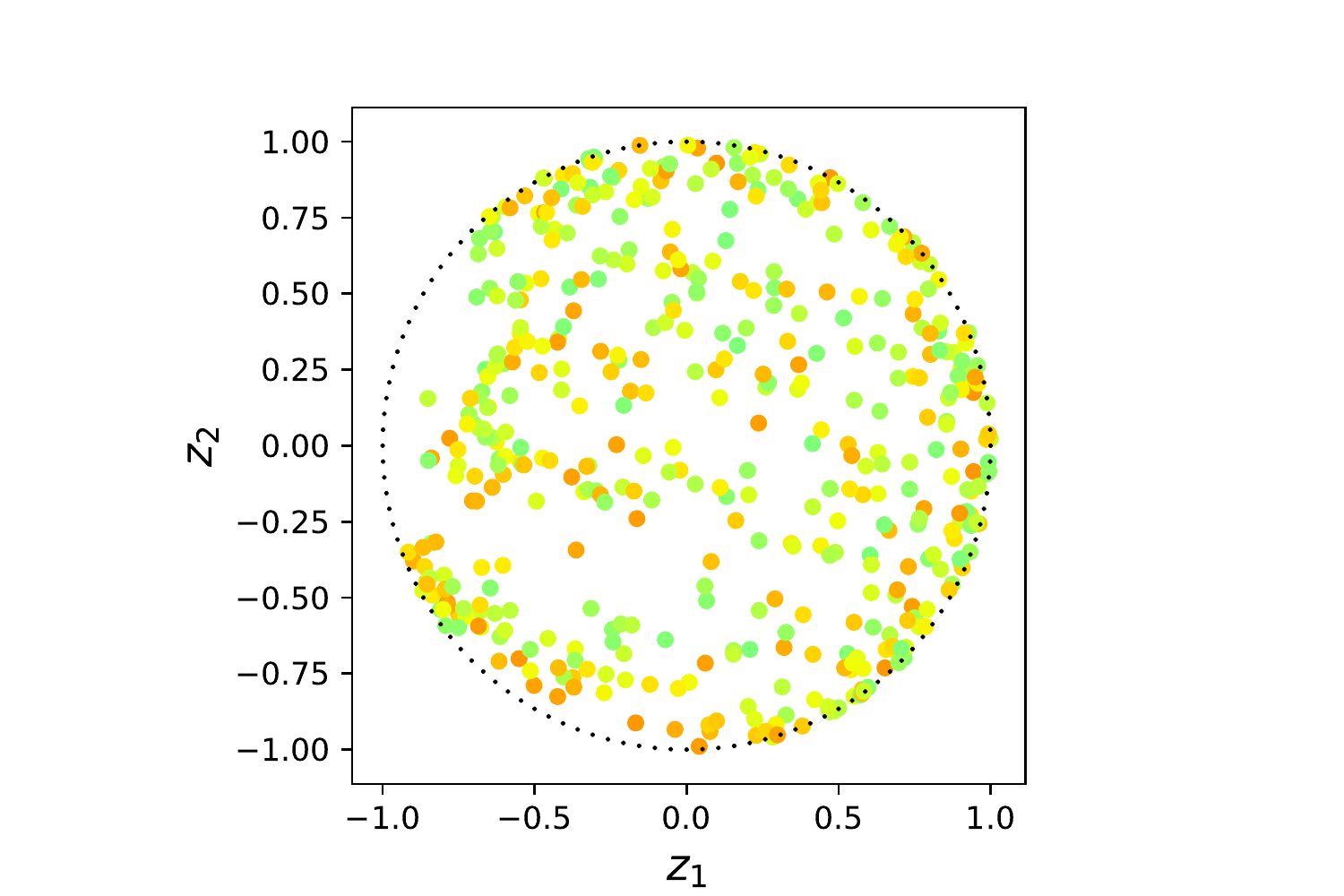}
				\par
				{(e) $z_3 \geq 0$ and $0.8 \leq C_{\rm L}<1.2$.}
			\end{center}
		\end{minipage}%
		\begin{minipage}[h]{0.25\textwidth}
			\begin{center}
				\includegraphics[width=1.2\linewidth]{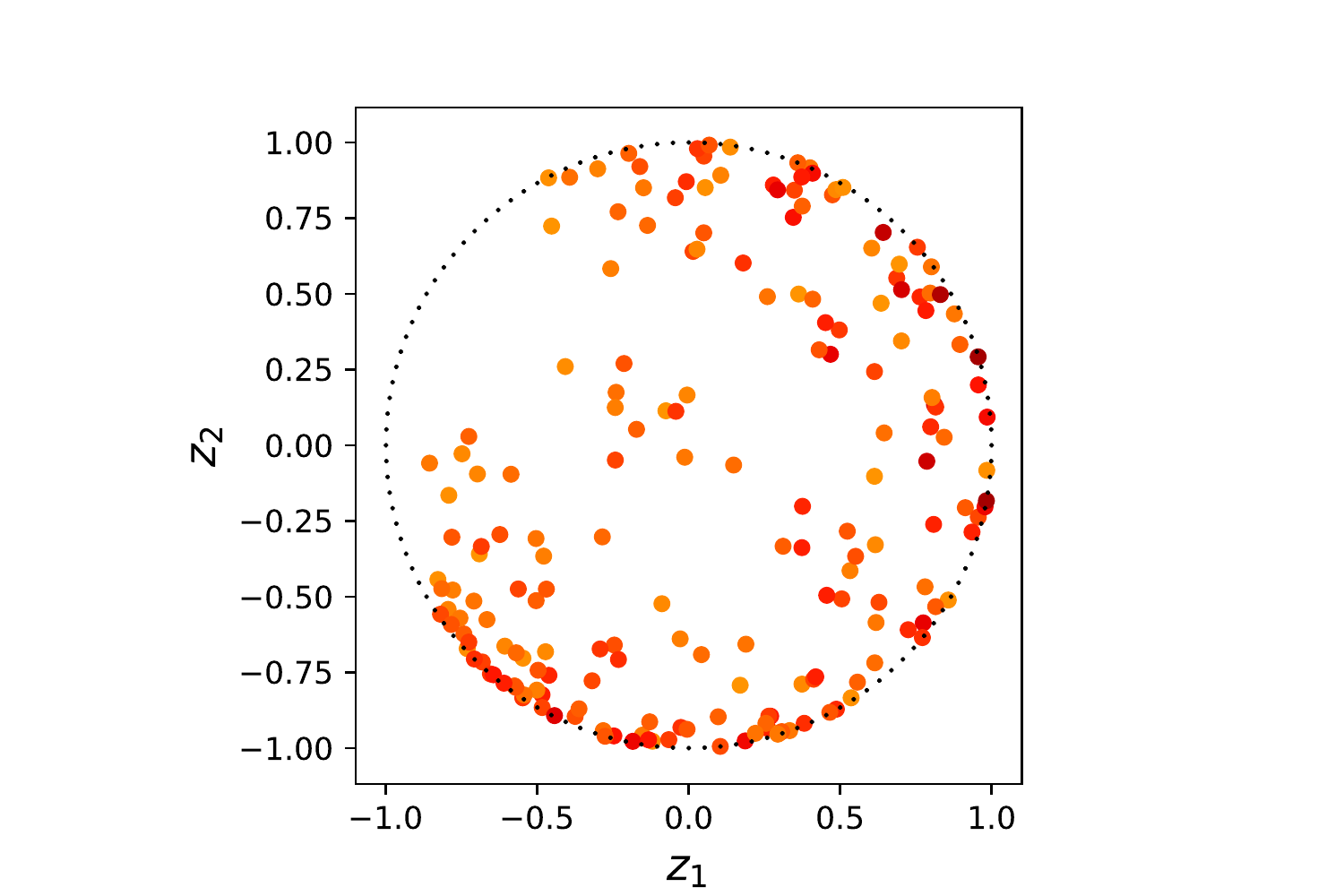}
				\par
				{(f) $z_3 \geq 0$ and $1.2\leq C_{\rm L}$.}
			\end{center}
		\end{minipage}%	
		\par
		\begin{minipage}[h]{0.25\textwidth}
			\begin{center}
				\includegraphics[width=1.2\linewidth]{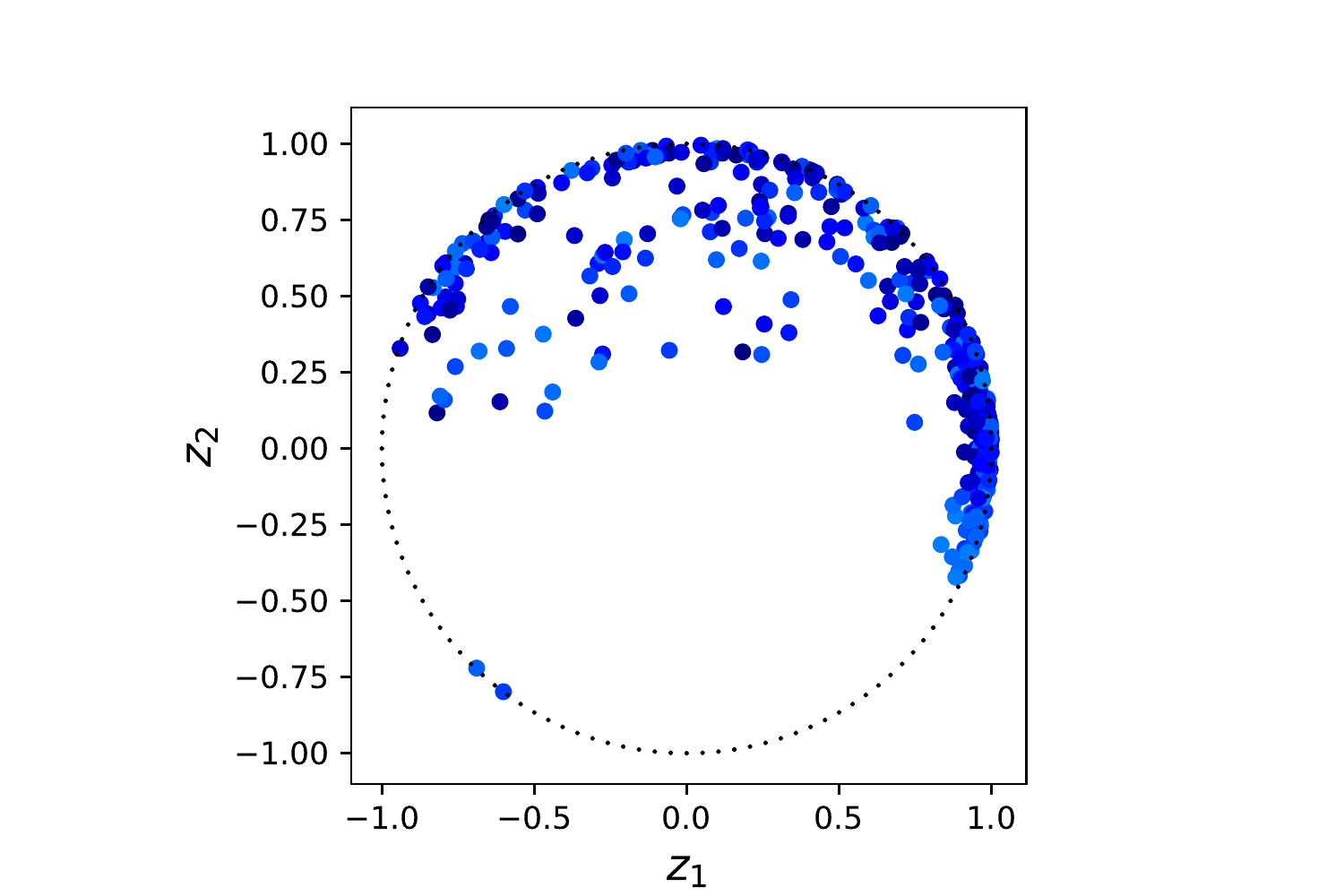}
				\par
				{(g) $z_3 < 0$ and $C_{\rm L}<0.4$.}
			\end{center}
		\end{minipage}%
		\begin{minipage}[h]{0.25\textwidth}
			\begin{center}
				\includegraphics[width=1.2\linewidth]{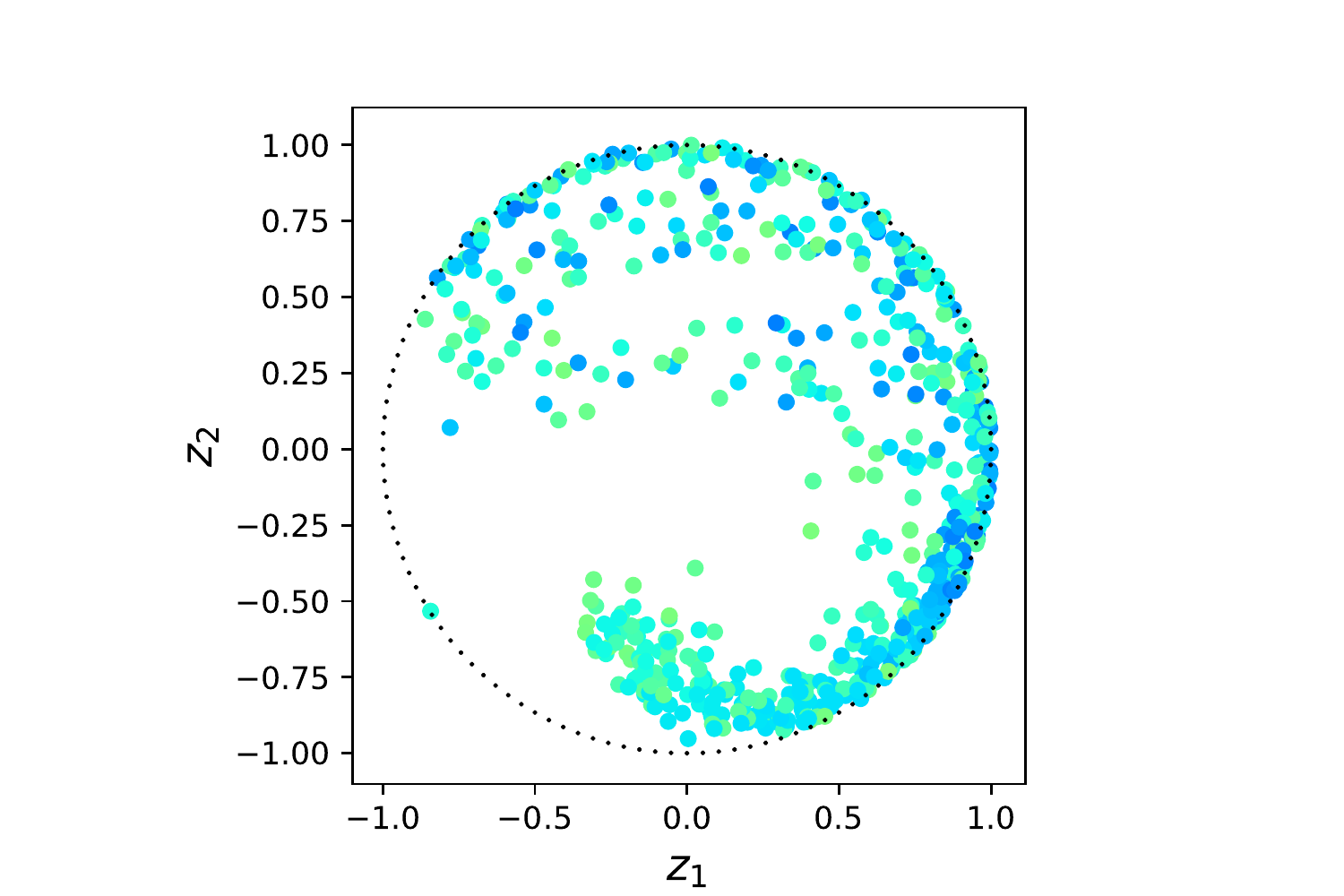}
				\par
				{(h) $z_3 < 0$ and $0.4\leq C_{\rm L}<0.8$.}
			\end{center}
		\end{minipage}%
		\begin{minipage}[h]{0.25\textwidth}
			\begin{center}
				\includegraphics[width=1.2\linewidth]{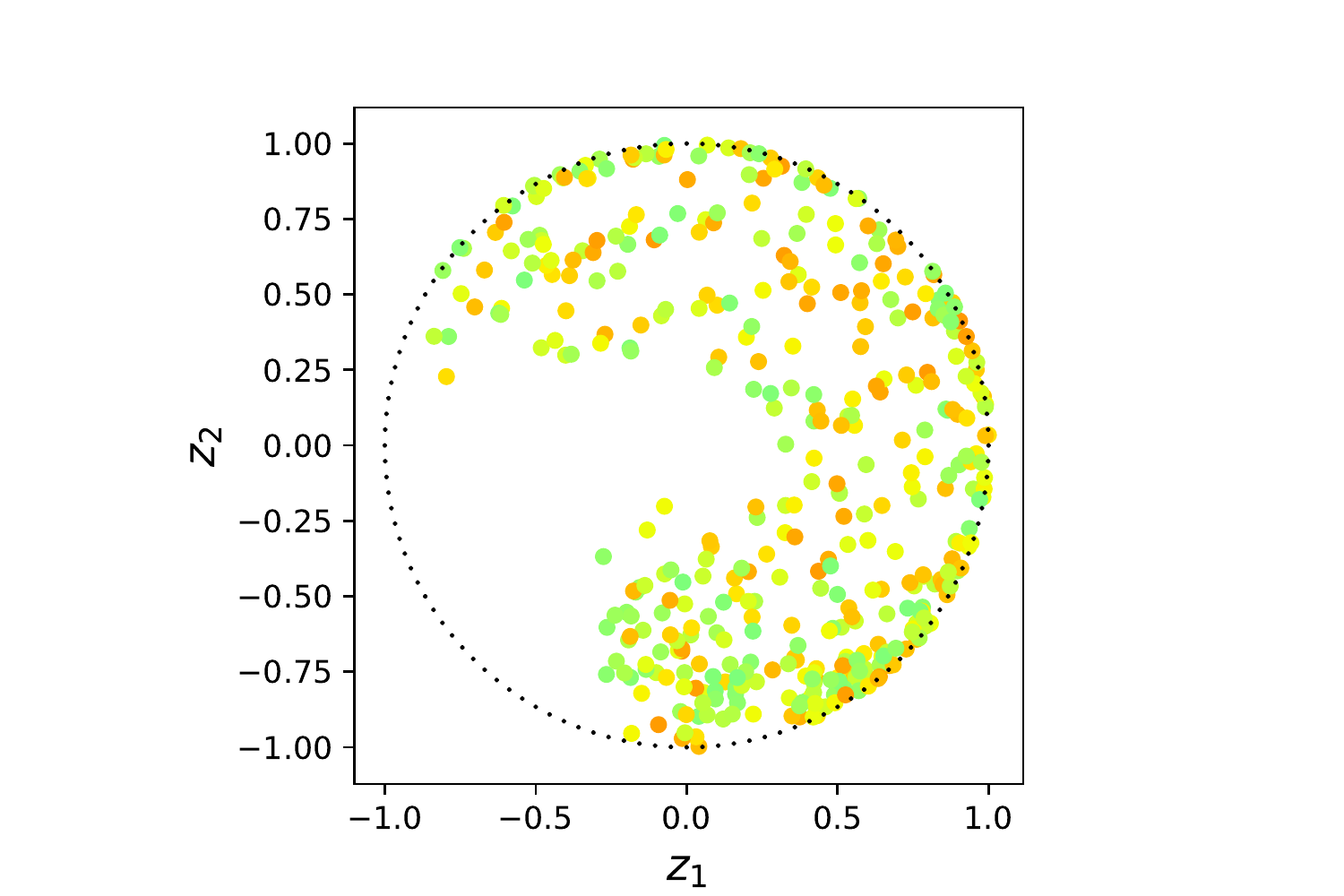}
				\par
				{(i) $z_3 < 0$ and $0.8 \leq C_{\rm L}<1.2$.}
			\end{center}
		\end{minipage}%
		\begin{minipage}[h]{0.25\textwidth}
			\begin{center}
				\includegraphics[width=1.2\linewidth]{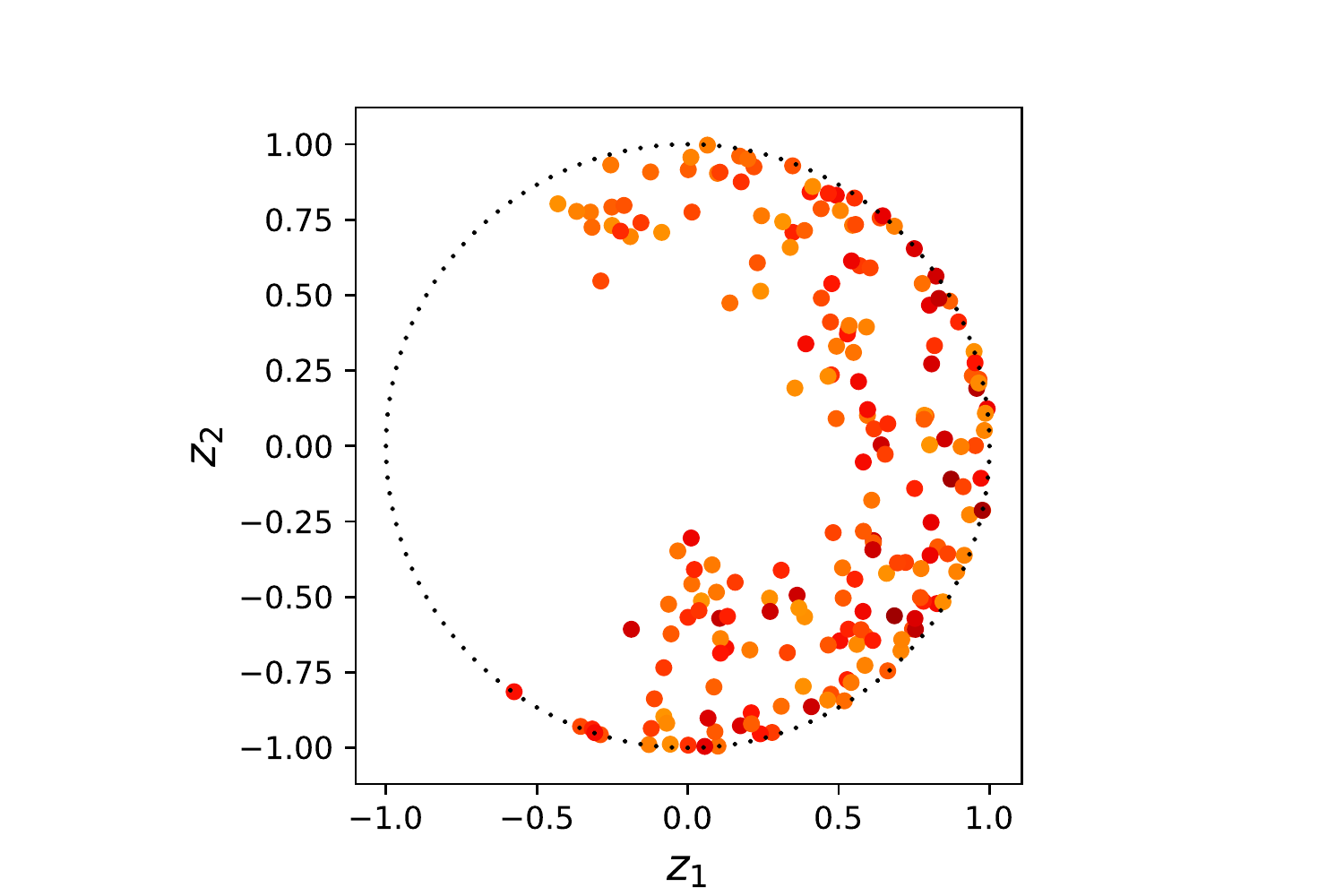}
				\par
				{(j) $z_3 < 0$ and $1.2\leq C_{\rm L}$.}
			\end{center}
		\end{minipage}%	
		\caption{Latent map of ${\mathcal S}$-CVAE.}
		\label{fig:Latentmap_HSCVAE}
	\end{center}
\end{figure}

The difference between the $\mathcal{S}$-CVAE and the $\mathcal{N}$-CVAE lies in the shape of the latent space and the KL divergence. 
In this section, the latent space is visualized and the difference between both the models is observed. 
In addition to the CVAE, the VAE model is also analyzed in this section for comparison. 
The architecture of the VAE model is nearly identical to that of the CVAE model, except that the label is not input into the model. 
The training data are fed into the encoder, and the mean latent vector $\bi{z}$ is plotted in \reffig{fig:VAE_lat_vis} -- \reffig{fig:Latentmap_HSCVAE} with a contour of $C_{\rm L}$. 
In the $\mathcal{N}$-VAE model (\reffig{fig:VAE_lat_vis} (a)), all the latent vectors shrink to $\bi{z}=(0,0)$, which indicates the KL collapse. In this case, all the data are encoded into the same area, and the decoder can output only one type of shape even though the training data has multiple shapes. 
The KL collapse does not occur in the $\mathcal{S}$-CVAE model (\reffig{fig:HSVAE_lat_vis_CL}). 
The latent vectors are widely distributed on the hypersphere. 
The input of the VAE model is only a set of shapes, and does not include the information on $C_{\rm L}$. However, $C_{\rm L}$ gradually changes in the latent space. 
This may be due to the strong relationship between the $C_{\rm L}$ and the airfoil geometry. 

Additionally, the KL collapse does not occur in the $\mathcal{N}$-CVAE and the $\mathcal{S}$-CVAE models (\reffig{fig:VAE_lat_vis} (b) and \reffig{fig:Latentmap_HSCVAE} (a) and (b), respectively) as well, as observed in Section 4.1. 
In the CVAE model, the input of the decoder includes the latent vector as well as the label, $C_{\rm L}$. Hence, the decoder processes the three-dimensional data in this case. 
To visualize the three-dimensional space, $\bi{z}$ is plotted with the label, $C_{\rm L}$ in a specific range (\reffig{fig:VAE_lat_vis} (c) -- (f) and \reffig{fig:Latentmap_HSCVAE} (c) -- (j)). 

In both the $\mathcal{N}$- and $\mathcal{S}$-CVAE models, the latent vectors are embedded in a similar area, even if the range of $C_{\rm L}$ is changed. 
This explains why, in Section 4.1, when one latent point is set and $C_{\rm L}$ is changed, the decoder outputs the appropriate airfoils. Conversely, for example, if one point is selected in the latent space, where the training data are embedded around the point in $C_{\rm L}<0.4$ and are not embedded in $1.2<C_{\rm L}$, then the decoder may output the wired shapes from the point for $1.2<C_{\rm L}$. 

When $\mathcal{N}$- and $\mathcal{S}$-CVAE models are compared (\reffig{fig:VAE_lat_vis} (b) and \reffig{fig:Latentmap_HSCVAE} (a) and (b)), the latent vectors are embedded in a narrower area in $\mathcal{N}$-CVAE when compared to $\mathcal{S}$-CVAE. 
As explained in the KL collapse in $\mathcal{N}$-VAE, if two training data are embedded in a very narrow area, the decoder cannot distinguish between these them, and outputs only one data even though the training data are different. 
$\mathcal{S}$-CVAE separates the training data and embeds it in a wider area. 
This difference in the latent space explains the difference in the shape variation in Section 4.1.

\afterpage{\clearpage}

\section{Combining the NACA and the Joukowski airfoils}
This section presents the utilization of the NACA and the Joukowski airfoil datasets. The aim is to generate novel shapes which are different from those of the NACA and the Joukowski airfoils. 

\subsection{Generated shapes}
\begin{figure}[htb]
	\begin{center}
		\begin{minipage}[h]{0.5\textwidth}
			\begin{center}
				\includegraphics[width=\linewidth]{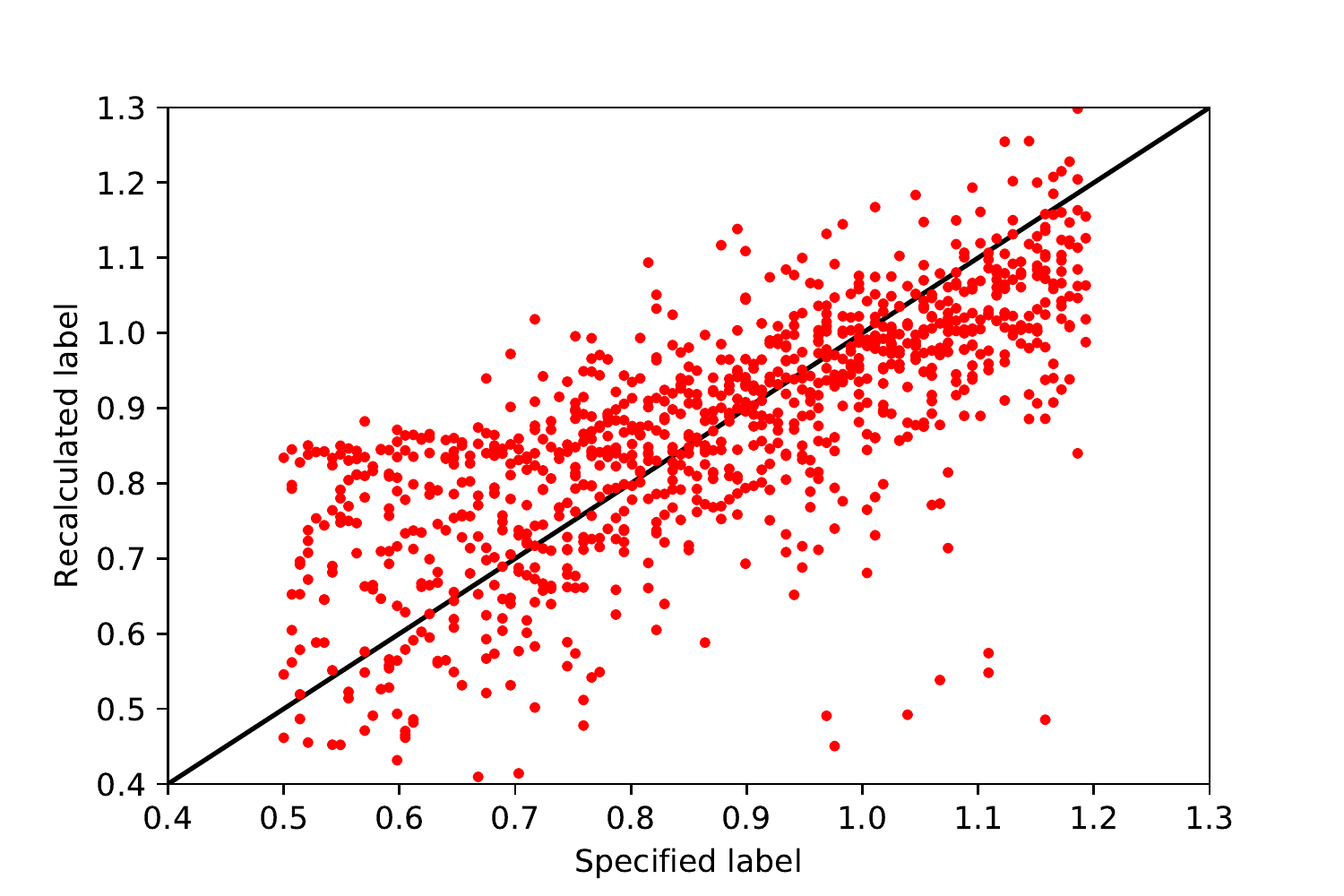}
				\par
				{(a) Error of $C_{\rm L}$ in $\mathcal{S}$-CVAE. ($\mathcal{L}_{C_{\rm L}}^{gen}=0.02371$)},
			\end{center}
		\end{minipage}%
		\begin{minipage}[h]{0.5\textwidth}
			\begin{center}
				\includegraphics[width=\linewidth]{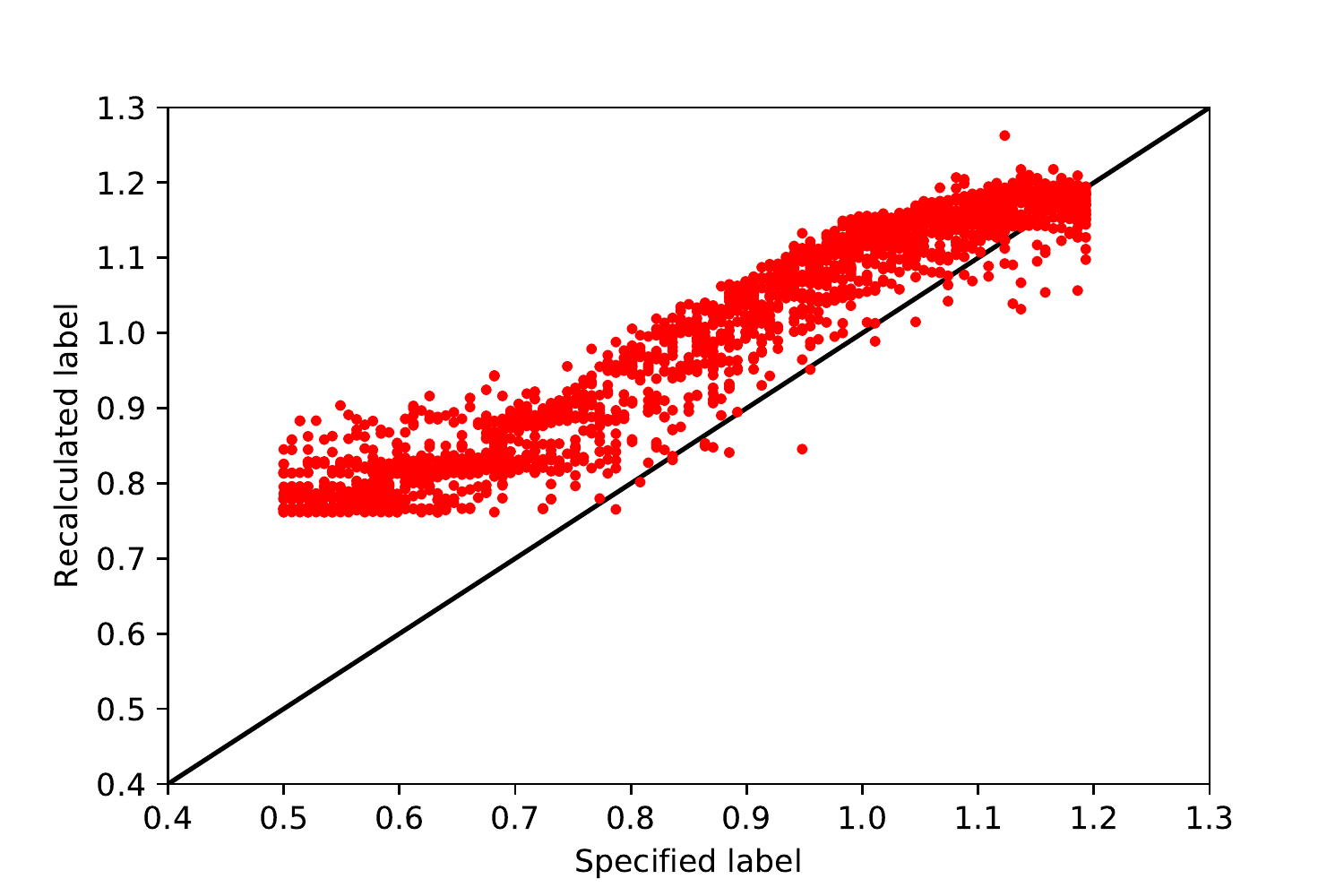}
				\par
				{(b) Error of $C_{\rm L}$ in $\mathcal{N}$-CVAE. ($\mathcal{L}_{C_{\rm L}}^{gen}=0.02341$)},
			\end{center}
		\end{minipage}%
		\caption{Comparison of error and $\mathcal{L}_{C_{\rm L}}^{gen}$.}
		\label{fig:error_NJ}
	\end{center}
\end{figure}

The NACA and the Joukowski datasets are combined into one dataset, and both the $\mathcal{N}$-CVAE and the $\mathcal{S}$-CVAE models are trained with a latent dimension of $d=2$. 
The new shapes are then generated by using the random latent vectors. 
The label and the recalculated $C_{\rm L}$ of the generated shapes are plotted in \reffig{fig:error_NJ}. 
The $\mathcal{L}_{C_{\rm L}}^{gen}$ presents a nearly identical value in both the models. 
However, in the $\mathcal{S}$-CVAE model, the result is scattered more broadly when compared to the $\mathcal{N}$-CVAE. 

\begin{figure}[htb]
	\begin{center}
		\begin{minipage}[h]{0.5\textwidth}
			\begin{center}
				\includegraphics[width=\linewidth]{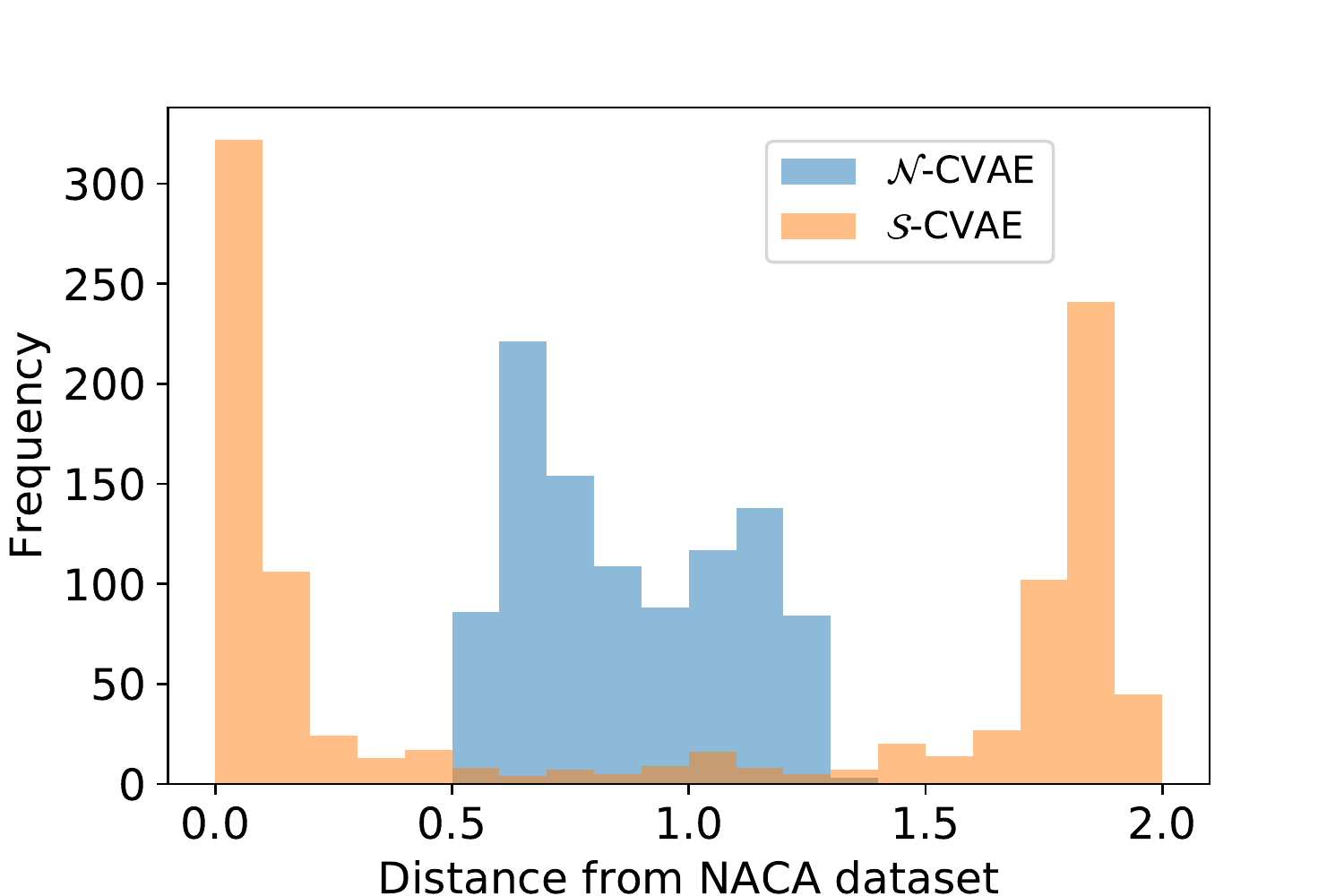}
				\par
				{(a) Histogram of the distance from the NACA dataset. }
			\end{center}
		\end{minipage}%
		\begin{minipage}[h]{0.5\textwidth}
			\begin{center}
				\includegraphics[width=\linewidth]{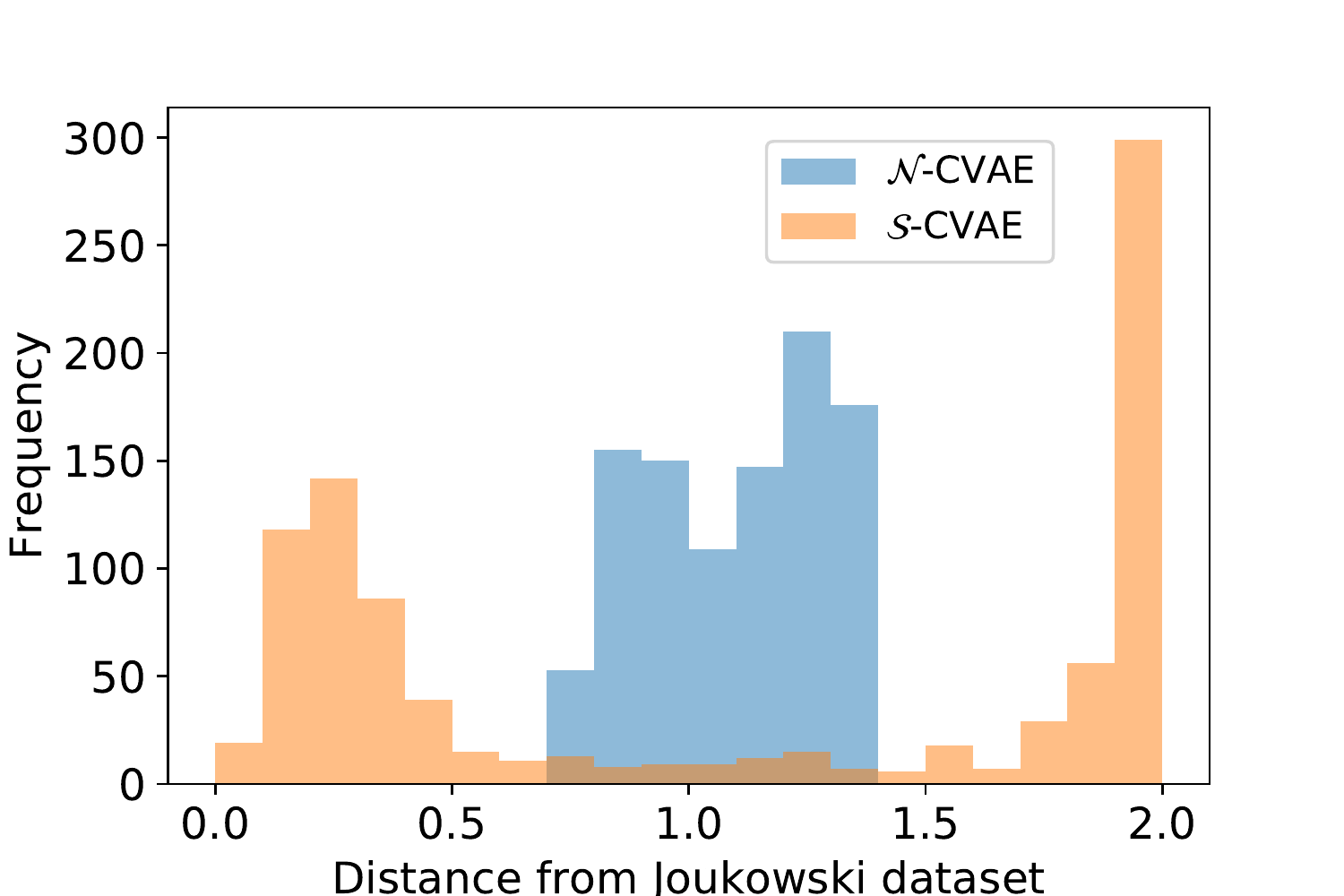}
				\par
				{(b) Histogram of distance from Joukowski dataset. }
			\end{center}
		\end{minipage}%
		\par
		\begin{minipage}[h]{1.0\textwidth}
			\begin{center}
				\includegraphics[width=0.5\linewidth, angle =270 ]{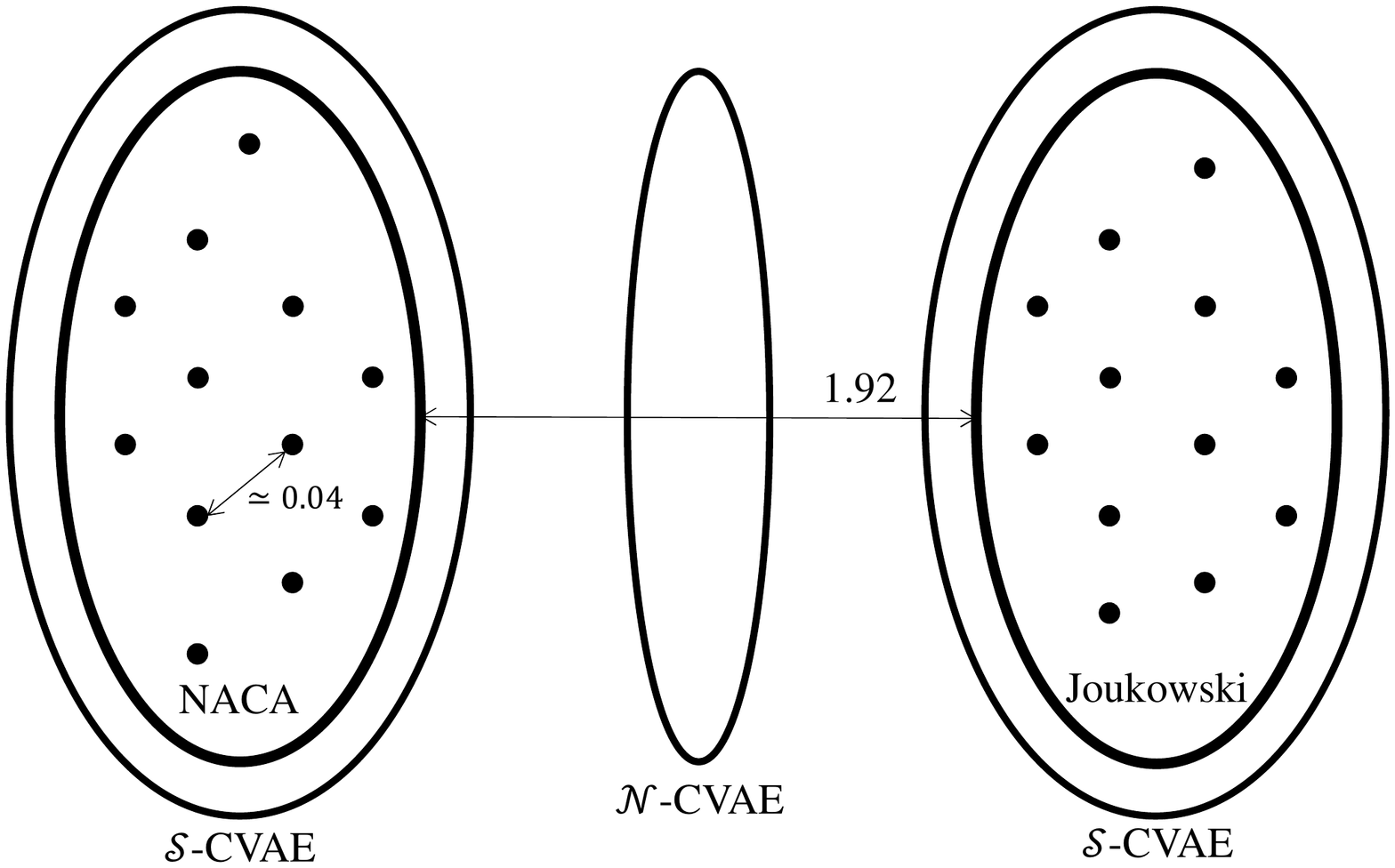}
				\par
				{(c) Sketch of distance between training data. }
			\end{center}
		\end{minipage}%
		\caption{Distance between generated shapes and train data.}
		\label{fig:dist}
	\end{center}
\end{figure}

The distance between each generated shape and the NACA and the Joukowski datasets are also calculated.
\reffig{fig:dist} (b) shows a sketch of the distance between a set of NACA airfoils and a set of Joukowski airfoils. The Euclidean distance between the two sets is $1.92$. 
The histograms are shown in \reffig{fig:dist}. 
The histogram of both models indicates that their distribution is completely different. 
In $\mathcal{S}$-CVAE, the histogram has two peaks around 0.0 and 2.0, which indicates that the generated shapes are similar to those of the NACA and the Joukowski airfoils, respectively. 
Conversely, the distance in $\mathcal{S}$-CVAE, is distributed in the range of $[0.5, 1.4]$. This implies that the generated shapes are located in the middle of the NACA and the Joukowski airfoils as illustrated in \reffig{fig:dist} (c). 
Consequently, the generated data in $\mathcal{N}$-CVAE are a combination of the NACA airfoils and the Joukowski airfoils.

\begin{figure}[htb]
	\begin{center}
		\begin{minipage}[h]{0.5\textwidth}
			\begin{center}
				\includegraphics[width=\linewidth]{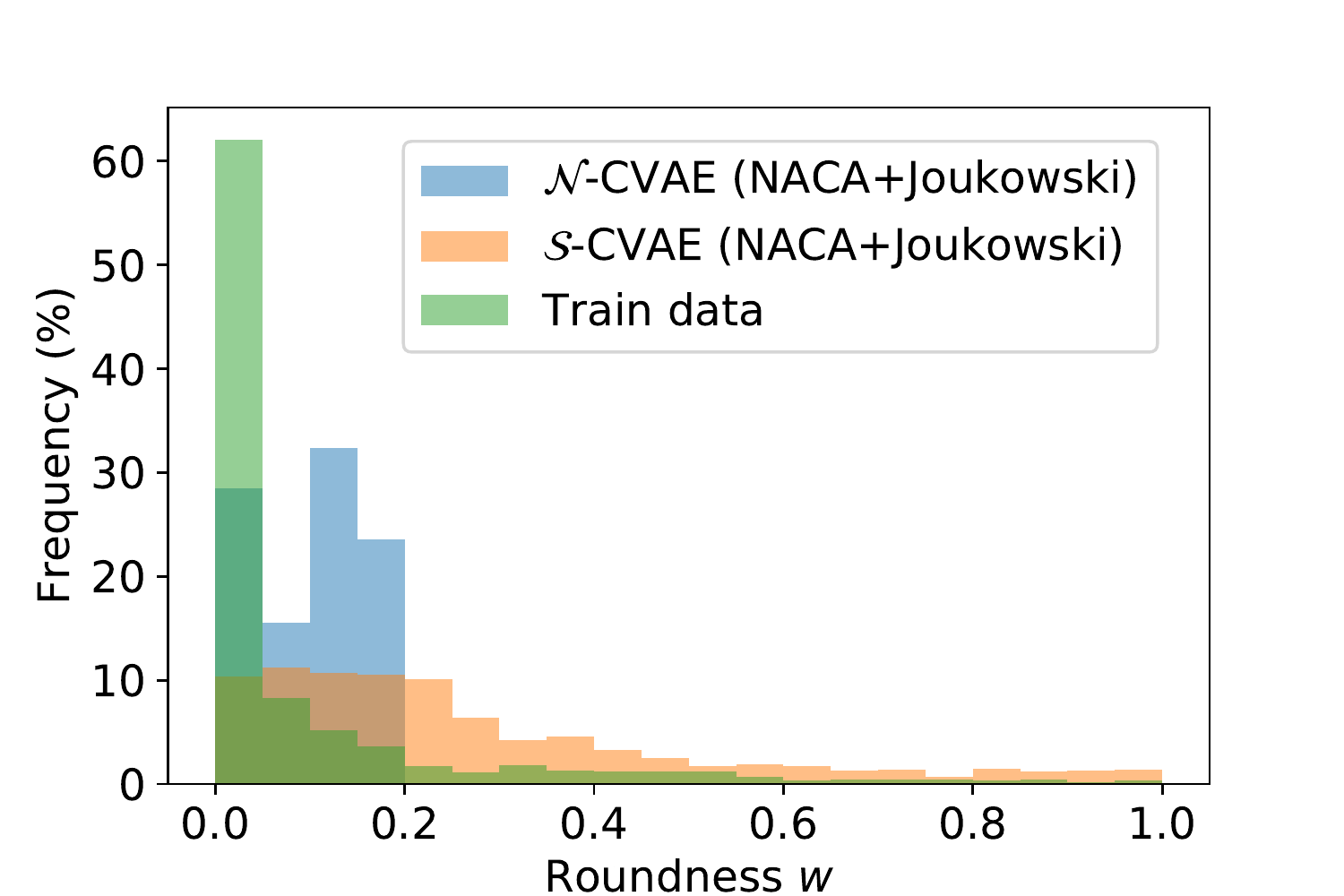}
				%\par
				%{(a) Error of $C_{\rm L}$ in $\mathcal{S}$-CVAE. MSE$=0.223798$.}
			\end{center}
		\end{minipage}%
		\caption{Histogram of roundness in Joukowski inverse transformation.}
		\label{fig:J_res}
	\end{center}
\end{figure}

The Joukowski inverse transformation is conducted on the generated shapes and the training data.
The result is shown in \reffig{fig:J_res}. 
The roundness, $w$, in the $\mathcal{S}$-CVAE model has a wide distribution, whereas $\mathcal{N}$-CVAE does not. 
Both the distance and the roundness results indicate that the variety of the generated shapes is large in $\mathcal{S}$-CVAE and small in $\mathcal{N}$-CVAE. 

\begin{figure}[htb]
	\begin{center}
		\begin{minipage}[h]{0.5\textwidth}
			\begin{center}
				\includegraphics[width=\linewidth]{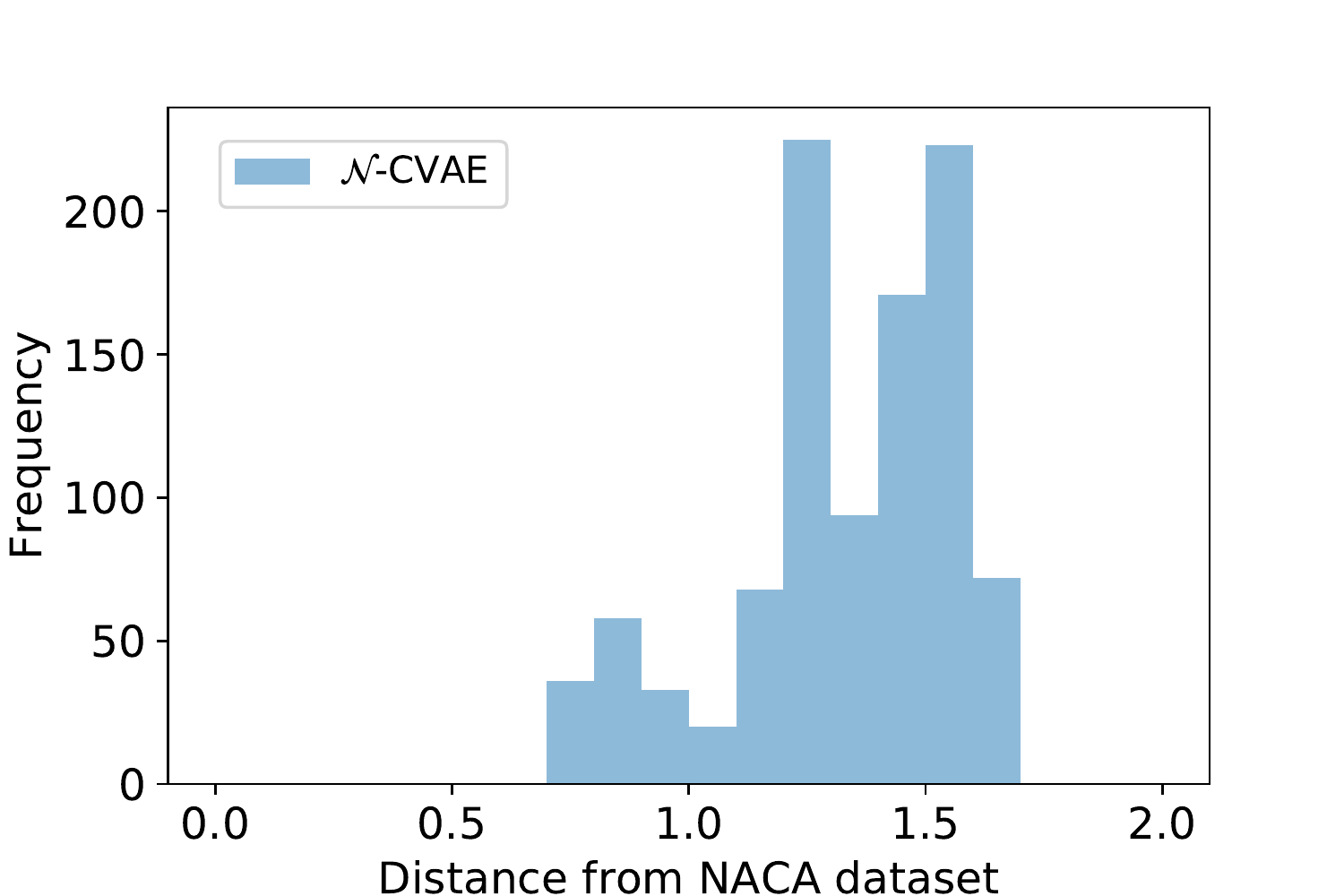}
				\par
				{(a) Histogram of the distance from the NACA dataset. }
			\end{center}
		\end{minipage}%
		\begin{minipage}[h]{0.5\textwidth}
			\begin{center}
				\includegraphics[width=\linewidth]{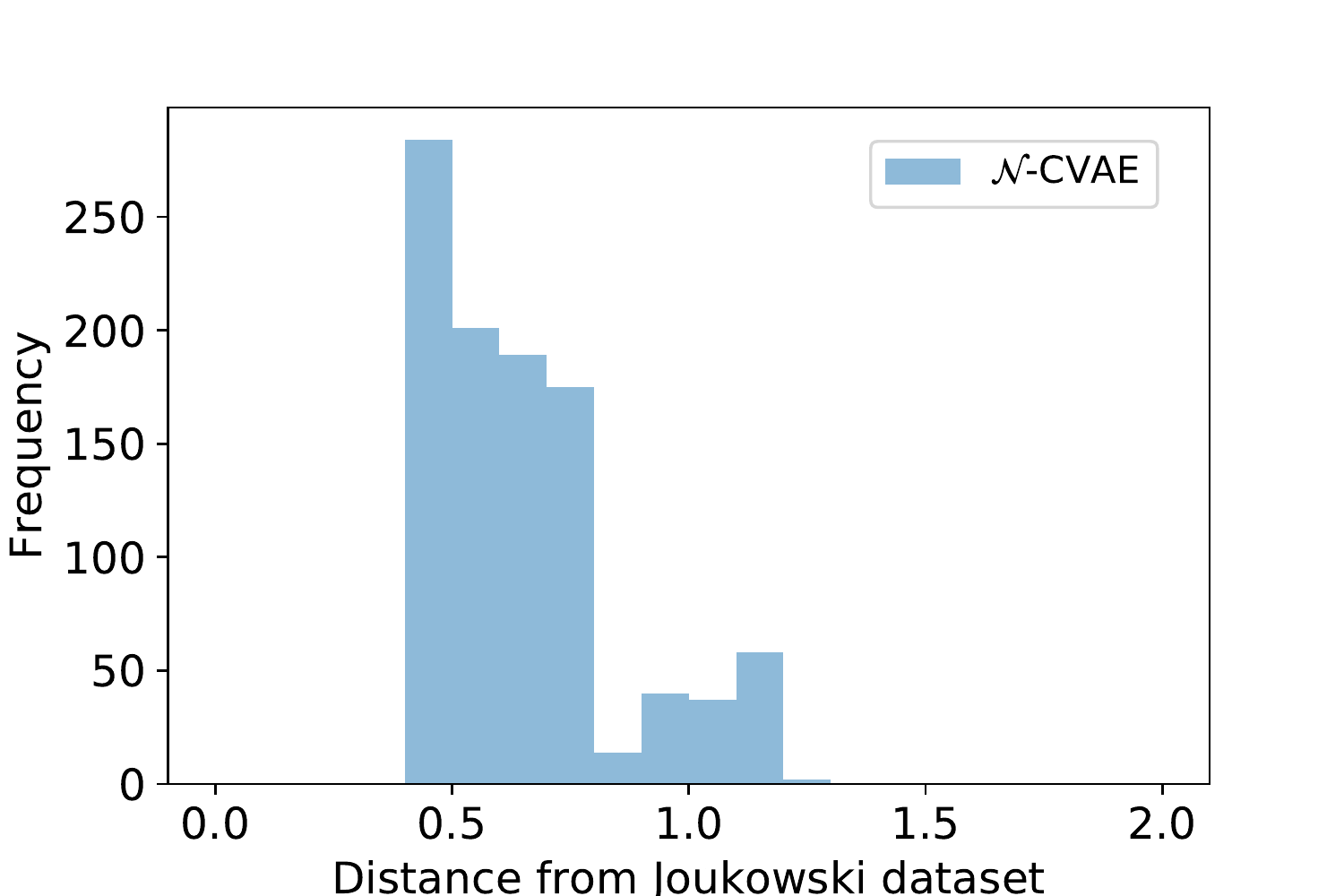}
				\par
				{(b) Histogram of distance from Joukowski dataset. }
			\end{center}
		\end{minipage}%
		\caption{Distance between generated shapes and train data ($3 \times{}$Joukowski case).}
		\label{fig:dist2}
	\end{center}
\end{figure}

\begin{figure}[htb]
	\begin{center}
		\begin{minipage}[h]{0.5\textwidth}
			\begin{center}
				\includegraphics[width=\linewidth]{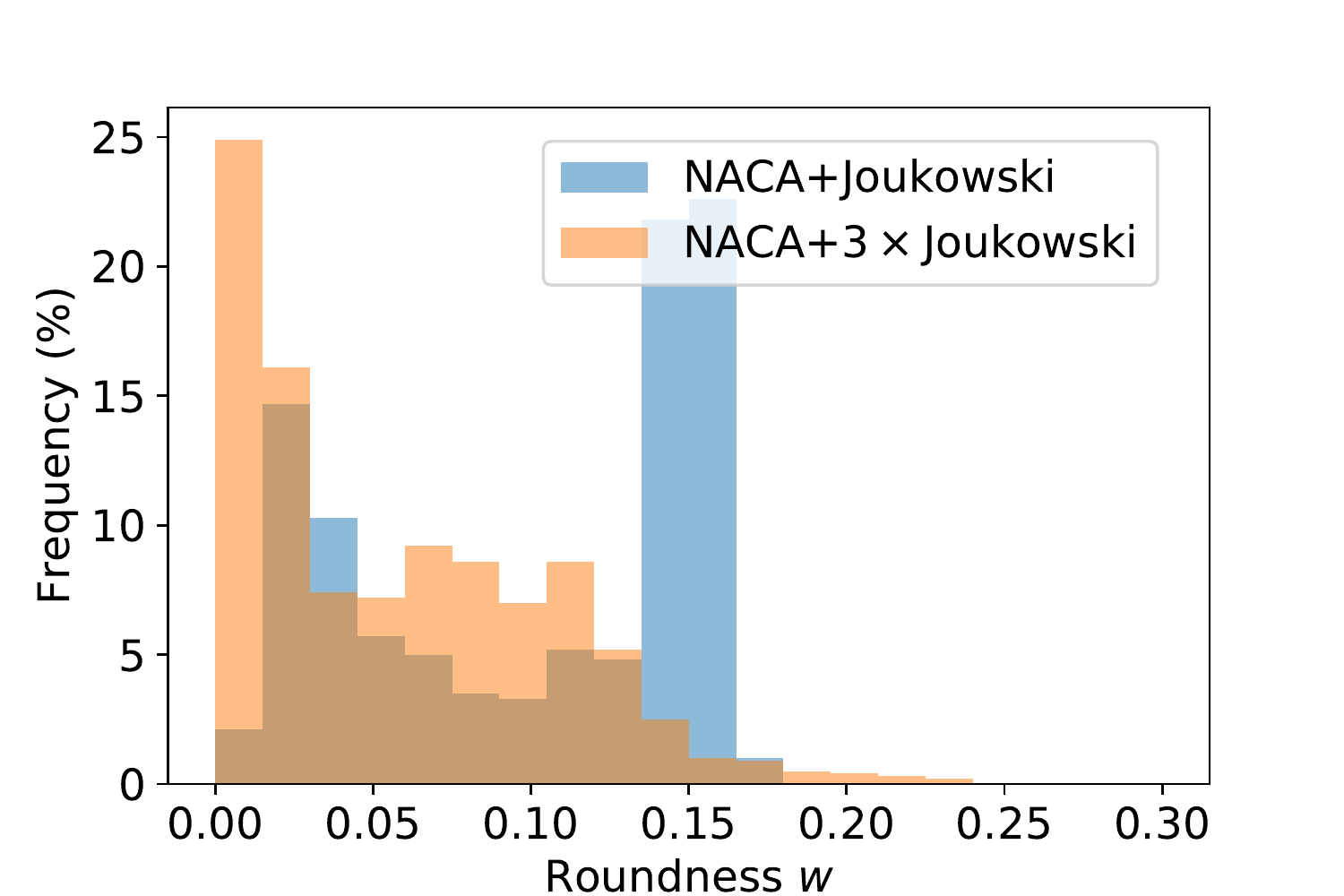}
				%\par
				%{(a) Error of $C_{\rm L}$ in $\mathcal{S}$-CVAE. MSE$=0.223798$.}
			\end{center}
		\end{minipage}%
		\caption{Histogram of roundness in Joukowski inverse transformation ($3 \times{}$Joukowski case).}
		\label{fig:J_res2}
	\end{center}
\end{figure}

The distance between generated shapes and train data can be controlled by changing the ratio of NACA and Joukowski data. We multiply Joukowski data by three by duplicating data, and trained $\mathcal{N}$-CVAE model ($3 \times{}$Joukowski case).
The distance and $w$ of the generated shapes are shown in \reffig{fig:dist2} and \reffig{fig:J_res2}. 
The generated shapes locate closer to Joukowski airfoils than \reffig{fig:dist}. 
$w$ also indicates smaller value than \reffig{fig:J_res2}. 
These result implies that the generated shapes can be controlled by changing the ratio of the dataset.

\subsection{Distribution in latent space}
\begin{figure}[htb]
	\begin{center}
		\begin{minipage}[h]{0.5\textwidth}
			\begin{center}
				\includegraphics[width=\linewidth]{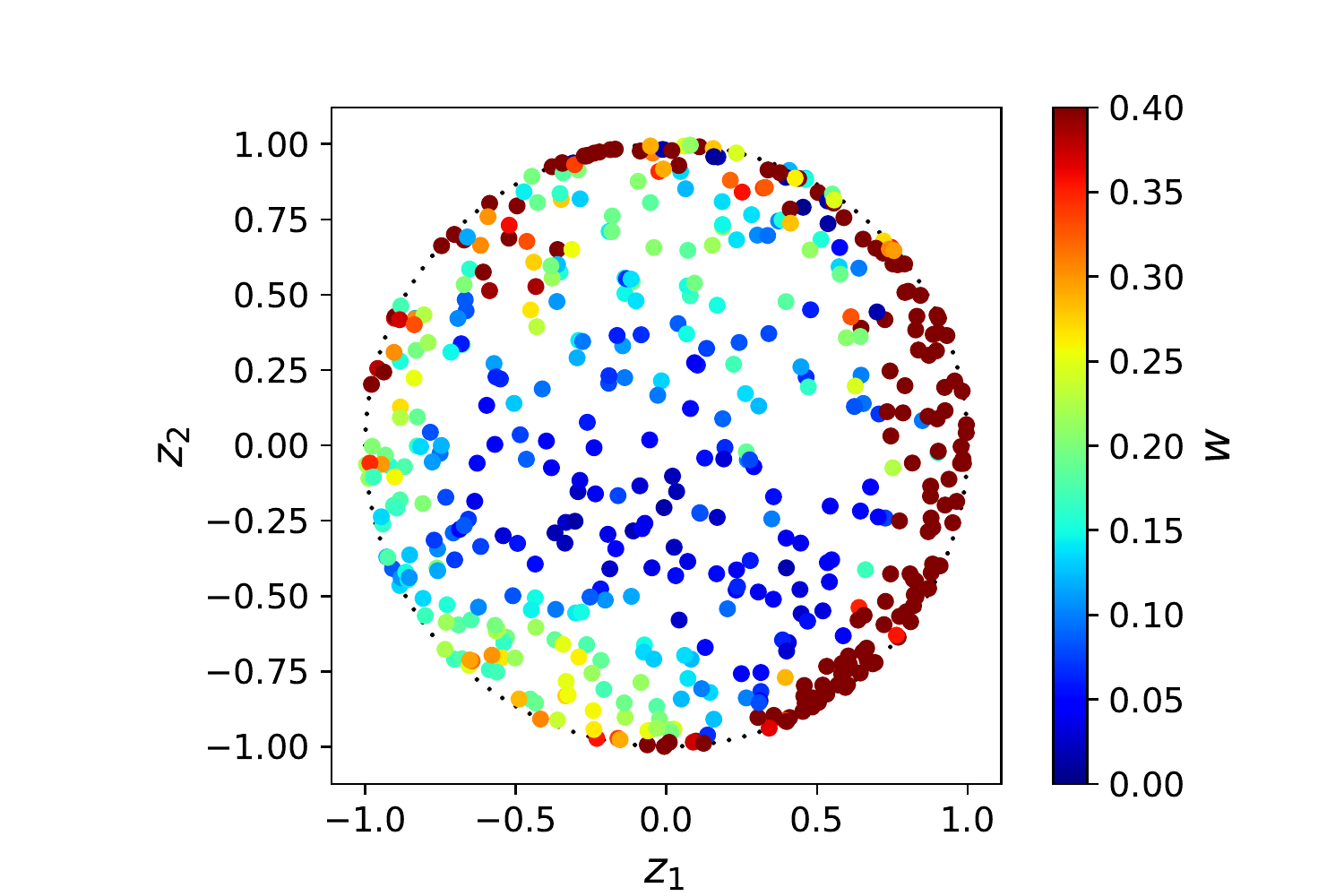}
				\par
				{(a) Plot of $w$ in $z_3\geq 0$. }
			\end{center}
		\end{minipage}%
		\begin{minipage}[h]{0.5\textwidth}
			\begin{center}
				\includegraphics[width=\linewidth]{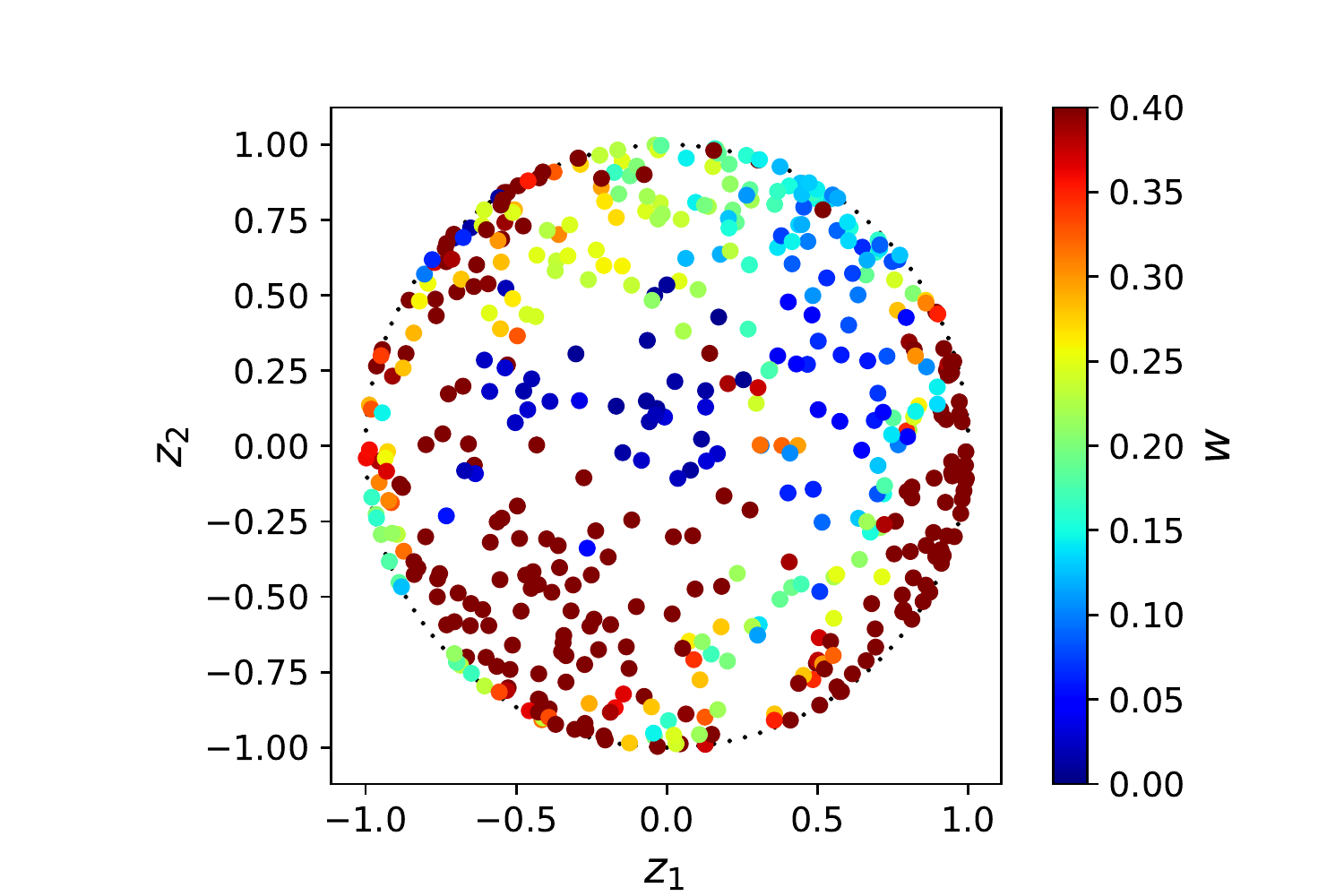}
				\par
				{(b) Plot of $w$ in $z_3<0$. }
			\end{center}
		\end{minipage}%
		\par
		\begin{minipage}[h]{0.5\textwidth}
			\begin{center}
				\includegraphics[width=\linewidth]{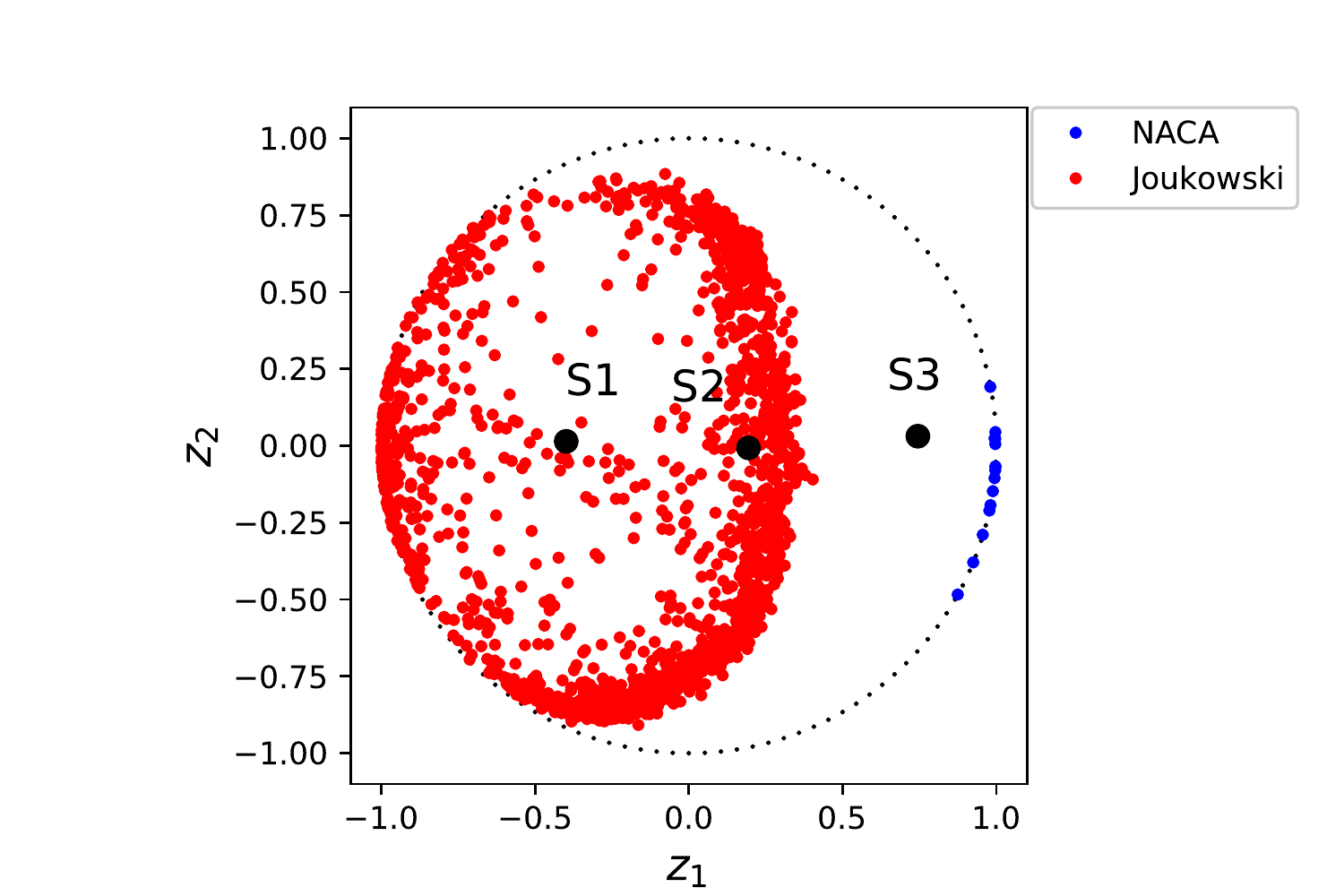}
				\par
				{(c) $z_3\geq 0$. }
			\end{center}
		\end{minipage}%
		\begin{minipage}[h]{0.5\textwidth}
			\begin{center}
				\includegraphics[width=\linewidth]{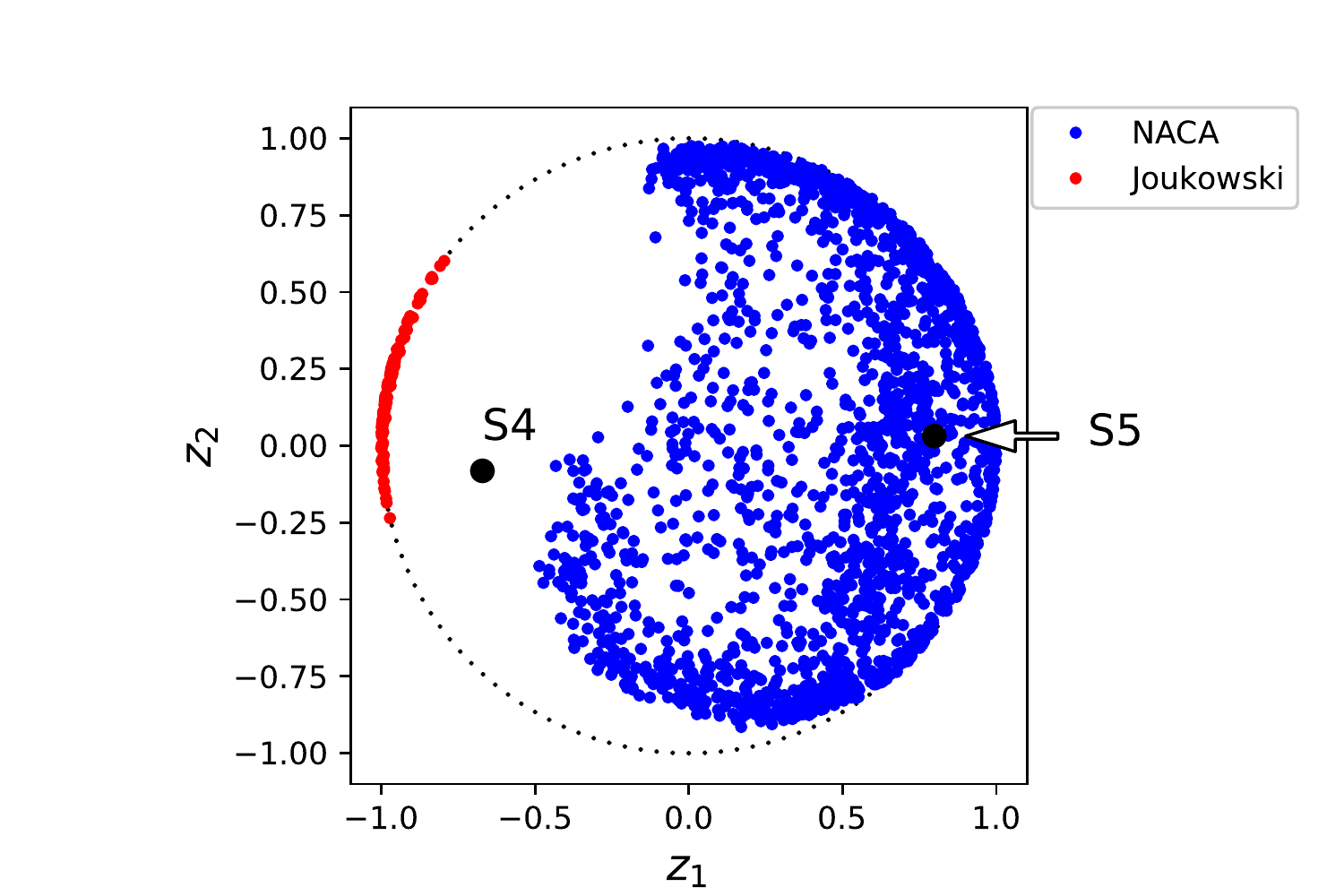}
				\par
				{(d) $z_3<0$. }
			\end{center}
		\end{minipage}%
		\caption{Plot of $w$ and embedded airfoil types ($\mathcal{S}$-CVAE).}
		\label{fig:latMap_mix}
	\end{center}
\end{figure}

\begin{figure}[htb]
	\begin{center}
		\begin{minipage}[h]{0.5\textwidth}
			\begin{center}
				\includegraphics[width=\linewidth]{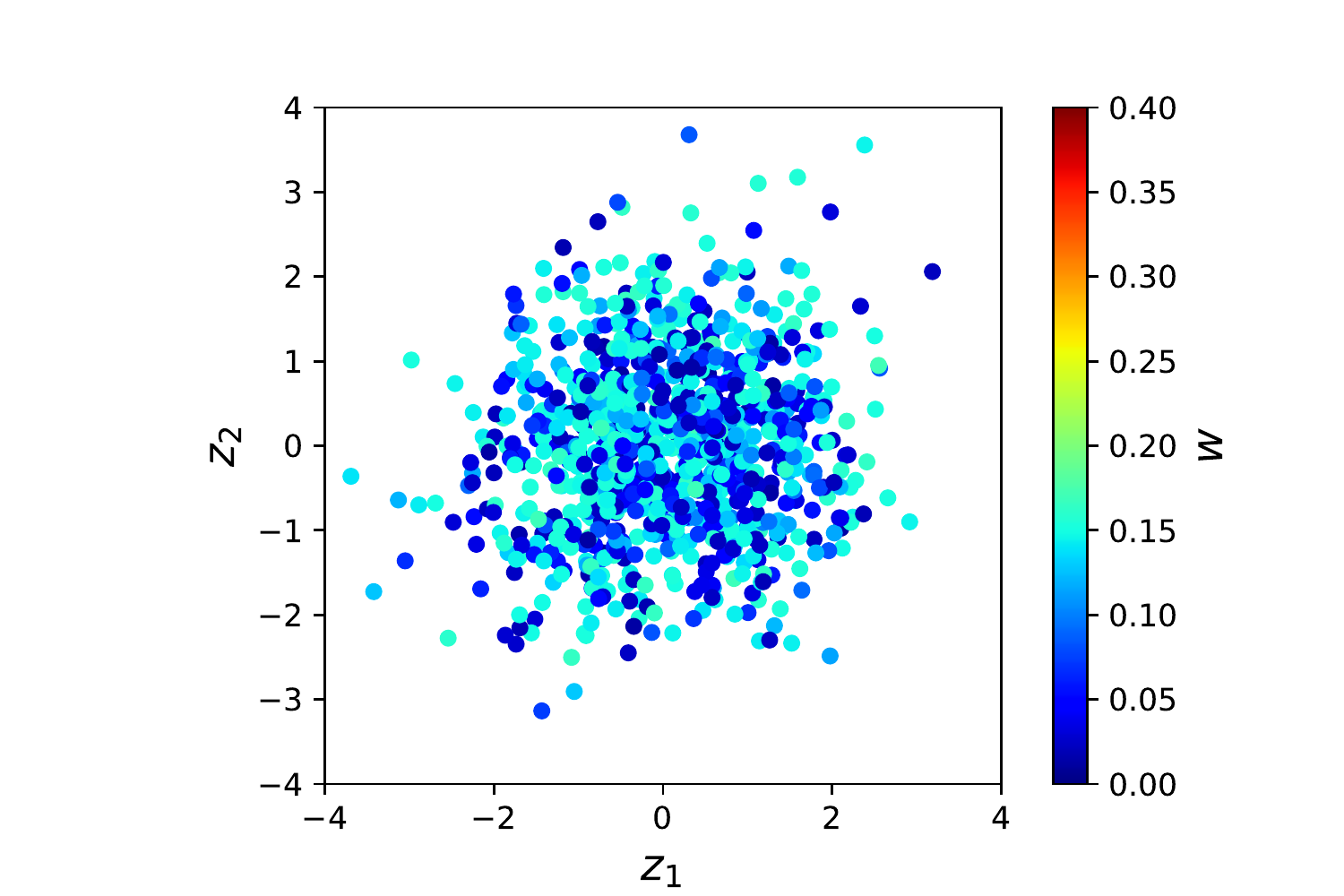}
				\par
				{(a) Plot of $w$ in latent space. }
			\end{center}
		\end{minipage}%
		\begin{minipage}[h]{0.5\textwidth}
			\begin{center}
				\includegraphics[width=\linewidth]{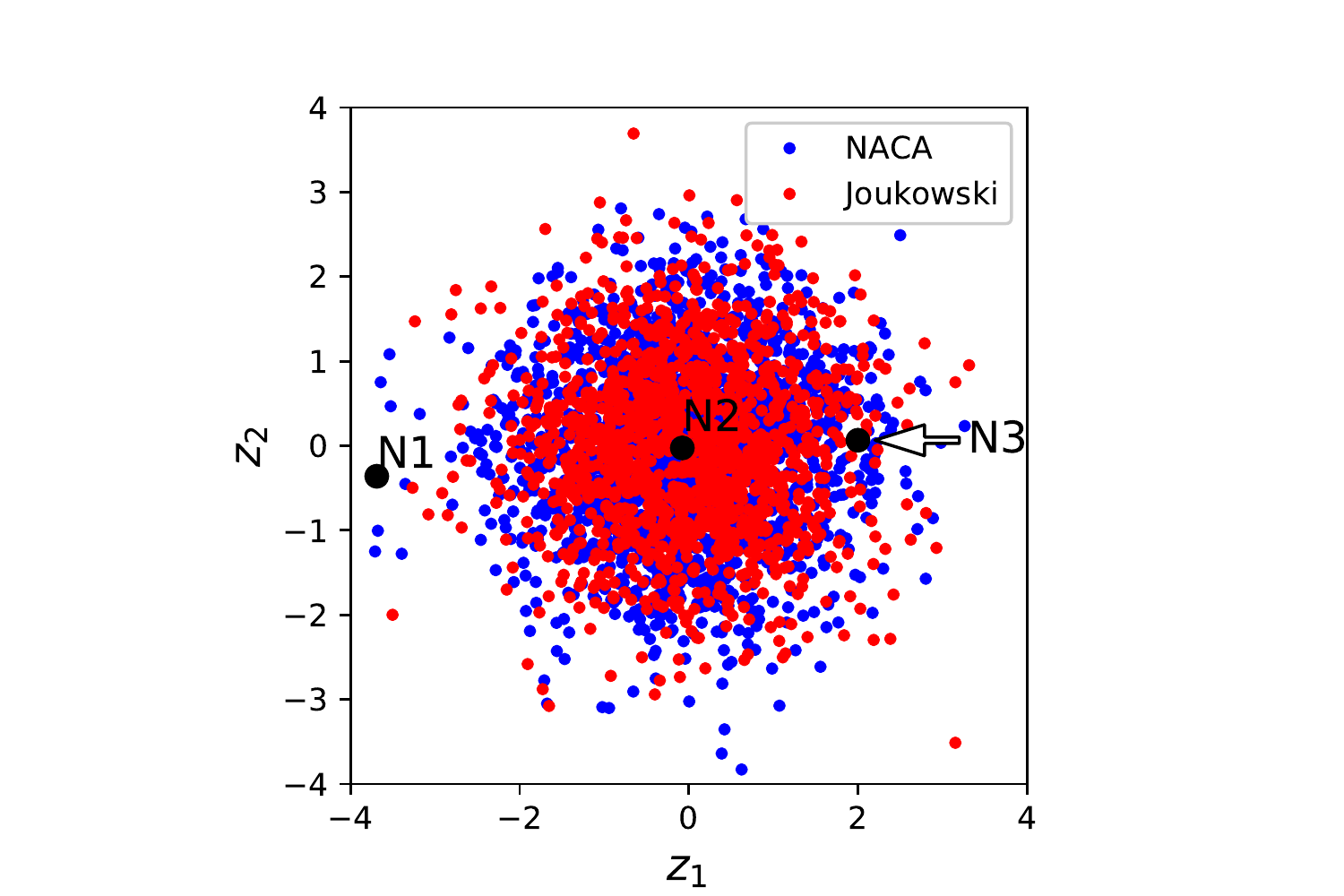}
				\par
				{(b) Embedded NACA and Joukowski data. }
			\end{center}
		\end{minipage}%
		\caption{Plot of $w$ and embedded airfoil types ($\mathcal{N}$-CVAE).}
		\label{fig:latMap_mix_N}
	\end{center}
\end{figure}

A plot of $w$ in the latent space is shown in \reffig{fig:latMap_mix} (a) and (b) and in \reffig{fig:latMap_mix_N} (a).
The points where the NACA and the Joukowski airfoils are embedded is shown in \reffig{fig:latMap_mix} (c) and (d) and \reffig{fig:latMap_mix_N} (b). 
In the $\mathcal{S}$-CVAE, the NACA airfoils and the Joukowski airfoils form clusters in the hypersphere. 
Conversely, in the $\mathcal{N}$-CVAE, the NACA and the Joukowski airfoils are combined. 
This difference corresponds to the capability of the $\mathcal{S}$-CVAE in the separation of the latent data, as observed in Section 4. 

\begin{figure}[htb]
	\begin{center}
		\begin{minipage}[h]{0.33\textwidth}
			\begin{center}
				\includegraphics[width=\linewidth]{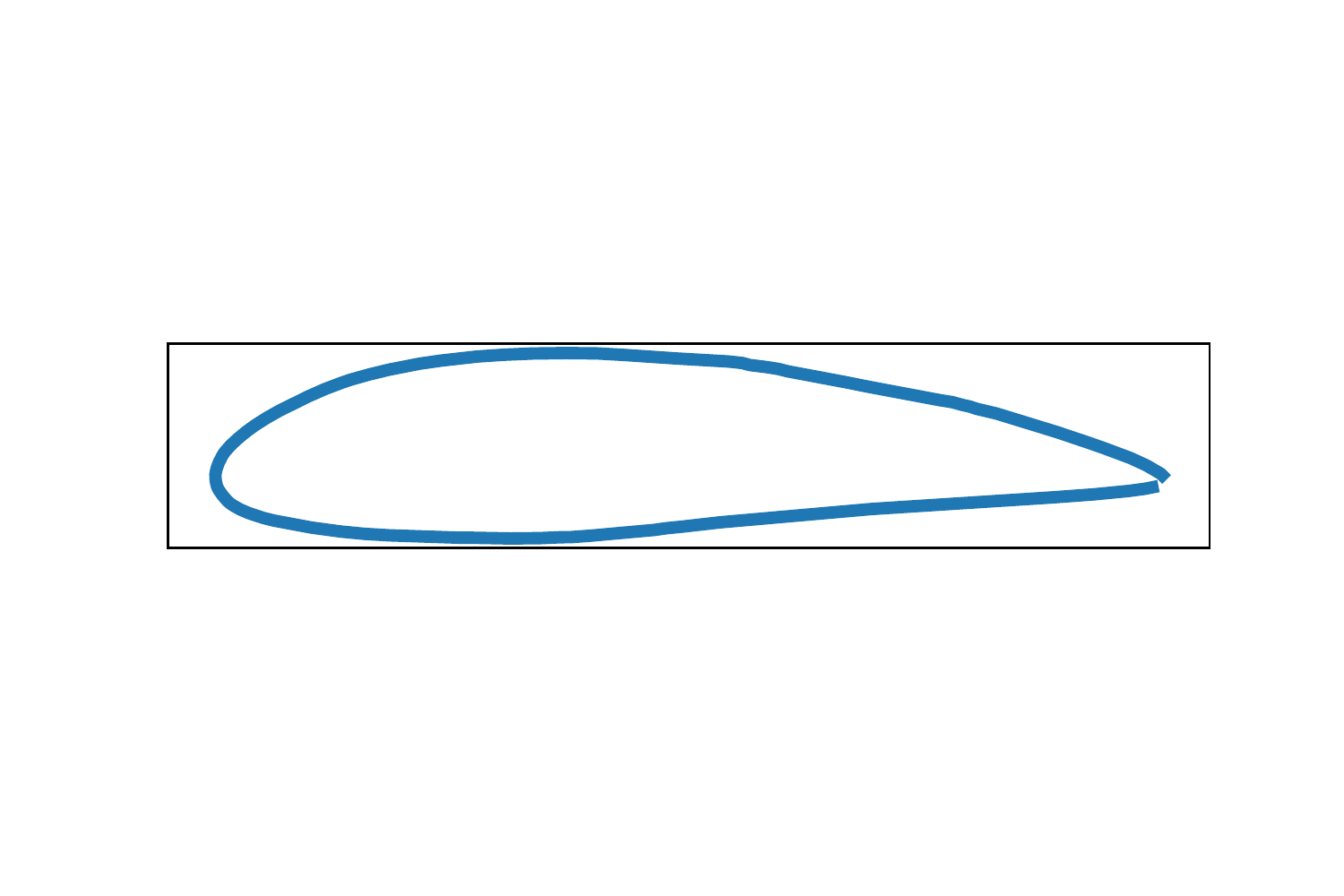}
				\par
				{(a) S1. }
			\end{center}
		\end{minipage}%
		\begin{minipage}[h]{0.33\textwidth}
			\begin{center}
				\includegraphics[width=\linewidth]{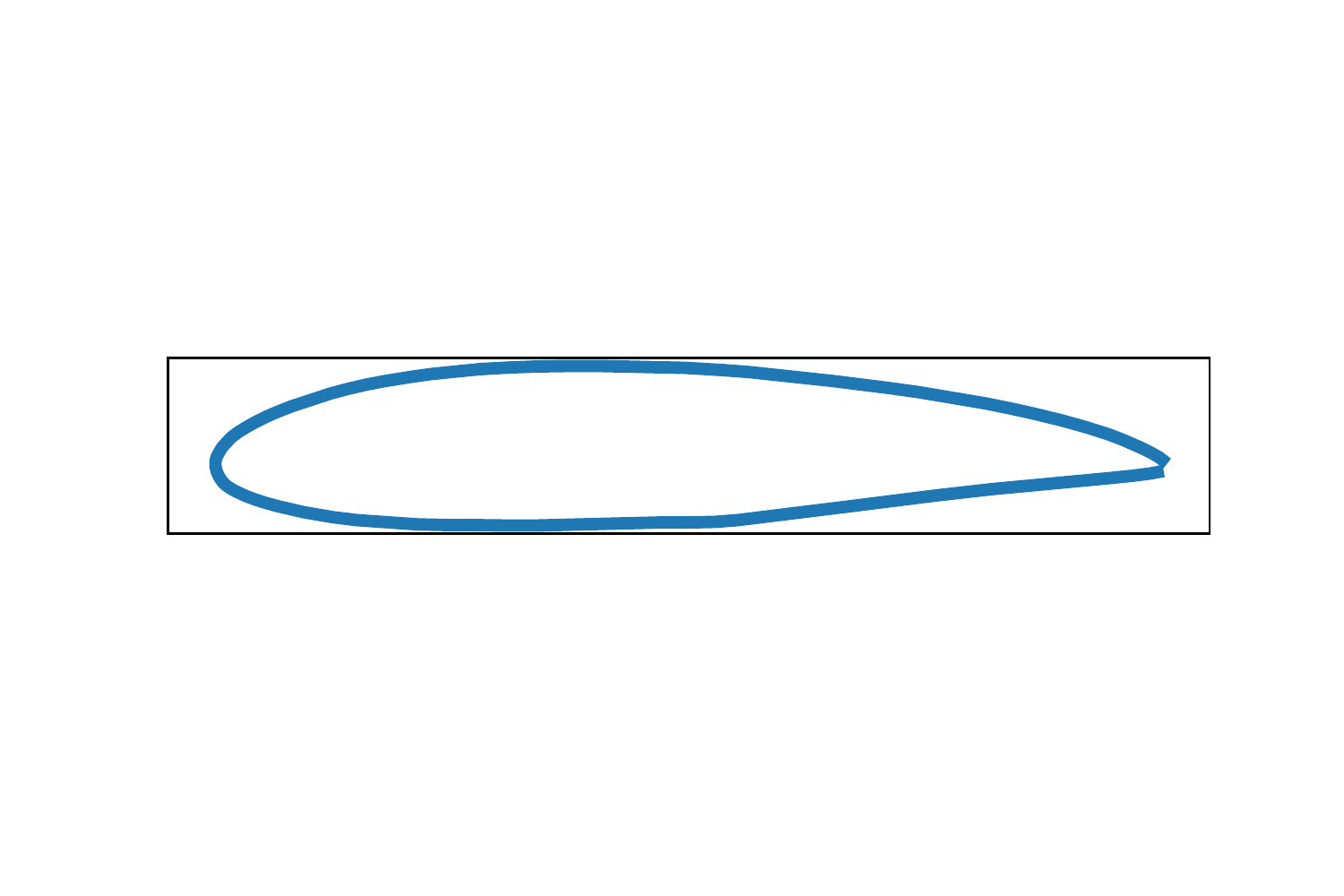}
				\par
				{(b) S2. }
			\end{center}
		\end{minipage}%
		\begin{minipage}[h]{0.33\textwidth}
			\begin{center}
				\includegraphics[width=\linewidth]{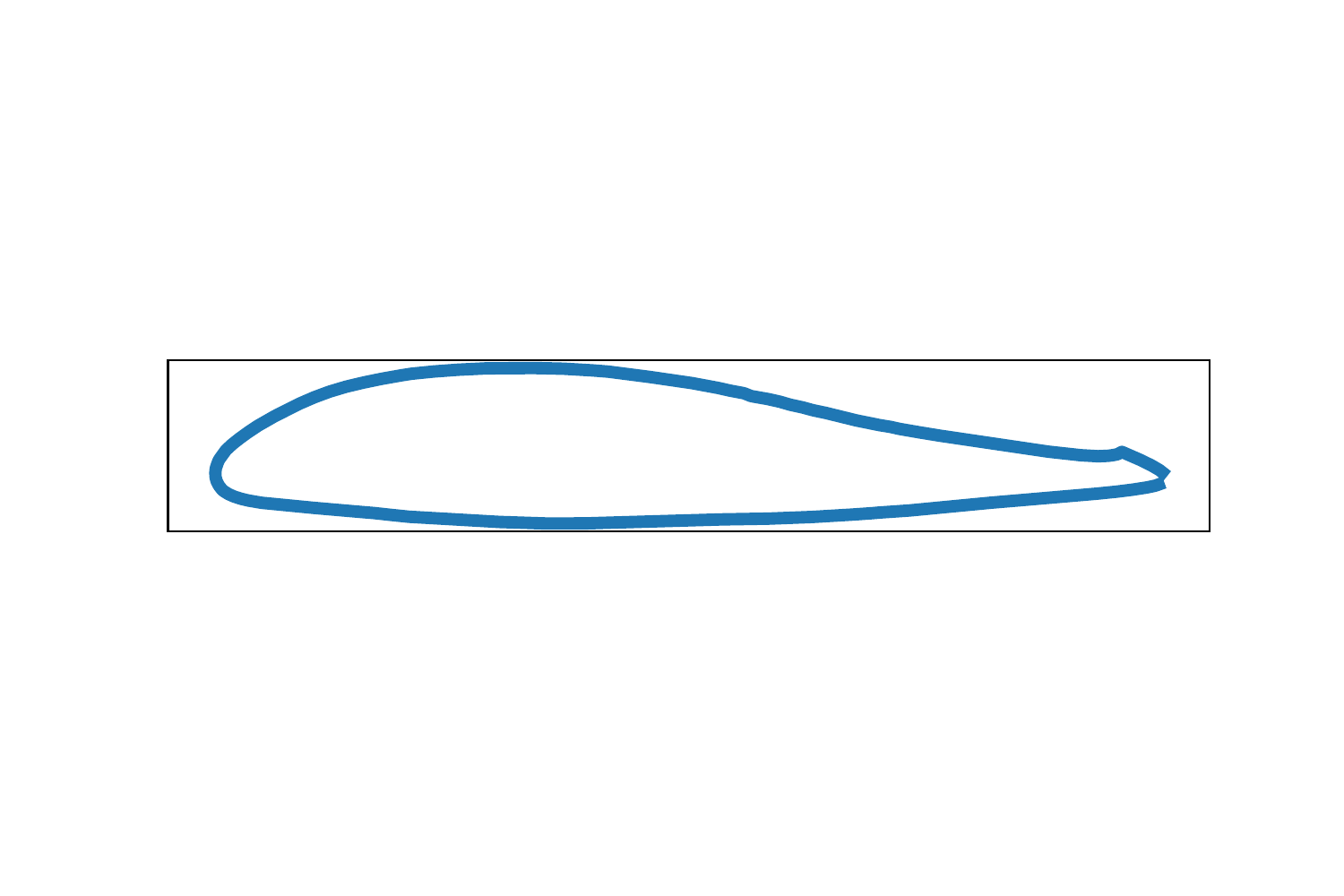}
				\par
				{(c) S3. }
			\end{center}
		\end{minipage}%
		\par
		\begin{minipage}[h]{0.3\textwidth}
			\begin{center}
				\includegraphics[width=\linewidth]{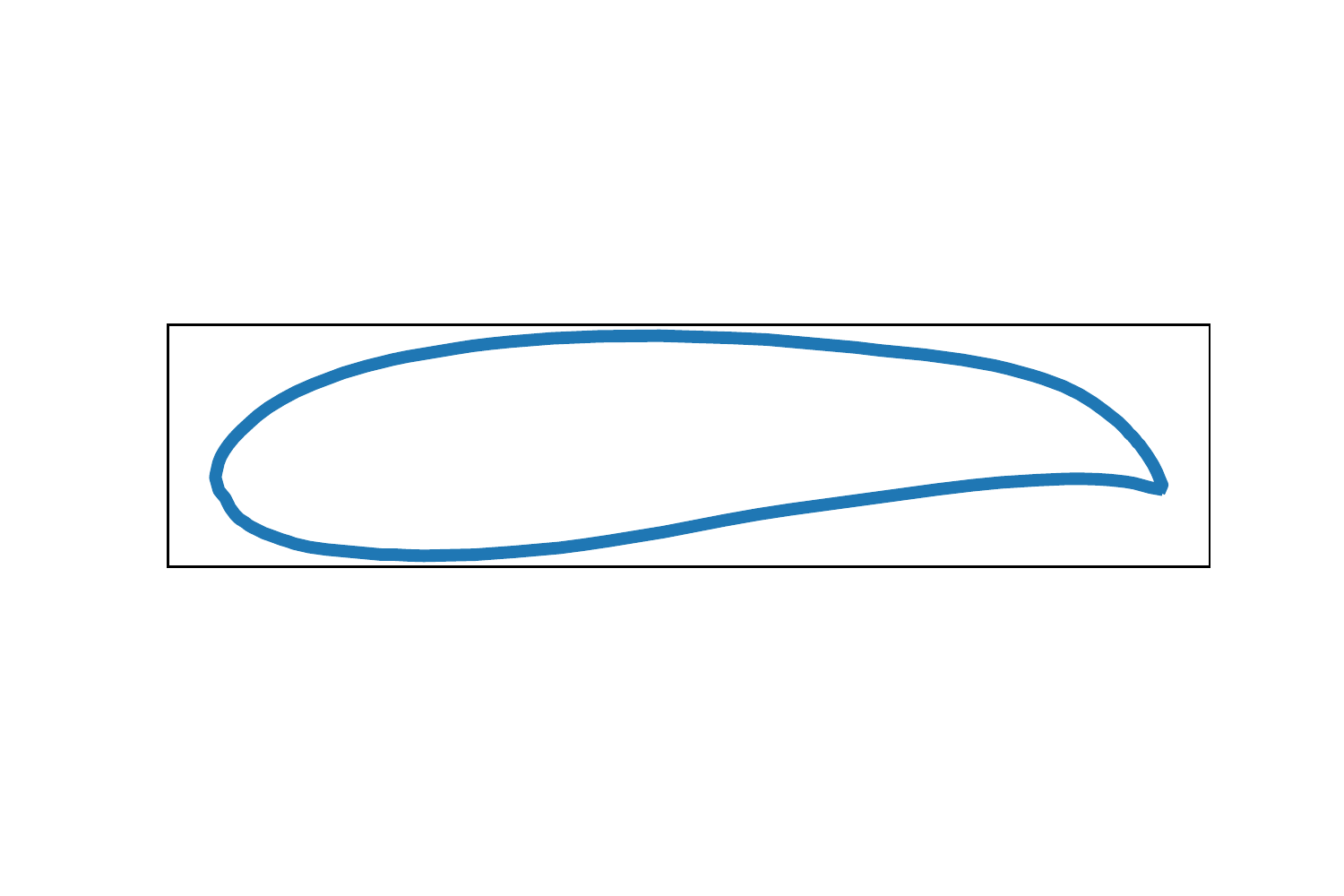}
				\par
				{(d) S4. }
			\end{center}
		\end{minipage}%
		\begin{minipage}[h]{0.3\textwidth}
			\begin{center}
				\includegraphics[width=\linewidth]{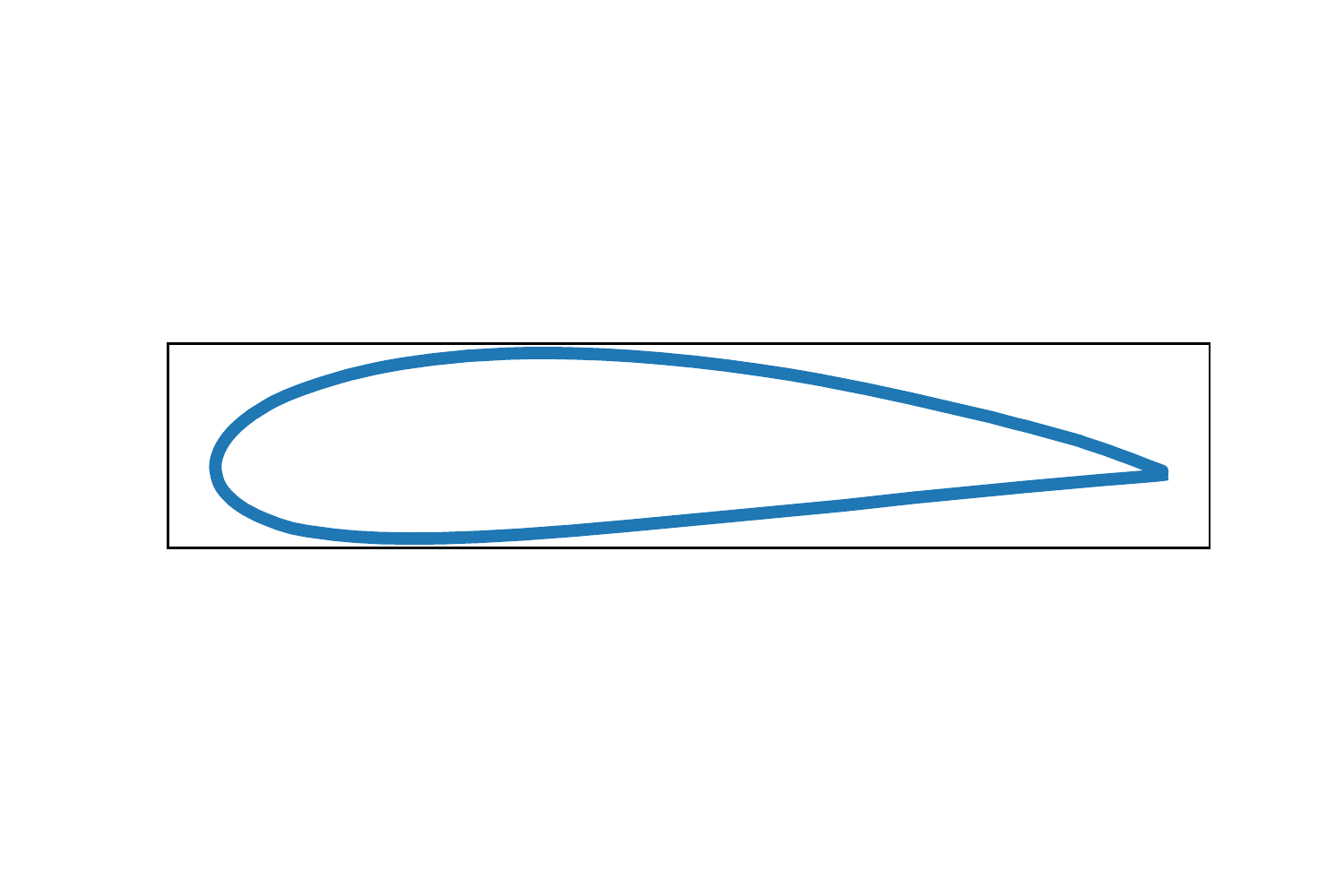}
				\par
				{(e) S5. }
			\end{center}
		\end{minipage}%
		\par
		
		\begin{minipage}[h]{0.33\textwidth}
			\begin{center}
				\includegraphics[width=\linewidth]{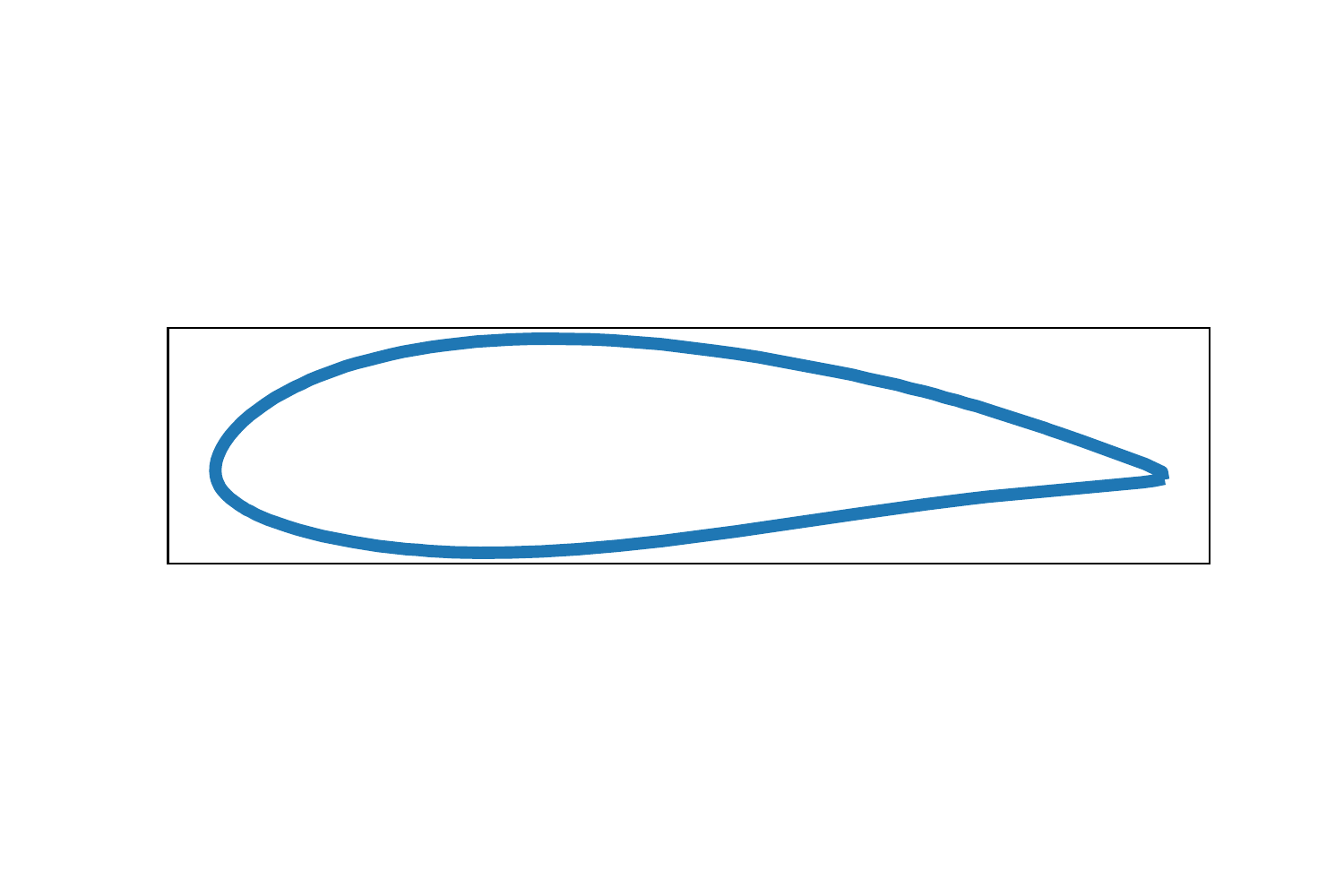}
				\par
				{(f) N1. }
			\end{center}
		\end{minipage}%
		\begin{minipage}[h]{0.33\textwidth}
			\begin{center}
				\includegraphics[width=\linewidth]{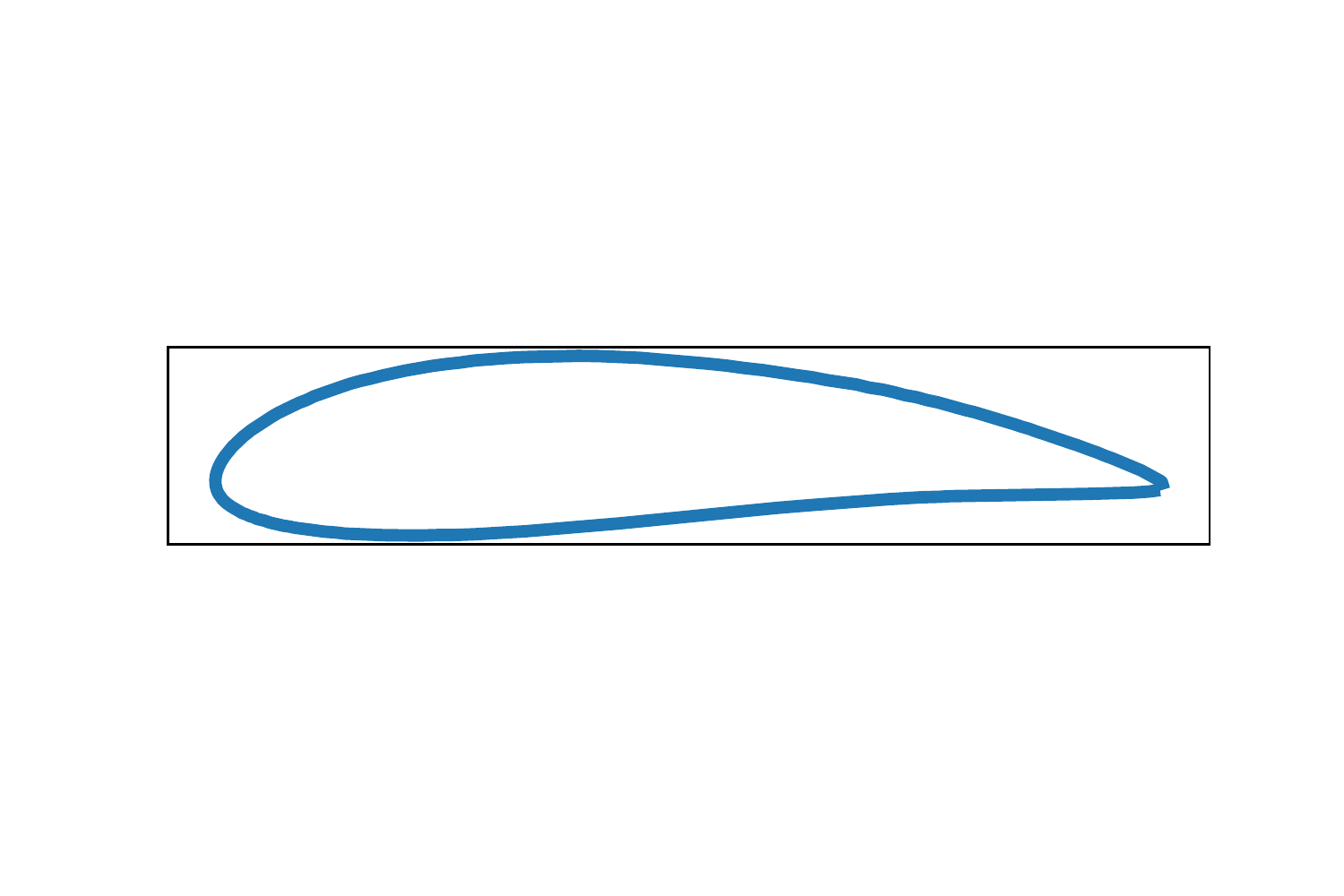}
				\par
				{(g) N2. }
			\end{center}
		\end{minipage}%
		\begin{minipage}[h]{0.33\textwidth}
			\begin{center}
				\includegraphics[width=\linewidth]{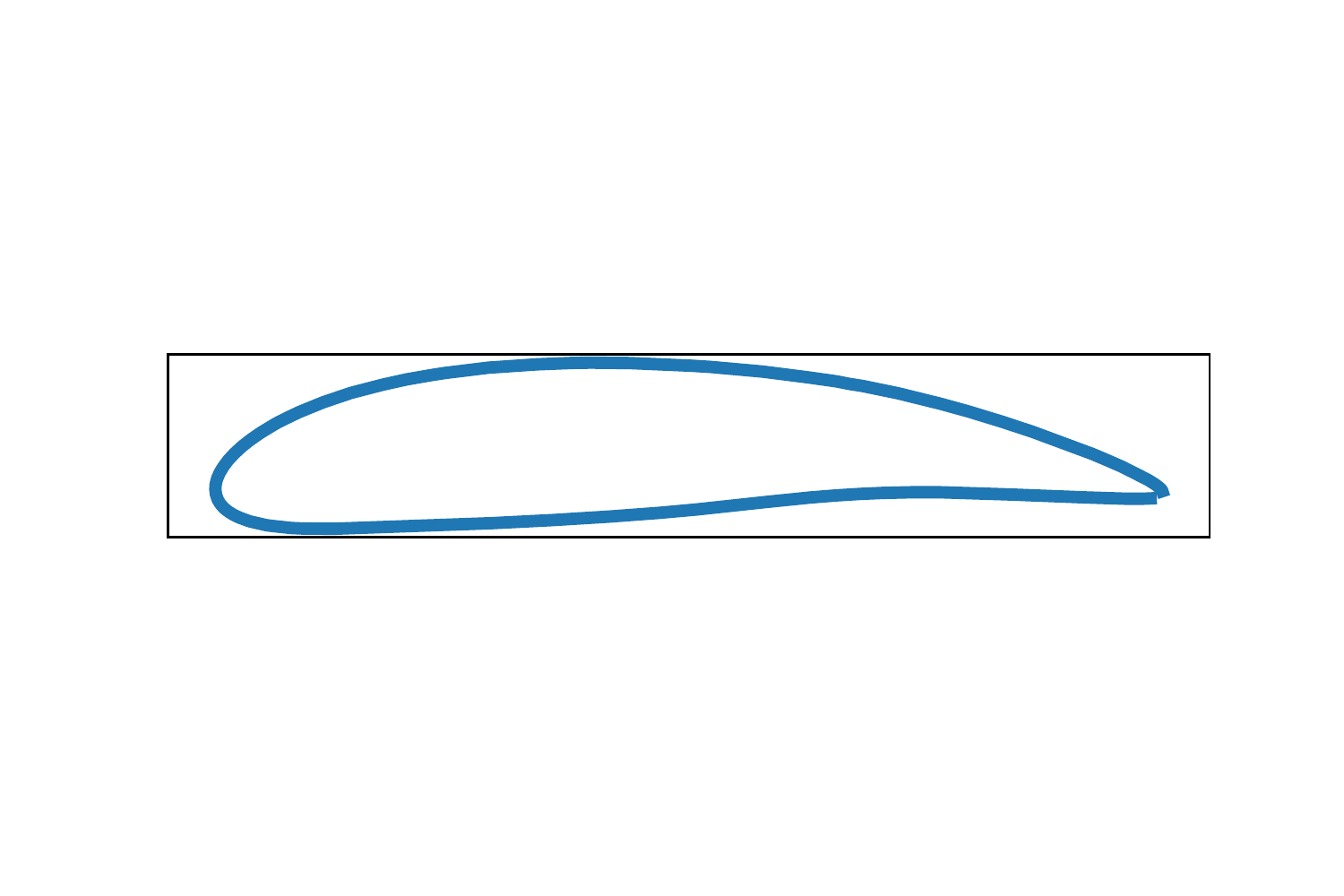}
				\par
				{(h) N3. }
			\end{center}
		\end{minipage}%
		\caption{Generated shapes from different latent vectors.}
		\label{fig:MixShapes}
	\end{center}
\end{figure}

In $\mathcal{S}$-CVAE, the roundness, $w$, is relatively high (i.e., larger than 0.4) in the area where neither the NACA nor the Joukowski airfoils are embedded, and relatively low, that is, smaller than 0.1, in the area where the Joukowski airfoils are embedded.
Conversely, $w$ is randomly distributed in the latent space in $ \mathcal{N}$ - CVAE. 
The shapes generated from the latent vectors S1--S5 and N1--N2 are shown in \reffig{fig:MixShapes}. The locations of the latent vectors are shown in \reffig{fig:latMap_mix} and \reffig{fig:latMap_mix_N}. 
S1 and S2 indicate a small value of $w$ because they are decoded from the Joukowski area, whereas S5 indicates a moderate value of $w$ because it is decoded from the NACA area. 
Both cases produce appropriate airfoil shapes. 
However, S3 and S4 form wired shapes because they are decoded from the areas where no training data are embedded. 
Therefore, N1--N3 produce appropriate shapes. 


Consequently, $\mathcal{S}$-CVAE is more capable of separating data, and hence it embeds the NACA and the Joukowski airfoils in separate areas. 
The generated shapes are also separated into two types: the NACA-like shapes and the Joukowski-like shapes. 
However, the $\mathcal{N}$-CVAE lacks the capability of data separation. Therefore, $\mathcal{N}$-CVAE embeds both the NACA and the Joukowski airfoils in the same area in the latent space, due to which, the decoder must generate intermediate shapes between the NACA and the Joukowski shapes.

\section{Conclusion}
This study proposed the CVAE based method to generate a variety of shapes. 
Two types of CVAE models, namely, the $\mathcal{S}$-CVAE and the $\mathcal{N}$-CVAE models, were analyzed to train and generate the airfoil data. 
Two tasks were set for this analysis. The first task involves the training and the generation of the NACA airfoil data.
The main objective is to generate a wide variety of NACA-like airfoils. 
The other task is to use the NACA and the Joukowski airfoils for the training and generation of novel shapes which are different from both the NACA and the Joukowski airfoils. 

The $\mathcal{S}$-CVAE model is superior to the $\mathcal{N}$-CVAE models when applied to single type of airfoil, as it separates the data in the latent space. 
However, if different types of data are to be combined, for example, the NACA airfoil and the Joukowski airfoil, $\mathcal{S}$-CVAE outputs only the NACA-like airfoils and the Joukowski-like airfoils, and does not output the feature-mixed airfoils. The generated shapes are nearly identical to the training data and are not novel shapes. 
Conversely, the $\mathcal{N}$-CVAE model combines different features and generates novel shapes. This is due to the lack of the data separation capability.

%\begin{acknowledgements}
%\end{acknowledgements}
\clearpage
%\afterpage{\clearpage}

\section*{A.1 Joukowski inverse transformation}
In this section, a Joukowski inverse transformation of the given airfoil shape, $\hat{\zeta}$.
The generated shapes by CVAE model is normalized by \eqref{eq.scale}. 
Hence, at first the shape is de-normalized as
\begin{align*}
	{\rm Re}(\zeta) = {\rm Re}(\hat{\zeta}) m+ \ell, \quad
	{\rm Im}(\zeta) = {\rm Im}(\hat{\zeta}) m
\end{align*}
The Joukowski inverse transformation is derived from Equation \refeq{eq.j} as 
\begin{align}
	\zeta =  z + \frac{c_{\rm J}^2}{z} 
	&\Leftrightarrow 
	z^2 - \zeta z + c_{\rm J}^2 = 0  \notag \\
	&\Leftrightarrow
	z = \frac{1}{2} (\zeta \pm \sqrt{\zeta^2 - 4 c_{\rm J} } ). \label{eq.inv}
\end{align}
In equation \eqref{eq.inv}, $c_{\rm J}$ is a function of $a$, $b$, and $r$, and cannot be determined from $\zeta$. Additionally, the scale factors, $\ell$ and $m$, in \refeq{eq.scale} are unknown. 
Hence, $c_{\rm J}$, $\ell$, and $m$, which define the most suitable inverse transformation, are searched. 

For a given set of points, $z_i$ $(i=1,\dots,N)$, the inverse transformation yields a set of points, $\zeta_i$. 
If the appropriate $c_{\rm J}$, $\ell$, and $m$, are chosen, $\zeta_i$ forms two circles. The Joukowski transformation generates the same airfoil from both the circles.
However, if $c_{\rm J}$, $\ell$, and $m$, $\zeta_i$ are not appropriately chosen, $z_i$ does not form the circles.
Hence, $c_{\rm J}$, $\ell$, and $m$ are optimized by minimizing the mean squared error from a circle denoted by $\omega$.
From $x_i$ and $y_i$, which are the real and imaginary parts of $z_i$, that is, $z_i = x_i+iy_i$, $\omega$ is defined by
\begin{align}
	\omega 
	&= \sum_i \left( \left( x_i -a \right)^2 + \left(y_i-b\right)^2 - r^2 \right)^2 \notag \\
	&= \sum_i \left( x_i^2 + y_i^2 + Ax_i + By_i + C\right) ,
\end{align} 
where $a$ and $b$ are the centers of the circles, and $r=1.1$ is the radius defined in Section \ref{sec.J}, $A = -2a$, $B=-2b$, and $C=a^2 + b^2 - r^2$. 
Because $\Omega$ is a convex function with respect to $a$, $b$, and $r$, the global optimal solution can be obtained by solving the stationary condition, $\frac{\partial \omega}{\partial A} = 2x_i \omega$, $\frac{\partial \omega}{\partial B} = 2y_i \omega$, $\frac{\partial \omega}{\partial C} = 2 \omega$. 
$A$, $B$, and $C$ are calculated from the stationary conditions as:
\begin{align*}
	\left(
	\begin{array}{c}
		A \\ B \\ C
	\end{array}
	\right)
	=
	\left(
	\begin{array}{ccc}
		\sum_i x_i^2 & \sum_i x_i y_i & \sum_i x_i \\ \sum_i x_i y_i & \sum_i y_i^2 & \sum_i y_i \\
		\sum_i x_i & \sum_i y_i & n
	\end{array}
	\right)^{-1}
	\left(
	\begin{array}{c}
		-\sum_i \left( x_i^3 - x_i y_i^2 \right) \\ 
		-\sum_i \left( x_i^2 y_i - y_i^3 \right) \\
		-\sum_i \left( x_i^2 - y_i^2 \right)
	\end{array}
	\right) .
\end{align*}
Therefore, $ \omega$ is calculated from $z$ and the calculation is denoted as $ \omega = f( z ) $. 

The parameters, $c_{\rm J}$, $\ell$, and $m$ which minimize $\omega$ are located.
The optimization problem is given as:
\begin{alignat*}{2}
	&{\rm given} &{\qquad}& \bi{\hat{\zeta}} \\
	&{\rm minimize}_{c_{\rm J},~\ell,~m,} && \omega = f( \bi{z}), \\
	&{\rm such~that} && {\rm Re}\left({\zeta_i} \right) = {\rm Re} \left(\hat{\zeta_i} \right)m + \ell, \\
	&&& {\rm Im} \left({\zeta_i} \right) = {\rm Im} \left(\hat{\zeta_i} \right) m, \\
	&&& z_i = \frac{1}{2} \left({\zeta_i} \pm \sqrt{{\zeta_i}^2 - 4 c_{\rm J} } \right). 
\end{alignat*}
$w$ is the minimum mean squared error of the inverse-transformed shapes from any circle. Hence, $w$ is equivalent to  roundness of the inverse-transformed shapes.


	\bibliographystyle{unsrt}  
	\bibliography{bib-topologyopt,bib-AM,bib-DDD}

\end{document}